\documentclass[nofootinbib,aps,prd,twocolumn,showpacs,notitlepage,amsmath,amssymb]{revtex4-2}

\usepackage[colorlinks=true,citecolor=blue,linkcolor=blue,urlcolor=blue]{hyperref}
\usepackage{cleveref}[2012/02/15]
\crefformat{footnote}{#2\footnotemark[#1]#3}

\newcommand{\2}{$\mathbb Z_2$}
\newcommand{\f}{{f\textrm{-}f}}
\newcommand{\e}{{e\textrm{-}e}}

\makeatletter
\def\l@subsubsection#1#2{}
\makeatother

\usepackage{amsmath,amssymb,bm}

\usepackage{tikz}
\usetikzlibrary{external}
\usetikzlibrary{decorations.pathreplacing}
\usetikzlibrary{decorations.text}
\usetikzlibrary{decorations.pathmorphing}
\usetikzlibrary{shapes}

\definecolor{newblue}{RGB}{112,178,255}
\definecolor{neworange}{RGB}{255,204,112}
\definecolor{blue2}{RGB}{120,0,255}
\definecolor{red2}{RGB}{255,0,120}
\definecolor{green2}{RGB}{0,130,130}

\tikzset{snake it/.style={decorate, decoration={snake,segment length=1mm, amplitude=0.5mm}}}

\definecolor{darkred}{RGB}{245,186,183}
\definecolor{lightred}{RGB}{249,217,215}

\usetikzlibrary{arrows,calc,external,shapes.geometric}

\tikzset{
	>=stealth',
	help lines/.style={dashed, thick},
	important line/.style={thick},
	connection/.style={thick, dotted},
}

\tikzstyle{A}=[circle,draw=red!50,fill=red!20,thick]
\tikzstyle{R}=[circle,draw=blue!50,fill=blue!20,thick]
\tikzstyle{U}=[circle,draw=green!50,fill=green!20,thick]
\tikzstyle{V}=[circle,draw=orange!50,fill=orange!20,thick]

\def\bra#1{\mathinner{\langle{#1}|}}
\def\ket#1{\mathinner{|{#1}\rangle}}

\usepackage[normalem]{ulem}

\begin{document}

\title{Unifying Kitaev magnets, kagom\'e dimer models and ruby Rydberg spin liquids}

\author{Ruben Verresen and Ashvin Vishwanath}

\affiliation{Department of Physics, Harvard University, Cambridge, MA 02138, USA}

\begin{abstract}
The exploration of quantum spin liquids (QSLs) has been guided by different approaches including the resonating valence bond (RVB) picture, deconfined lattice gauge theories and the Kitaev model. More recently, a spin liquid ground state was numerically established on the ruby lattice, inspired by the Rydberg blockade mechanism. Here we unify these varied approaches in a single parent Hamiltonian, in which local fluctuations of anyons stabilize deconfinement. The parent Hamiltonian is defined on kagom\'e triangles---each hosting four RVB-like states---and includes only Ising interactions and single-site transverse fields. In the weak-field limit, the ruby spin liquid and exactly soluble kagom\'e dimer models are recovered, while the strong-field limit reduces to the Kitaev honeycomb model, thereby unifying three seemingly different approaches to QSLs. We similarly obtain the chiral Yao-Kivelson model, honeycomb toric code and a new spin-1 quadrupolar Kitaev model. The last is shown to be in a QSL phase by a non-local mapping to the kagom\'e Ising antiferromagnet. We demonstrate various applications of our framework, including (a) an adiabatic deformation of the ruby lattice model to the exactly soluble kagom\'e dimer model, conclusively establishing the QSL phase in the former; 
and (b) demystifying the dynamical protocol for measuring off-diagonal strings in the Rydberg implementation of the ruby lattice spin liquid. More generally, we find an intimate connection between Kitaev couplings and the repulsive interactions used for emergent dimer models. For instance, we show how a spin-1/2 XXZ model on the ruby lattice encodes a Kitaev honeycomb model, providing a new route toward realizing the latter in cold-atom or solid-state systems.
\end{abstract}

\date{\today}

\maketitle

\tableofcontents

\newpage

\section{Introduction \label{sec:introduction}}

The study of quantum spin liquids lies at the intersection of quantum magnetism, lattice gauge theories and quantum computing and has key implications for all three fields. Beginning with the seminal idea of resonating valence bond (RVB) liquids \cite{Anderson}, early theoretical work  connected this picture to the notion of a gapped spin liquid \cite{WenBook} and made connections to parallel work on \2 lattice gauge theories \cite{Wegner71,FradkinShenker}. Such \2 spin liquids and related states have been studied as models of frustrated magnetism  \cite{SpinLiquidsReviewBalents, Read91,Wen,Sachdev_Triangle,ChakrabortyRead} and other strongly correlated materials \cite{Anderson87,BASKARAN,Kivelson_1987,Baskaran88,Affleck88b,Sachdev_1991,Read89,RK,Kivelson_1987,Balents98,SenthilFisher}. A complementary viewpoint emphasizes the long-range quantum entanglement and topological order \cite{Witten89, Wen89, Wen_90,Kitaev_2003,Hamma05,Kitaev06b,Levin06} of gapped quantum liquids. This has important implications for  quantum error correction and the quest to build a fault-tolerant quantum computer \cite{Nayak_RMP,Terhal_2015}. In fact, the degenerate ground states of the $\mathbb Z_2$ topological order underlie the `toric code' \cite{Kitaev_2003,Kitaev06} and `surface code' \cite{Fowler_2012} models for topologically protected quantum memory. 

However, despite these strong motivations and  years of theoretical and experimental effort, and notwithstanding multiple promising candidate materials \cite{SpinLiquidsReviewBalents,Broholmeaay0668}, a clear-cut realization of \2 topological order has yet to be demonstrated in a solid-state material. Recently, progress has been reported by approaching this problem from an entirely different angle, utilizing synthetic quantum platforms such as Rydberg atom arrays \cite{Endres16,Browaeys_2020} and near-term quantum devices to engineer \2 topological order \cite{Verresen21,Semeghini21,Satzinger21}. While these and other proposals \cite{Glaetzle_2014,Samajdar_2021,PhysRevX.10.021057} continue to be explored, new theoretical approaches for constructing topologically ordered states \cite{Myerson_Jain_2022,measurement_rv,Slagle22,Cuadra22,RS-SS,Cheng21,Giudici22dyn,Z3} 
are also being actively pursued. This renewed interest in \2 spin liquids and their realization in new platforms calls for a deeper understanding of the landscape of models stabilizing these phases. 

{\it Landscape of models.} Focusing on models realizing \2 spin liquids, there are four approaches worth recalling. First is the paradigmatic toric code model \cite{Kitaev_2003}, which however requires multi-body interactions and is challenging to directly realize. Next are dimer models on non-bipartite lattices \cite{fradkin_2013,Sachdev92}, such as the kagom\'e \cite{Misguich02} or triangular \cite{Moessner_2001} lattice. The dimers themselves may either represent singlets between spins on neighboring sites, or, be an intrinsic degree of freedom where the dimer constraint is enforced \cite{RK,Moessner_2001} or emerges from interactions \cite{BFG,Sheng05,Roychowdhury15}. Third, there are 
Kitaev spin models \cite{Kitaev06} including the honeycomb lattice $S=1/2$ model and related constructions \cite{Yao07,Baskaran09,Yao09,Yao11,Chua11,Whitsitt12,Natori16,Natori17,deCarvalho18,Natori18,Seifert20,Farias20,Natori20,Chulliparambil20,Ray21,Jin21,Chulliparambil21}, which can be solved by mapping to non-interacting fermions. In all these previous cases, one is guaranteed a spin liquid phase from analytical arguments. However, in some instances, not just the ground state but every excited state can be analytically obtained (toric code \cite{Kitaev_2003} or kagom\'e dimer liquid \cite{Misguich02}) while in other instances (triangular dimer model at the Roksar-Kivelson point \cite{RK,Moessner_2001}) only the ground state is  analytically accessible.  Finally, there are a handful of models that require full-fledged  numerics to establish their phase diagrams. This includes the ruby lattice spin liquid with Ising or PXP type interactions \cite{Verresen21} and the spin-1 Kitaev model \cite{Baskaran08,Rousochatzakis_2018,Chen_2022}. 

These various models may seem unrelated---not only because they reside on different lattices, but also because the approach to solving them varies substantially. Furthermore, one typically attributes the spin liquid in these different models to different kinds of frustration---`geometrical' frustration, in the case of non-bipartite dimer models, in contrast to spin-orbit induced `exchange' frustration in Kitaev magnets \cite{Kimchi_Review}. Despite these apparent differences, we show that there are a number of hidden relations between these various models that are exposed by embedding them in a larger parent model. 

\begin{figure}
    \centering
    \includegraphics[scale=0.65]{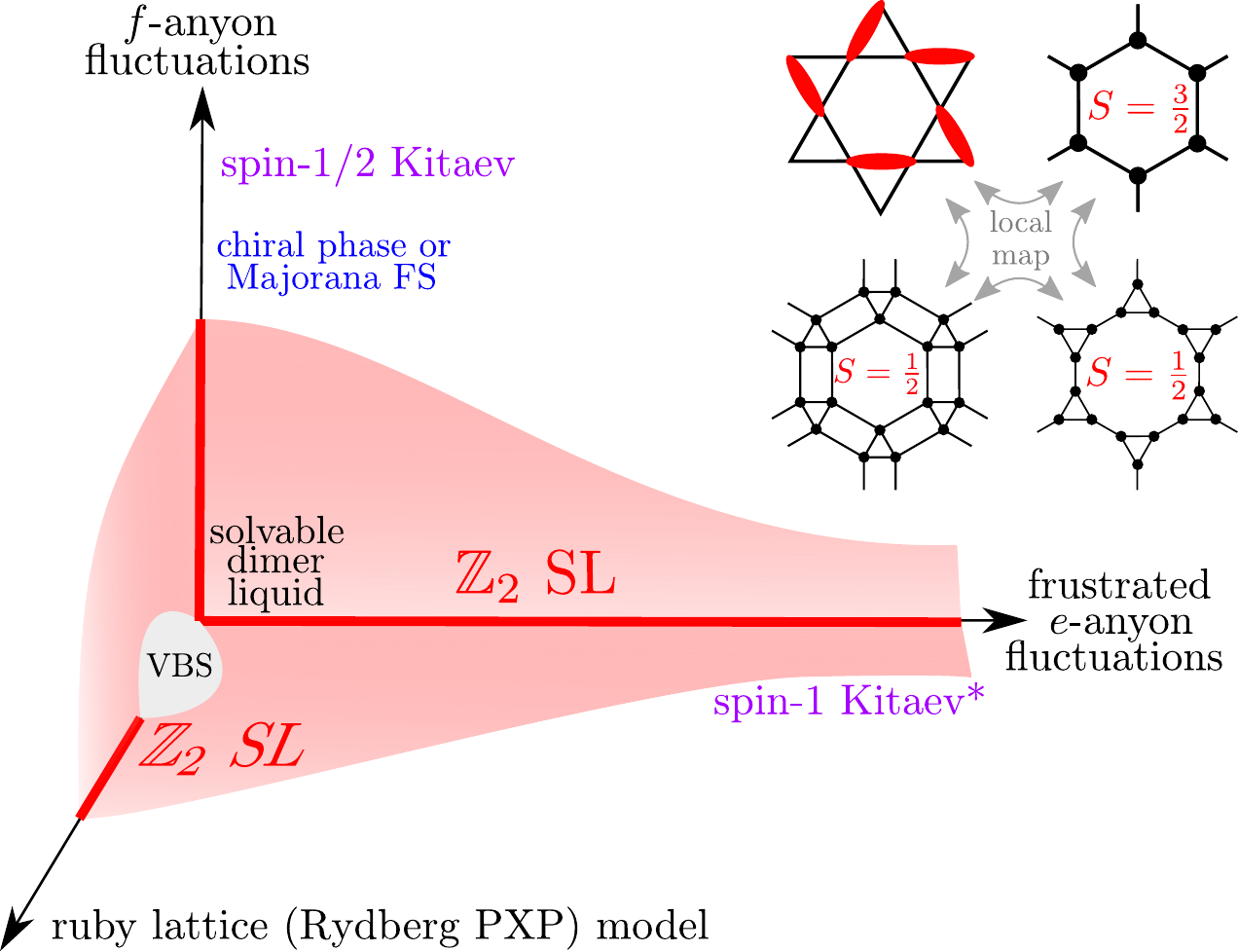}
    \caption{\textbf{Schematic phase diagram.} We consider an emergent dimer model on the kagom\'e lattice which can locally be written as a 4-state model on the honeycomb lattice (Sec.~\ref{sec:model}), or two-body spin-1/2 models on the ruby or star lattices (Sec.~\ref{sec:equivalence}). Weak $e$- or $f$-anyonic fluctuations generate a solvable dimer liquid similar to the toric code \cite{Misguich02}. Strong fluctuations give effective Kitaev-type Hamiltonians: in the case of $f$-fluctuations this is the paradigmatic spin-$1/2$ Kitaev honeycomb model, whereas for (frustrated) $e$-fluctuations we discover a new spin-1 quadrupolar Kitaev liquid. The $f$-fluctuating model is free-fermion solvable and contains a non-Abelian phase or Majorana Fermi surface (depending on a choice of sign); the $e$-fluctuating model is dual to the frustrated Ising model on the kagom\'e lattice, showing that the $\mathbb Z_2$ spin liquid (SL) persists to the spin-1 Kitaev limit. Finally, our parent model also contains the ruby lattice Ising model with Rydberg blockade \cite{Verresen21}, and we numerically confirm that its SL is adiabatically connected to the aforementioned solvable dimer liquid.}
    \label{fig:schematic}
\end{figure}

\emph{Deconfinement from fluctuations.} A common theme running through this paper is that anyon fluctuations, in certain settings, provide a route to deconfinement. This may seem paradoxical, since it is known that anyon fluctuations beyond a threshold lead to anyon condensation (and hence confinement of all other anyons that braid non-trivially with it). However, in a recent work of a model of Rydberg atoms on the ruby lattice \cite{Verresen21}, an intermediate strength of $e$-anyon fluctuations was shown to {\em stabilize} the toric code topological order by melting a valence bond solid. Indeed, the present work can be viewed as exploring other contexts where such fluctuations can be helpful, and how the outcomes depend on the choice of the fluctuating anyon. In fact, the parent model described below has a fixed two-body interaction term that selects the relevant constrained Hilbert space, but the different transverse magnetic field terms correspond to choosing distinct anyon fluctuations.

\emph{Parent model.} We study a simple emergent dimer model on the kagom\'e lattice. We treat triangles as the elementary unit, each of which can host four distinct states: $\raisebox{-2pt}{\includegraphics[scale=0.47]{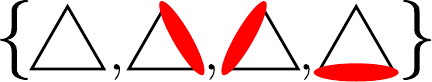}}$.
As a concrete implementation, such an effective 4-level state, or `spin-3/2', can naturally emerge from, e.g., blockade interactions \cite{Jaksch00,Lukin01,Gaetan09,Urban09}, such as in the recent Rydberg atom array experiment \cite{Semeghini21} where each red dimer represents an excited atom\footnote{This 4-level approximation is particularly justified in such a setting since the distance-dependent $\sim 1/r^6$ interactions are exceedingly large within triangles.}. (We discuss possible underlying spin-1/2 models in Sec.~\ref{sec:equivalence}.) Our Hamiltonian is then Ising-like: diagonal two-body interactions energetically enforce the dimer constraint between kagom\'e triangles, and a single-site field introduces ultralocal fluctuations of the anyons in a dimer model. In fact, the previously-studied case of the ruby lattice spin liquid \cite{Verresen21} corresponds to the following choice of transverse field in this 4-state system:
\begin{equation}
\mathcal X_\textrm{PXP on ruby lattice} = \left( \begin{array}{cccc}
0 & 1 & 1 & 1 \\
1 & 0 & 0 & 0 \\
1 & 0 & 0 & 0 \\
1 & 0 & 0 & 0
\end{array}\right). \label{eq:Xintroruby}
\end{equation}
Its reported spin liquid ground state was based on numerical simulations. However, in this work we show that a simple modification,
\begin{equation}
\mathcal X_\textrm{$e$-anyon fluctuations} = \left( \begin{array}{cccc}
0 & 1 & 1 & 1 \\
1 & 0 & 1 & 1 \\
1 & 1 & 0 & 1 \\
1 & 1 & 1 & 0
\end{array}\right), \label{eq:Xintroe}
\end{equation}
leads to an {\em exactly-solvable} dimer liquid for small field, which we connect to the aforementioned model. Remarkably, its spin liquid phase is robust as we take the field strength to $+\infty$, where we find that the dimer-enforcing interactions effectively become spin-1 Kitaev-like interactions. We also consider another choice of field\footnote{I.e., {\scriptsize
$\mathcal X_\textrm{$f$-anyon fluctuations} = \left( \begin{array}{cccc}
0 & -i & -i & -i \\
i & 0 & -i & i \\
i & i & 0 & -i \\
i & -i & i & 0
\end{array}\right).$}}
which is integrable and provides a link to the spin-1/2 Kitaev model. Thus it appears that the peculiar Kitaev couplings can be motivated, not just by the abstract requirement of exact solubility, but also from a geometric RVB picture. Some of the connections are summarised in Fig.~\ref{fig:schematic}.

\emph{Outline.} In Sec.~\ref{sec:model}, we start by further motivating our parent Hamiltonian by introducing convenient operators and their graphical notation, after which we summarize the key physical results. Sec.~\ref{sec:monomer} discusses in more detail the case of monomer fluctuations \eqref{eq:Xintroe} in the emergent dimer model, including how its exactly solvable dimer model in the weak-field limit adiabatically connects to a novel spin-1 quadrupolar Kitaev liquid in the large-field limit. 
Sec.~\ref{sec:spinon} studies the case for fermionic (monomer-flux composite) anyon fluctuations, which turns out to be exactly solvable and connects to the spin-1/2 Kitaev magnet; here we introduce a Majorana representation to aid the analysis alongside the graphical notation.
In Sec.~\ref{sec:applications} we explore applications of some of these insights, including: interpolating the spin liquid of the ruby lattice PXP model \cite{Verresen21} to the exactly solvable dimer liquid found in Sec.~\ref{sec:monomer}; and elucidating the `Hadamard' rotation used in recent Rydberg experiments \cite{Semeghini21} to rotate off-diagonal string operators to diagonal ones. Finally, Sec.~\ref{sec:equivalence} highlights how our effective 4-state system can arise from spin-1/2 systems in two different ways, before concluding in Sec.~\ref{sec:conclusions}.

\begin{figure}
	\centering
	\begin{tikzpicture}
	\node at (0,0) {\includegraphics[scale=0.64]{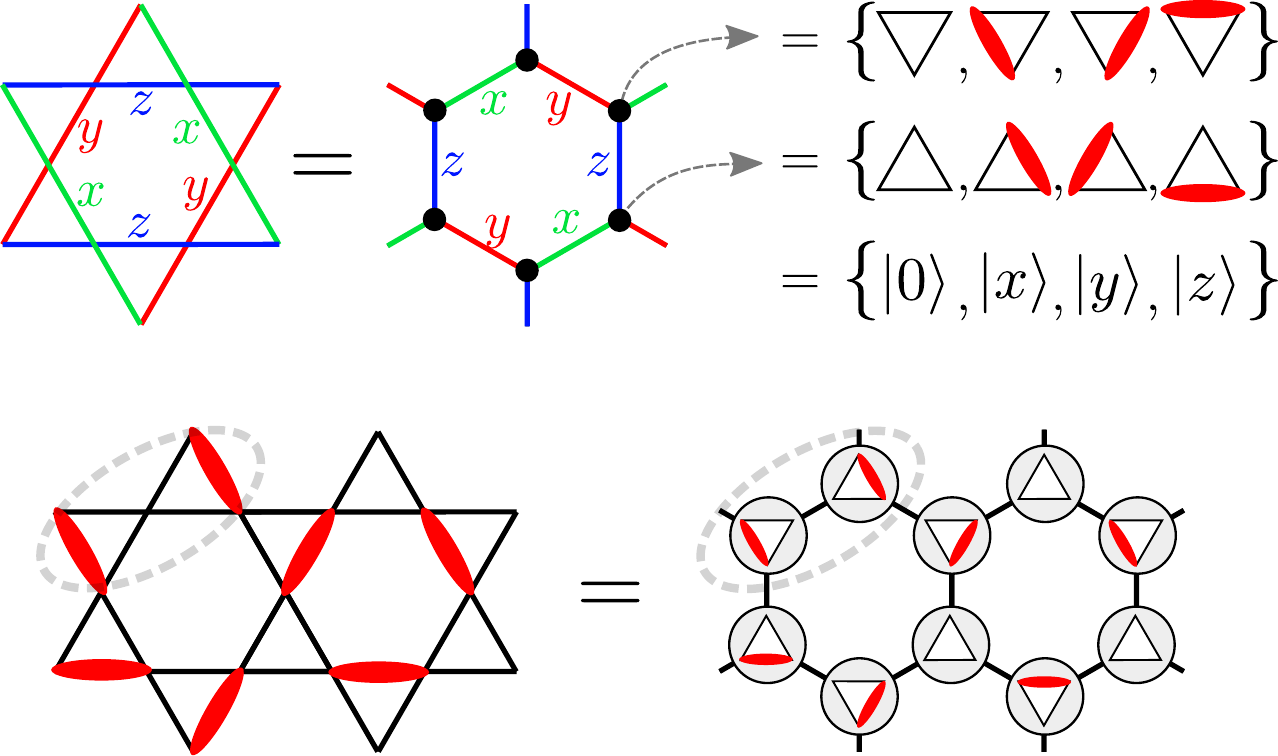}};
	\node at (-3.95,2.4) {(a)};
	\node at (-3.95,-0.3) {(b)};
	\end{tikzpicture}
	\caption{\textbf{Emergent dimer model on the kagom\'e lattice as a spin-$3/2$ model on the honeycomb lattice.} (a) Each triangle of the kagom\'e lattice is effectively a four-level state as shown, making a kagom\'e lattice dimer model into a spin-$3/2$ model on the honeycomb lattice. Labeling the three different types of bonds of the honeycomb lattice as $x,y,z$ \cite{Kitaev06} also induces a labeling on the kagom\'e lattice as shown. Correspondingly, we can label the basis states according to whether the dimer occupies the $x,y,z$ bond of the kagom\'e lattice. (b) An example of how a dimer state on the kagom\'e lattice is encoded in a spin-$3/2$ state on the honeycomb lattice. This tensor product Hilbert space not only contains all dimer coverings but also defects: the dashed line highlights a monomer ($e$-anyon) excitation (i.e., a vertex not touched by a dimer). Directionally-dependent two-body interactions energetically penalize such configurations to obtain the emergent dimer constraint in Eq.~\eqref{eq:gauss} (see also Fig.~\ref{fig:stabilizers}).}
	\label{fig:fromkagometohoneycomb}
\end{figure}

\section{Model \label{sec:model}}

\subsection{Emergent dimer model in a tensor product Hilbert space \label{sec:modelhilb}}

We study a quantum dimer model on the kagom\'e lattice. Somewhat uncommon for dimer models, we will define it in a tensor product Hilbert space. One option is to treat each triangle as a four-level system: either the triangle is empty or one of the three bonds has a dimer. This way, the dimer model is encoded in an effective 4-state or ``spin-$3/2$'' model on the \emph{honeycomb lattice}, whose sites represent the triangles of the kagom\'e lattice; see Fig.~\ref{fig:fromkagometohoneycomb}. (To see how this can emerge from a spin-$1/2$ model, see Sec.~\ref{sec:equivalence}.)

\subsubsection{Parity string operator and dimer constraint}

By construction, this tensor product Hilbert space ensures that a given triangle cannot contain more than one dimer. However, it does not ensure the full dimer constraint, namely, that every vertex of the kagom\'e lattice is touched by \emph{exactly one} dimer. We will enforce this constraint energetically by coupling two neighboring triangles, giving rise to an emergent dimer model at low energies.

To this end, it is very convenient to introduce diagonal string operators which measure the parity of the dimers that it intersects. Visually\footnote{We read off that its matrix representation is {\tiny $\left( \begin{array}{cccc} 1& 0 & 0 & 0 \\ 0 & -1 & 0 & 0 \\ 0 & 0 & -1 & 0 \\ 0 & 0 & 0 & 1 \end{array} \right)$} in the basis ordering shown in Fig.~\ref{fig:fromkagometohoneycomb}. See Appendix~\ref{app:spinthreehalf} for explicit matrices for all the operators we discuss.}, we depict this as an orange dashed line:
\begin{align}
&\raisebox{-3pt}{\includegraphics[scale=0.54]{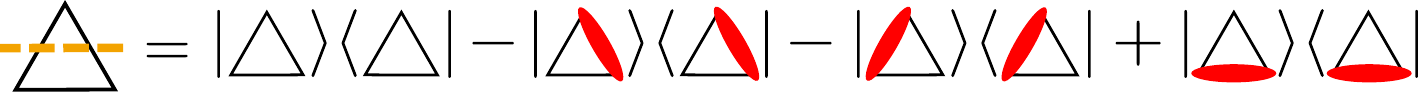}} \label{eq:Zdef} \\
&\quad \quad= |0\rangle \langle 0| - |x\rangle \langle x| - |y\rangle \langle y| + |z\rangle \langle z|. \label{eq:explicitstates}
\end{align}
We will find it convenient to work with the graphical notation as in line \eqref{eq:Zdef}.
However, for transparency and completeness, we included line \eqref{eq:explicitstates} where the basis states are labeled by the type of bond of the kagom\'e lattice supporting the dimer (see Fig.~\ref{fig:fromkagometohoneycomb} for the $x,y,z$ labels of the three directions). We denote the above operator as $\mathcal Z^{\alpha}$ with $\alpha=x,y,z$ depending on the orientation of the line cutting through the triangle:
\begin{equation}
\raisebox{-7pt}{\includegraphics[scale=0.56]{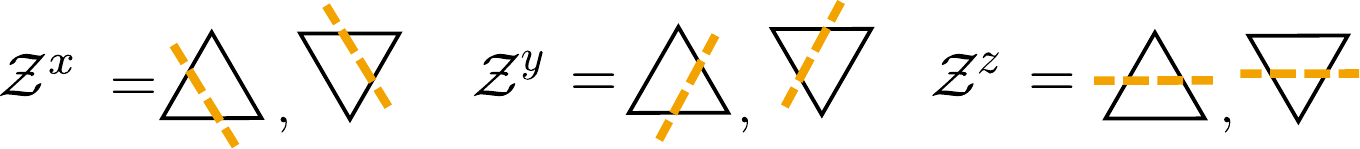}}
\end{equation}
Note that $\mathcal Z^x \mathcal Z^y \mathcal Z^z = 1$. Equivalently, the product of any two equals the third, which we can represent graphically as $\mathcal Z^x \mathcal Z^z \; \; \raisebox{-8pt}{\includegraphics[scale=0.55]{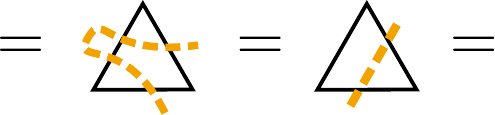}} \; \; \mathcal Z^y$.

The dimer constraint for any two touching triangles of the kagom\'e lattice is then enforced by including the `Ising' penalty term $\sum_{\alpha\in\{x,y,z\}} \sum_{\langle i,j\rangle_\alpha}\mathcal Z^\alpha_i \mathcal Z^\alpha_j$ on the honeycomb lattice. For instance, for a $z$-bond of the honeycomb lattice, the Gauss law is:
\begin{equation}
G_{i,j} := \mathcal Z^z_i \mathcal Z^z_j = \raisebox{-12pt}{\includegraphics[scale=0.53]{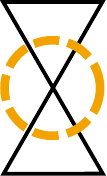}} = -1 \label{eq:gauss}.
\end{equation}
Indeed, given the definition \eqref{eq:Zdef}, this is satisfied if and only if there is exactly one dimer touching this vertex\footnote{In terms of the basis states labeled by $\{ \ket{0},\ket{x},\ket{y},\ket{z} \}$, the Gauss law corresponds to, say, the $z$-bond being adjacent to (exactly) one $\ket{x}$ or $\ket{y}$ state.}.

Upon enforcing the Gauss law everywhere, the resulting ground state manifold is extensively degenerate, given by all dimer configurations of the kagom\'e lattice. We will let the local pair creation of charges and/or fermions lift this degeneracy (see Sec.~\ref{sec:modelham} for the Hamiltonian). Crucially, such anyon fluctuations do not commute with the aforementioned energetic dimer constraint: this is key to being able to write down an ultra-local Hamiltonian which still gives rise to an effective resonating dimer Hamiltonian for small fluctuations. Before concretely defining the Hamiltonian, we need to briefly discuss what the excitations are and what operators create them.

\begin{figure}
    \centering
    \begin{tikzpicture}
    \node at (0,0) {\includegraphics[scale=0.57]{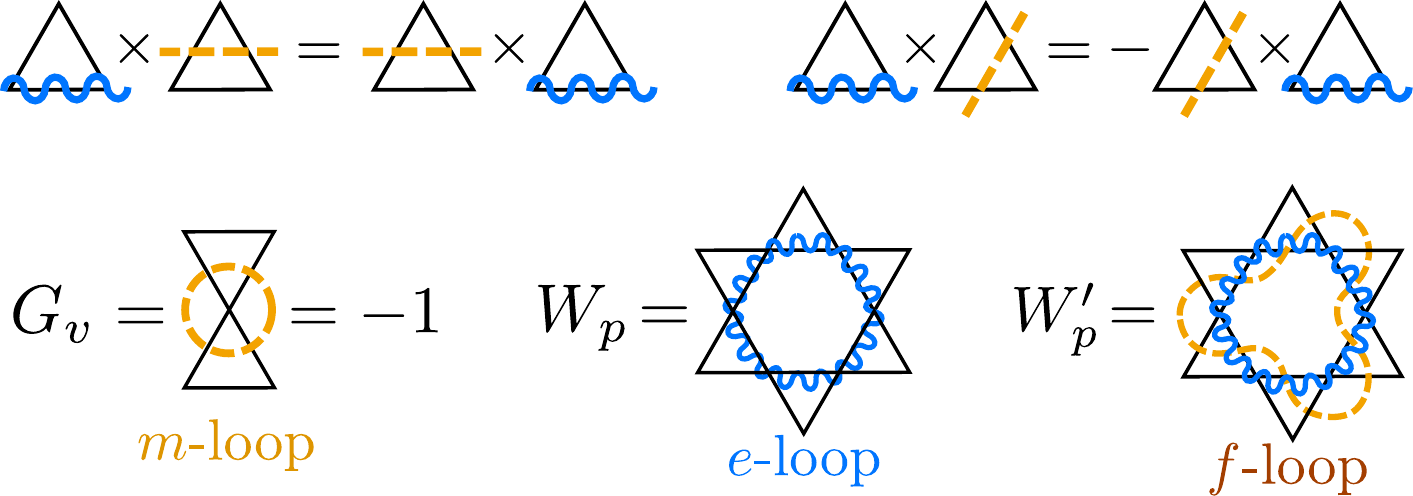}};
    \node at (-4.1,1.5) {(a)};
    \node at (-4.1,0.2) {(b)};
    \end{tikzpicture}
    \caption{\textbf{Topological string operators and stabilizers for dimer liquids.} We define diagonal $\mathcal Z^\alpha$ (orange; Eq.~\eqref{eq:Zdef}) and off-diagonal $\mathcal X^\alpha$ (blue; Eq.~\eqref{eq:Xdef}) operators in the spin-3/2 Hilbert space; these respectively create pairs of $m$- and $e$- anyons in our emergent dimer model. (a) These string operators anticommute when they intersect, encoding the mutual semionic statistics of $e$- and $m$-anyons.
    (b) Closed loops of these operators define commuting stabilizers. We add the diagonal loop (which is a two-body operator in the spin-3/2 space) as an energetic constraint in the model \eqref{eq:model}; at low energies $G_v=-1$ enforces the dimer constraint. Single-site $e$-anyon ($f$-anyon) fluctuations perturbatively generate the $W_p$ ($W_p'$) stabilizers, which can be interpreted as worldlines of these anyons around a plaquette. The Hamiltonian for the fixed-point dimer liquid on the kagom\'e lattice is simply $H = \sum_v G_v - \sum_p W_p$. }
    \label{fig:stabilizers}
\end{figure}

\subsubsection{Monomers and fluxes in dimer models: $e$- and $m$-anyons}

It is known that a dimer model (or $\mathbb Z_2$ gauge theory more generally) admits three types of topological excitations \cite{Kivelson_1987,Kitaev_2003}, commonly called the $e$-, $m$-, and $f$-anyons. The first is the `\emph{electric charge}' of the emergent gauge theory; in the dimer model this corresponds to violations of the Gauss law \eqref{eq:gauss}. A vertex with $G_v=1$ has either \emph{none} or \emph{two} dimers touching---indeed, since our Hilbert space only allows for one dimer per triangle, there are no states with more than two dimers touching per vertex. These two types of charges can be called a monomer and double-dimer respectively, although we will sometimes refer to both as a monomer for simplicity, which is a harmless abuse of language since both are in the same superselection sector. The $m$-anyon is the `\emph{magnetic flux}' of the gauge theory (sometimes also call a vortex or vison); in a dimer model this is encoded in the phase information of the wavefunction, as we will discuss. The $e$- and $m$-anyons are bosonic, but they have mutual semionic statistics; this is the $\mathbb Z_2$ analogue of the Aharonov-Bohm effect \cite{AB59}. This statistics implies that their composite is a fermion ($f=e\times m$).

By definition, a local operator can never create a single topological excitation; instead, they are created at the end of stringlike operators. Fluxes are created by an open string of $\mathcal Z^\alpha$ operators (which we soon discuss in more detail). Similarly, we can define an \emph{off-diagonal} string operator which shuffles dimers around and creates monomers (i.e., $e$-anyons) at its endpoints:
\begin{equation}
\raisebox{-5pt}{\includegraphics[scale=0.54]{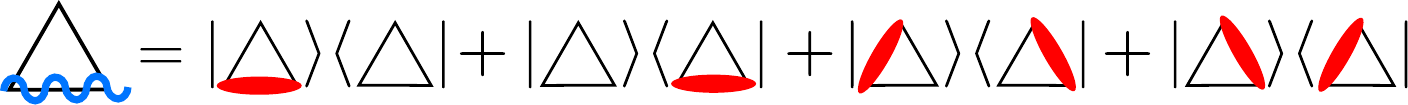}} . \label{eq:Xdef}
\end{equation}
We denote this blue wiggly line by $\mathcal X^\alpha$ depending on which type of bond (equivalently, in which direction) it acts:
\begin{equation}
\raisebox{-7pt}{\includegraphics[scale=0.55]{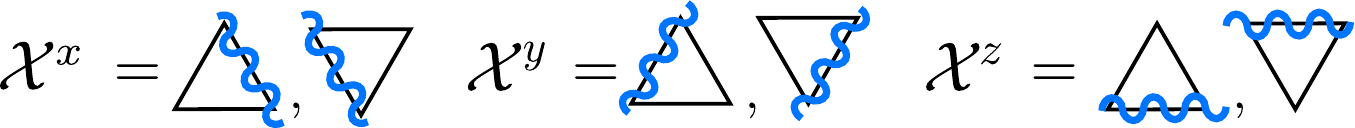}}
\end{equation}
(We note that $\mathcal X^x \mathcal X^y \mathcal X^z = 1$.)
The key algebraic property is that the diagonal string $\mathcal Z$ and off-diagonal string $\mathcal X$ anticommute when they intersect, otherwise they commute\footnote{I.e., $\mathcal X^{\alpha} \mathcal Z^{\beta} + (-1)^{\delta_{\alpha,\beta}} \mathcal Z^{\beta} \mathcal X^{\alpha} = 0$. See Appendix~\ref{app:spinthreehalf} for more algebraic properties, where we also link these operators to a Majorana representation of spin-$3/2$, which helps with, e.g., constructing a Hadamard transformation $\mathcal X^{\alpha} \leftrightarrow \mathcal Z^{\alpha}$.}:
\begin{equation}
\raisebox{-8pt}{\includegraphics[scale=0.53]{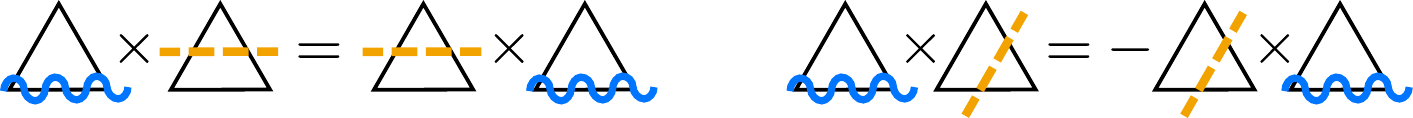}}. \label{eq:semionic}
\end{equation}
This can be interpreted as the mutual semionic statistics between $e$- and $m$-anyons. Indeed, it implies that if one acts with a string of $\mathcal X^\alpha$ on a state that satisfies the Gauss law (i.e., a valid dimer covering of the kagom\'e lattice), the resulting state will violate the Gauss law at the two ends of the string (i.e., measuring the Gauss operator $G_v$ \eqref{eq:gauss} around those vertices will give $+1$ instead of $-1$, indicating the presence of charges).

This is in agreement with the claim that fluxes are created at the endpoint of a $\mathcal Z$-string, such that one can interpret the Gauss operator \eqref{eq:gauss} as a tiny worldline of a flux excitation. Indeed, the latter measures the charge content of the enclosed vertex by virtue of the mutual statistics. The fact that $G_v = -1$ for a dimer state (rather than $G_v=1$) thus implies that every vertex hosts a `background' $e$-anyon; it is for this reason that a dimer model is sometimes called an \emph{odd} $\mathbb Z_2$ gauge theory. When we speak of an $e$-anyon excitation, it is usually relative to this background. This can be likened to a Mott insulator, where we might describe the act of removing a background charge as adding a hole excitation (and note that for a $\mathbb Z_2$ gauge theory there is no difference between charges and holes since $-1=1 \mod 2$).

Similarly, closed $\mathcal X$-loops measure the parity of fluxes enclosed. Indeed, this is the worldline of an $e$-anyon (or monomer) which picks up the presence of a flux on that plaquette due to the Aharonov-Bohm effect. The smallest such loop is the `Wilson'\footnote{In the gauge theory literature, the worldline of a charged particle is called a Wilson line \cite{Wegner71,Wilson74,FradkinShenker}.} plaquette operator:
\begin{equation}
W_p = \prod_{\alpha \in p} \mathcal X^\alpha = \raisebox{-18pt}{\includegraphics[scale=0.6]{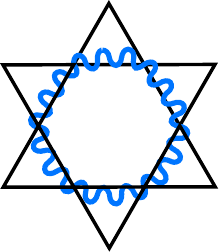}}. \label{eq:Wp}
\end{equation}
Note that such a \emph{closed} loop operator respects the Gauss law and thus shuffles \emph{within} the manifold of valid dimer coverings; in fact it resonates $2^5=32$ distinct dimer patterns around the hexagon, making its action equivalent to the operator defined by Misguich, Serban and Pasquier in Ref.~\onlinecite{Misguich02}. We say $W_p =1$ ($W_p=-1$) labels the absence (presence) of a magnetic flux on this plaquette, which can be toggled by the endpoint of a $\mathcal Z^\alpha$-string (due to Eq.~\eqref{eq:semionic}). Relatedly, the mathematical identity $\mathcal X^x \mathcal X^y \mathcal X^z = 1$ can thus be physically interpreted as saying that fluxes cannot live on the triangles of the kagom\'e lattice---only on the hexagons.

It is a special property of the kagom\'e lattice (or more generally, lattices of corner-sharing-triangles) that one can even define such an off-diagonal string operator! For dimer models on more generic lattices (including the square, triangular and honeycomb lattices), the dimer shuffles needed to create charges usually depend on the particular state on which one acts\footnote{I.e., for any given dimer state, one is then forced to choose a path of alternating dimers and empty bonds.}. The absence of such a string operator for those lattices prevents the definition of a stabilizer\footnote{To wit, a stabilizer Hamiltonian is a sum of commuting terms, each term squaring to one. The many-body spectrum can be labeled by the eigenvalues of these stabilizers, and the ground state is then the special case where each energy term is minimized. \label{footnote:stabilizer}} dimer liquid Hamiltonian---indeed, the solvable Rokhsar-Kivelson models \cite{RK} do not have a solvable spectrum. Only the ground state is exactly known. In contrast, lattices of corner-sharing triangles do admit such stabilizer liquids \cite{Misguich02}. In fact, the stabilizers defining the fixed-point dimer liquid on the kagom\'e lattice are exactly the commuting loop operators $G_v$ \eqref{eq:gauss} and $W_p$ \eqref{eq:Wp}, as we discuss in Sec.~\ref{subsec:weakmonomer}. (Note that while the toric code Hamiltonian is also a stabilizer Hamiltonian, its space of states does not encode a dimer Hilbert space; although see Sec.~\ref{subsec:toric}.)

\subsubsection{Fermionic composite: $f$-anyons}

The emergent fermion in a dimer model is the composite of the monomer and flux. Hence, one way of creating a pair of fermions is by the product of the above string operators, e.g.:
\begin{equation}
\mathcal X^z \mathcal Z^z = \raisebox{-6pt}{\includegraphics[scale=0.5]{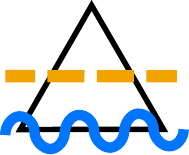}}
\end{equation}
Indeed this creates an `$e \times m$' pair on both sides of the triangle. However, this is not the ideal `string operator' for creating a fermion pair separated by an arbitrary extent. To see why, we need only consider a few more triangles:
\begin{equation}
\mathcal X^z_1 \mathcal Z^z_1 \cdot \mathcal X^z_2 \cdot \mathcal X^z_3 \mathcal Z^z_3 = \raisebox{-14pt}{\includegraphics[scale=0.55]{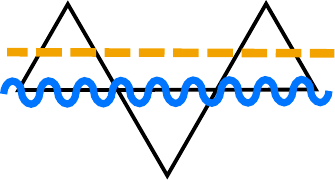}}
\end{equation}
Note that while we indeed act with the product operator $\mathcal X^z \mathcal Z^z$ on the left and right triangle, we acted with $\mathcal X^z$ on the middle triangle. The reason for this is conceptually simple: charges live on the vertices of the kagom\'e lattice, whereas fluxes live at the centers of hexagons (i.e., vertices of a triangular lattice). The spacing of these lattices differs by a factor of two, hence, whenever the flux moves by one step, the charge needs to move by two steps. This is seemingly an obstruction to defining a local operator whose product gives the fermionic string operator. We note that defining, say, $\mathcal X^\alpha \mathcal Z^\alpha$ for up-triangles and $\mathcal X^\alpha$ for down-triangles does not solve this issue, since then the fermion cannot make certain turns.

There is a simple way of avoiding this issue: we let the flux move `sideways', at a 60 degrees angle relative to the motion of the $e$-anyon, so that the flux effectively moves $2 \times \cos(\pi/3) = 1$ steps in that direction. Visually:
\begin{equation}
\mathcal X^z \mathcal Z^x = \raisebox{-6pt}{\includegraphics[scale=0.6]{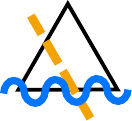}} \label{eq:spinonfluctuation}
\end{equation}
with all of its rotations, and with the mirror image for downward-facing triangles. Indeed, products of these give us the movement of a fermion:
\begin{equation}
\mathcal X^z_1 \mathcal Z^x_1 \cdot \mathcal X^z_2 \mathcal Z^y_2 \cdot \mathcal X^z_3 \mathcal Z^x_3 = \raisebox{-14pt}{\includegraphics[scale=0.6]{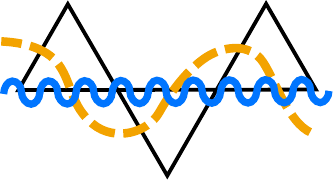}}
\end{equation}

There is also a more fundamental and general perspective on why these are the appropriate fermion hopping operators\footnote{We thank Steve Kivelson for illuminating discussions on this matter.}.
The key point is that fermions---being a bound state of a charge and a flux---are `framed': upon placing a charge on a vertex of the kagom\'e lattice, we have a choice of putting the flux on one of the two neighboring plaquettes. Hence, we first need to choose a `framing', i.e., for each $e$-anyon we have to choose which $m$-anyon we want to pair with it to define our fermion $f$; once this choice is made, one also directly obtains a notion of `fermion hopping operator' for this choice of fermions living on the kagom\'e lattice. This choice is equivalent to fixing an orientation of the bonds of the honeycomb lattice: $e$-anyons live on the bonds of the honeycomb lattice (i.e., the vertices of the kagom\'e lattice), and one can then associate them to $m$-anyons on the right-hand side of the arrow. A particularly simple choice is to e.g. choose arrows which point from the $A$ sublattice to the $B$ sublatice of the honeycomb lattice, or equivalently, from the up triangles to the down triangles of the kagom\'e lattice. With this choice of fermions, the hopping operator is indeed given by Eq.~\eqref{eq:spinonfluctuation}! Moreover, all choices of orientations are unitarily related\footnote{More precisely, the choice of orientation of a given bond $\alpha$ of the honeycomb lattice is toggled by a discrete unitary rotation generated by $\mathcal Z^\alpha$ on the two spin-$3/2$'s neighboring that bond. Hence, we do not need to explicitly choose a Kasteleyn orientation.}.

\subsection{Hamiltonian \label{sec:modelham}}

We now have all the pieces to define our emergent dimer Hamiltonian in the above Hilbert space. It takes the form of an Ising model in terms of the above ``spin-$3/2$'' operators on the honeycomb lattice (see Fig.~\ref{fig:fromkagometohoneycomb}):
\begin{equation}
\boxed{ H_a = J \sum_{\alpha=x,y,z} \sum_{\langle i,j\rangle_\alpha} \mathcal Z^\alpha_i \mathcal Z^\alpha_j + h_t \sum_i \mathcal X_{a,i} - h_l \sum_i \mathcal Z_i } \label{eq:model}
\end{equation}
with a `longitudinal' field $\mathcal Z = \mathcal Z^x + \mathcal Z^y + \mathcal Z^z$ which functions as a chemical potential for dimers. As discussed above, $J>0$ energetically enforces the dimer constraint (imposing the Gauss law \eqref{eq:gauss} for large $J$). Note, the $J$ term has, in some ways, the flavor of a Kitaev interaction \cite{Kitaev06}, with terms like $\mathcal Z^z_i \mathcal Z^z_j$ on the vertical bonds and rotated indices for the other two directions of bonds. In fact, we will see that in the presence of a large field, this \emph{diagonal} interaction---implementing the dimer constraint---can indeed morph into an effective \emph{off-diagonal} Kitaev coupling (see Secs.~\ref{subsec:strongmonomer} and \ref{subsec:strongspinon}). This points to how `geometric frustration' and `exchange frustration' are two sides of the same coin.

The only quantum term is a `transverse field' $\mathcal X_a$ corresponding to anyon pair creation. We consider three choices (which we define in the following paragraphs), corresponding to $e$-anyon pair-fluctuations ($a=\e$), fermionic fluctuations ($a=\f$), and the previously-studied model given by the ruby lattice Ising `PXP' model ($a=\textrm{ruby}$) \cite{Verresen21}. For the first two we will always set the longitudinal field $h_l=0$; only the ruby model has $h_l \neq 0$ (assigning different energy costs to monomers and double-dimers) to help stabilize the topological liquid phase.

\subsubsection{$H_\e$: $e$-anyon fluctuations}

The first choice of transverse field is the fluctuation of $e$-anyon pairs:
\begin{equation}
\boxed{ \mathcal X_\e := \mathcal X^x + \mathcal X^y + \mathcal X^z =  \raisebox{-6pt}{\includegraphics[scale=0.55]{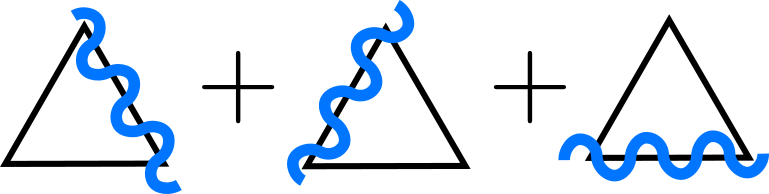}} } \label{eq:Xmono}
\end{equation}
(and the same for down-triangles); this corresponds to the matrix \eqref{eq:Xintroe} mentioned in the introduction. As discussed in the previous subsection, this indeed creates pairs of $\mathbb Z_2$ charges. Moreover, since the $\mathcal X^\alpha$ operators mutually commute, $W_p$ defined in Eq.~\eqref{eq:Wp} is a local conserved quantity for all $J,h_t$ (where $h_l=0$):
\begin{equation}
[H_\e,W_p] = 0. \label{eq:Wpsym}
\end{equation}
As discussed above, $W_p$ can physically be interpreted as the worldline of an $e$-anyon wrapped around a hexagon, and consequently the value $W_p = \pm 1$ measures the presence/absence of a flux excitation on the plaquette $p$. (Later we will see that the ground state of $H_\e$ is flux-free: $W_p = 1$.)

\begin{figure}
    \centering
\begin{tikzpicture}
\node at (-0.06-0.04-0.23,-1.4*2){
    \begin{tikzpicture}
    \node at (-1.4,0.1) {$H_\e$};
    \node at (-1.4,-0.3) {\scriptsize ($h_t>0$, $h_l=0$)};
    \draw[->] (0,0) -- (5,0) node[right] {$\frac{h_t}{|J|}$};
    \node[red] at (0,0.5) {\scriptsize fixed-point};
    \node[red] at (0,0.25) {\scriptsize dimer liquid};
    \node[red] at (4.5,0.5) {\scriptsize spin-1 Kitaev};
    \node[red] at (4.5,0.25) {\scriptsize model$^*$};
    \draw[-,red,line width=2] (0,0) -- (4.5,0) node[above,midway] {$\mathbb Z_2$ spin liquid};
    \filldraw[black] (0,0) circle (2pt) node[below] {$0$};
    \filldraw[red] (4.5,0) circle (2pt) node[below] {$\infty$};
    \node[red] at (3.6,0) {/};
    \node[red] at (3.68,0) {/};
    \end{tikzpicture}
};
\node at (-0.06-0.23,-1.4){
    \begin{tikzpicture}
    \node at (-1.4,0.1) {$H_\e$};
    \node at (-1.4,-0.3) {\scriptsize ($h_t<0$, $h_l=0$)};
    \draw[->] (0,0) -- (5,0) node[right] {$\frac{h_t}{|J|}$};
    \node[blue] at (4.5,0.5) {\scriptsize product};
    \node[blue] at (4.5,0.25) {\scriptsize state};
    \draw[-,red,line width=2] (0,0) -- (0.34*4.5,0) node[above,midway,xshift=15] {\scriptsize $\mathbb Z_2$ SL};
    \draw[-,blue,line width=2] (0.34*4.5,0) -- (4.5,0) node[midway,above,xshift=-5] {trivial};
    \filldraw[black] (0.34*4.5,0) circle (2pt) node[below] {$-0.34$};
    \filldraw[black] (0,0) circle (2pt) node[below] {$0$};
    \node[red] at (0,0.5) {\scriptsize fixed-point};
    \node[red] at (0,0.25) {\scriptsize dimer liquid};
    \filldraw[blue] (4.5,0) circle (2pt) node[below] {$-\infty$};
    \node[blue] at (3.6,0) {/};
    \node[blue] at (3.68,0) {/};
    \end{tikzpicture}
};
\node at (-0.2+0.15,0){
    \begin{tikzpicture}
    \node at (-1.4+0.05,0.1) {$H_\textrm{ruby}$};
    \node at (-1.4+0.05,-0.3) {\scriptsize $(h_l \to J>0)$};
    \draw[->] (0,0) -- (5,0) node[right] {$\frac{|h_t|}{J-h_l}$};
    \node[blue] at (4.5,0.5) {\scriptsize product};
    \node[blue] at (4.5,0.25) {\scriptsize state};
    \draw[-,cyan,line width=2] (0*4.5,0) -- (0.48*4.5,0) node[above,midway] {VBS};
    \draw[-,red,line width=2] (0.48*4.5,0) -- (0.68*4.5,0) node[above,midway] {$\mathbb Z_2$ SL};
    \draw[-,blue,line width=2] (0.68*4.5,0) -- (4.5,0);
    \filldraw[black] (0.48*4.5,0) circle (2pt) node[below] {$0.5$}; 
    \filldraw[black] (0.68*4.5,0) circle (2pt) node[below] {$0.7$}; 
    \filldraw[black] (0,0) circle (2pt) node[below] {$0$};
    \filldraw[blue] (4.5,0) circle (2pt) node[below] {$\infty$};
    \node[blue] at (3.6,0) {/};
    \node[blue] at (3.68,0) {/};
    \end{tikzpicture}
};
\node at (-0.1-0.2,-1.4*3){
    \begin{tikzpicture}
    \node at (-1.4,0.1) {$H_\f$};
    \node at (-1.4,-0.3) {\scriptsize achiral $(h_l=0)$};
    \draw[->] (0,0) -- (5,0) node[right] {$\frac{|h_t|}{|J|}$};
    \node[red] at (0,0.5) {\scriptsize fixed-point};
    \node[red] at (0,0.25) {\scriptsize dimer liquid};
    \draw[-,red,line width=2] (0,0) -- (0.58*4.5,0) node[above,midway,xshift=10] {$\mathbb Z_2$ SL};
    \draw[-,magenta,line width=2] (0.58*4.5,0) -- (4.5,0) node[above,midway,xshift=0] {Maj.~FS};
    \filldraw[black] (0.58*4.5,0) circle (2pt) node[below] {$1/\sqrt{3}$};
    \filldraw[black] (0,0) circle (2pt) node[below] {$0$};
    \filldraw[black] (4.5,0) circle (2pt) node[above] {$\infty$};
    \node[magenta] at (3.6,0) {/};
    \node[magenta] at (3.68,0) {/};
    \end{tikzpicture}
};
\node at (-0.1-0.18,-1.4*4+0.2){
    \begin{tikzpicture}
    \node at (-1.4,0.1) {$H_\f$};
    \node at (-1.4,-0.3) {\scriptsize chiral $(h_l=0)$};
    \draw[->] (0,0) -- (5,0) node[right] {$\frac{|h_t|}{|J|}$};
    \node[purple] at (0,0.5) {\scriptsize fixed-point};
    \node[purple] at (0,0.25) {\scriptsize dimer liquid};
    \node[black] at (4.5,0.5+0.3) {\scriptsize spin-$1/2$ Kitaev};
    \node[black] at (4.5,0.25+0.3+0.05) {\scriptsize model};
    \node[black!80] at (4.85,0.28) {\tiny AFM};
    \node[black!80] at (4.85,1.2) {\tiny FM};
    \draw[-,purple,line width=2] (0,0) -- (0.58*4.5,0) node[above,midway,xshift=10] {$\mathbb Z_2$ SL'};
    \draw[-,green,line width=2] (0.58*4.5,0) -- (4.5,0) node[above,midway,xshift=0] {Ising SL};
    \filldraw[black] (0.58*4.5,0) circle (2pt) node[below] {$1/\sqrt{3}$};
    \filldraw[black] (0,0) circle (2pt) node[below] {$0$};
    \filldraw[black] (4.5,0) circle (2pt) node[below] {$\infty$};
    \node[green] at (3.6,0) {/};
    \node[green] at (3.68,0) {/};
    \draw[->] (4.5,1) -- (4.5,1.3);
    \draw[->] (4.5,0.45) -- (4.5,0.15);
    \end{tikzpicture}
};
\end{tikzpicture}
    \caption{\textbf{Phase diagrams for emergent dimer models on kagom\'e lattice with fluctuating anyons.} We consider the spin-$3/2$ Ising model on the honeycomb lattice, defined in Eq.~\eqref{eq:model} with the three different choices of transverse field in Eqs.~\eqref{eq:Xmono}, \eqref{eq:Xspinon} and \eqref{eq:Xruby}. Setting $h_t=0$ gives a degenerate space of dimer coverings on the kagom\'e lattice, which is lifted in different ways by $h_t \neq 0$. The model $H_\textrm{ruby}$ has been studied previously as a spin-$1/2$ Ising model in the context of Rydberg blockade, and numerical studies indicate that monomer ($e$-anyon) fluctuations stabilize a $\mathbb Z_2$ spin/dimer liquid (\2 SL) \cite{Verresen21}. $H_\e$ is a simplified model with the same mechanism, where we report an exactly-solvable dimer liquid \eqref{eq:perturbation} in the limit $h_t\to 0$, which we moreover show to be adiabatically connected to the ruby model. If the fluctuations are strong and unfrustrated (i.e., negative transverse field), we find a trivial phase. In contrast, frustrated $e$-anyon fluctuations ($h_t>0$) stabilize the spin liquid up to $h_t\to \infty$ where we find a spin-1 quadrupolar Kitaev model \eqref{eq:Hquadrupolar}. We also consider the case with $f$-anyon (monomer-flux composites) fluctuations, where the small-field case is similar, although in the case of chiral fluctuations it leads to a distinct SET phase in the presence of translation symmetry: \2 SL (\2 SL') has a background $e$-anyon ($f$-anyon). Larger fields stabilize a Majorana Fermi surface (FS) for achiral fluctuations (which moreover has a flux transition; see Sec.~\ref{subsec:intermediatespinon}), and a gapped non-Abelian Ising topological order (Ising SL) in the chiral case (the latter corresponds to a sector of the Yao-Kivelson model \cite{Yao07}). The $|h_t|\to \infty$ limit recovers the spin-$1/2$ Kitaev honeycomb model.}
    \label{fig:phasediagram}
\end{figure}
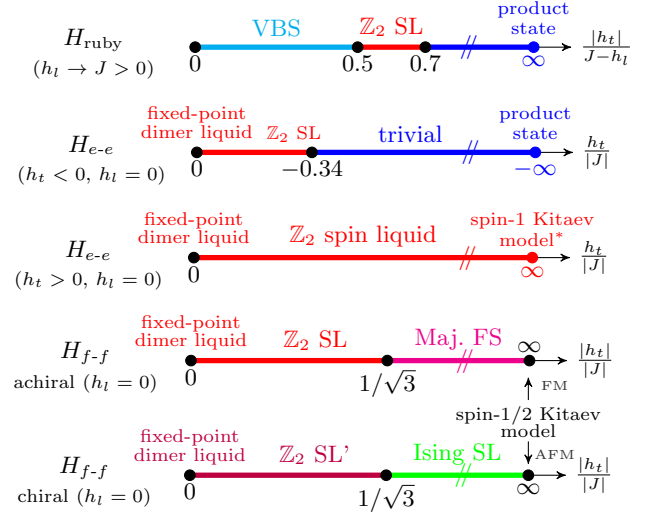

\subsubsection{$H_\f$: $f$-anyon fluctuations}

The second case is the fluctuation of fermion pairs. As we saw in Eq.~\eqref{eq:spinonfluctuation}, the right way to generate (framed) $f$-anyon fluctuations/dynamics is:
\begin{equation}
\boxed{ \mathcal X_\f := \raisebox{-6pt}{\includegraphics[scale=0.55]{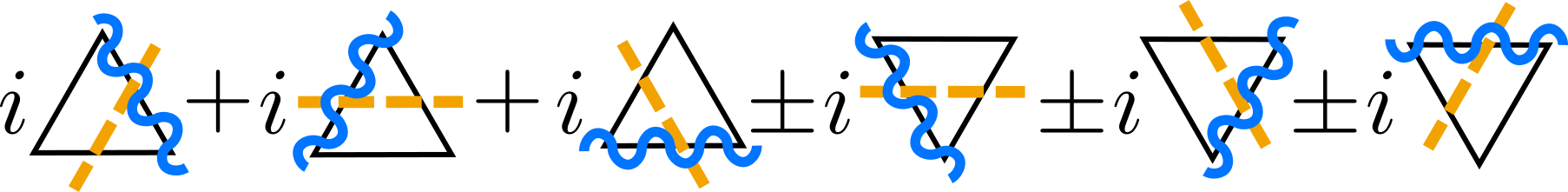}} } \label{eq:Xspinon}
\end{equation}
We will use the convention that we first act with the diagonal (orange dashed line) operator. Explicitly (in the order of the above pictures): $\mathcal X_\f = i \mathcal X^x \mathcal Z^y + i \mathcal X^y \mathcal Z^z + i\mathcal X^z \mathcal Z^x$ for up-triangles and $\mathcal X_\f = \pm \left(i \mathcal X^x \mathcal Z^z + i\mathcal X^y \mathcal Z^x + i \mathcal X^z \mathcal Z^y\right)$ for down-triangles. Note that the factor of $i$ is necessary to make the Hamiltonian hermitian (since the $\mathcal X$ and $\mathcal Z$ operators anticommute when they intersect; see Eq.~\eqref{eq:semionic}). We see that we have a choice of relative sign $\pm$ in Eq.~\eqref{eq:Xspinon}. The choice $+$ is \emph{achiral}\footnote{E.g., we see that point-centered inversion exchanges the two triangles and is thus a symmetry of the Hamiltonian. Similarly one can define an anti-unitary symmetry.} whereas the choice of $-$ is \emph{chiral}, and we will see that they lead to very different phase diagrams for strong fluctuations (and in fact even for small fluctuations they lead to distinct symmetry-enriched topological (SET) order for the deconfined $\mathbb Z_2$ gauge theory). We will henceforth refer to these two choices as the (a)chiral fermion-fluctuating models.

Similar to the previous case, this model also has a conserved plaquette operator: 
\begin{equation}
[H_\f,W_p'] = 0 \quad \textrm{with } W_p' := \raisebox{-18pt}{\includegraphics[scale=0.4]{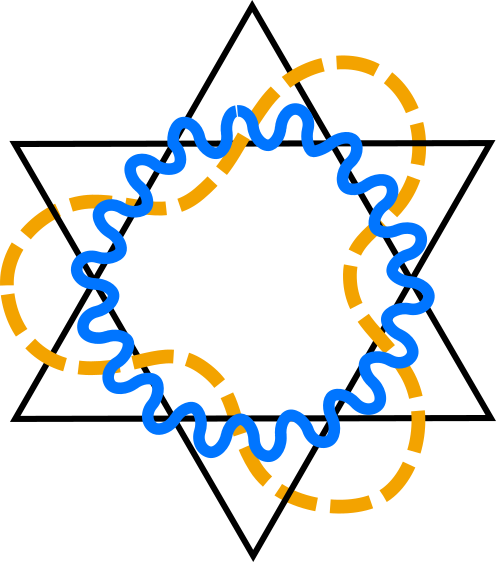}} \label{eq:Wpprime}
\end{equation}
This can be interpreted as an $f$-anyon traveling around a hexagon\footnote{Mathematically, this corresponds to this string operator defining an \emph{anomalous} 1-form symmetry \cite{Gaiotto_2015,McGreevy22}. This is key to being able to `fermionize' \cite{Verstraete05ferm,Kitaev06,Whitfield16,Chen18ferm,Chen20ferm,Po21,Chen22ferm} the system and solve it using free fermions (Sec.~\ref{subsec:intermediatespinon}). In contrast, $W_p$ in $H_\e$ corresponds to a \emph{non-anomalous} 1-form symmetry, which allows us to perform a Kramers-Wannier transformation (Sec.~\ref{subsec:monomerduality}). We thank Nat Tantivasadakarn for discussions on this topic.}. We will find that in the ground state, $W_p'=1$ for the chiral model, whereas the achiral model undergoes a flux transition ($W_p'=-1$ for small fields, $W_p'=1$ for large fields; see Sec.~\ref{subsec:intermediatespinon}). 

For both $H_\e$ and $H_\f$ (where we set $h_l=0$) it can be shown that the sign of $J$ can be toggled by a unitary transformation\footnote{Using the notation of Appendix~\ref{app:spinthreehalf}: for $H_\e$ one uses strings of $\prod \mathcal X^\alpha$ whereas for $H_\f$ we use $\prod \mathcal S^\alpha$.}. The physical interpretation of $J<0$ is as a self-avoiding loop model on the kagom\'e lattice (without any sharp turns), i.e., a toric code model with additional constraints. We henceforth set $J>0$.

\subsubsection{$H_\textrm{ruby}$: ruby lattice Ising model}

The third choice of transverse field is given by:
\begin{align}
\boxed{ \mathcal X_\textrm{ruby} } &:= \sum_{\alpha=x,y,z} \mathcal X^\alpha \left(1 + \mathcal Z^\alpha \right) \notag \\
&= \boxed{ \raisebox{-6pt}{\includegraphics[scale=0.55]{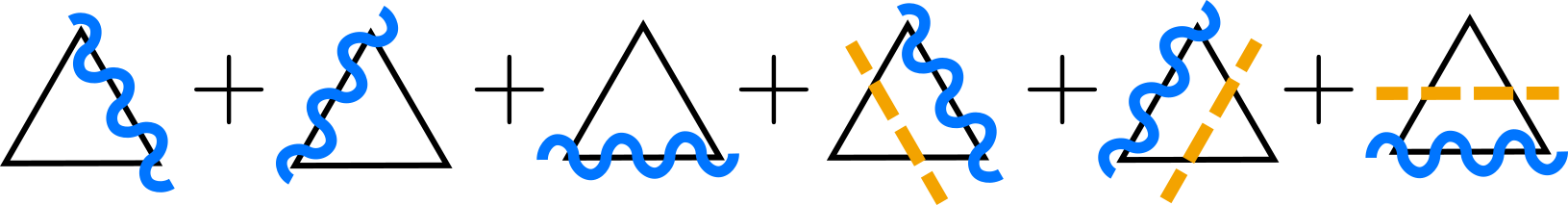}} } \label{eq:Xruby}
\end{align}
This term naturally arises in the spin-$1/2$ Ising model on the ruby lattice \cite{Verresen21} and corresponds to Eq.~\eqref{eq:Xintroruby} mentioned in the introduction (up to a factor of two). More precisely, as explained in Sec.~\ref{subsec:ruby}, we can equate a limit of $H_\textrm{ruby}$ in Eq.~\eqref{eq:model} with the PXP model on the ruby lattice $H_\textrm{PXP} = \frac{1}{2} \sum_i \left( \Omega P \sigma^x_i P - \delta \sigma^z_i \right)$ introduced in Ref.~\onlinecite{Verresen21}. Here $P$ projects into the subspace where an up-spin blockades the six nearest spins from being up, which physically can arise from a Rydberg blockade (this corresponds to the $U,V \to \infty$ limit in Sec.~\ref{subsec:ruby}). This PXP model was numerically argued to have a spin liquid for $0.5 \lessapprox |\Omega|/\delta \lessapprox 0.7$. The dictionary (see Sec.~\ref{subsec:ruby}) is that in the limit that our longitudinal field $h_l \to J$, more precisely $0<J-h_l\ll J$, we recover the PXP model with
\begin{equation}
\Omega = 4h_t \quad \textrm{and} \quad \delta = 4(J-h_l)  . \label{eq:rubydictionary}
\end{equation}

\subsection{Brief summary of the phase diagram(s)}

In Fig.~\ref{fig:phasediagram} we summarize the phase diagrams of Eq.~\eqref{eq:model}. While we discuss the phase diagram in more detail throughout this note, let us briefly discuss some of the salient features. This also serves as an outline for the rest of the paper.

\subsubsection{$H_\textrm{ruby}$}

As we have already mentioned, the model $H_\textrm{ruby}$ corresponds to a previously-studied spin-$1/2$ Ising model on the ruby lattice (see Sec.~\ref{subsec:ruby}). Using the dictionary in Eq.~\eqref{eq:rubydictionary}, the known phase diagram of the PXP model \cite{Verresen21} implies the spin liquid regime shown in Fig.~\ref{fig:phasediagram}. Evidently, the anyon fluctuations in $\mathcal X_\textrm{ruby}$ in Eq.~\eqref{eq:Xruby} lead to deconfinement; it is this observation that motivates much of the present study.

\subsubsection{$H_\e$}

Our charge-fluctuating model arises by simplifying the transverse-field of $\mathcal X_\textrm{ruby}$ in Eq.~\eqref{eq:Xruby} to $\mathcal X_\e$ in Eq.~\eqref{eq:Xmono}. Moreover, we set $h_l=0$; this simply assigns the same energy cost to monomers and double-dimers.

In Sec.~\ref{subsec:weakmonomer}, we show how the limit of small transverse field gives an exactly-solvable stabilizer\cref{footnote:stabilizer} Hamiltonian for the dimer liquid---a novel reincarnation of the kagom\'e lattice model introduced by Misguich, Serban and Pasquier \cite{Misguich02}. In Sec.~\ref{subsec:connectruby} we adiabatically connect this solvable point to the aforementioned ruby lattice spin liquid.

The fate for large field (i.e., strong charge fluctuations) depends on the sign of $h_t$. If $h_t<0$, the charges eventually condense, leading to a trivial phase. However, for $h_t>0$, the charge fluctuations are frustrated, and we find that the \2 spin liquid is robust as $h_t \to +\infty$. This claim turns out to be dual to the known disordered phase of the spin-1/2 transverse-field Ising model on the kagom\'e lattice (Sec.~\ref{subsec:monomerduality}).

In Sec.~\ref{subsec:strongmonomer} we show how this frustrated strong-field limit gives a novel spin-1 quadrupolar Kitaev model on the honeycomb lattice. Its interactions naturally descend from the $J$-couplings which enforce the dimer Gauss law. We find that its ground state is a remarkably robust \2 spin liquid with very short correlation length.

\subsubsection{$H_\f$}

If we instead consider fermionic fluctuations (corresponding to charge-flux bound states), we still find that these stabilize a \2 spin liquid for small field, but the phase diagram is richer.

Firstly, both the chiral and achiral models give rise to an exactly solvable dimer liquid for weak fields. However, only the achiral model is in the same phase as the solvable liquid we encountered in the discussion of $H_\e$ (i.e., the Misguich-Serban-Pasquier model \cite{Misguich02}). The chiral model leads to a distinct symmetry-enriched topological (SET) phase: it is `double odd' $\mathbb Z_2$ gauge theory, whereas conventional dimer liquids on the kagom\'e lattice (with one dimer per vertex) are only `odd'. Indeed, in the presence of translation symmetry that does not permute anyons, one can assign a `background' anyon to each unit cell \cite{Barkeshli_2019,Tarantino_2016}; the choice of trivial anyon corresponds to even gauge theory, while the `$e$'-anyon corresponds to odd; the double-odd corresponds to a fermion in each unit cell (see Sec.~\ref{subsec:weakspinon}).

Second, we elucidate the phase diagram for arbitrary fields by showing the models are solvable using the methods introduced by Kitaev \cite{Kitaev06} (Sec.~\ref{subsec:intermediatespinon}). The chiral model can be related to the Yao-Kivelson model (see Sec.~\ref{subsec:star}) which is known to have both a $\mathbb Z_2$ spin liquid as well as a non-Abelian chiral spin liquid \cite{Yao07}. We find a similar phase diagram for the achiral case, where the latter phase is replaced by a Majorana Fermi surface.

Lastly, the large-field limit transforms the $J$-term in Eq.~\eqref{eq:model} to the spin-1/2 Kitaev honeycomb model (see Sec.~\ref{subsec:strongspinon}).

\section{Quantum liquids from $e$-anyon fluctuations \label{sec:monomer}}

Let us consider $H_\e$ (i.e., the parent Hamiltonian in Eq.~\eqref{eq:model} with $\mathcal X_\e$ in Eq.~\eqref{eq:Xmono}). Using our graphical notation:
\begin{equation}
H = J \sum_\textrm{\tiny vertices} \raisebox{-12pt}{\includegraphics[scale=0.53]{gausslaw.pdf}} + h_t \sum_\textrm{\tiny triangles} \left( \raisebox{-6pt}{\includegraphics[scale=0.53]{Xe.png}} \right).
\end{equation}

We will show that in the weak-field limit, leading-order perturbation theory gives us a stabilizer Hamiltonian for a topological dimer liquid on the kagom\'e lattice (Sec.~\ref{subsec:weakmonomer}). (Later, in Sec.~\ref{subsec:connectruby} we adiabatically connect this solvable point to the spin-$1/2$ ruby lattice Ising model, i.e., Eq.~\eqref{eq:model} with Eq.~\eqref{eq:Xruby}.)
In Sec.~\ref{subsec:monomerduality} we show that for generic values of $h_t$, we can effectively solve for the ground state properties of $H_\e$ via a duality to a previously-studied model with a known phase diagram. This suggests that the spin liquid is very robust.
The strong-field limit of frustrated monomer fluctuations connects to a spin-1 Kitaev model (Sec.~\ref{subsec:strongmonomer}).

\subsection{Weak-field limit: dimer liquid stabilizer model \label{subsec:weakmonomer}}

We have already discussed in Sec.~\ref{sec:modelham} that the Ising term in Eq.~\eqref{eq:model} enforces the dimer constraint on the kagom\'e lattice (presuming the `antiferromagnetic' case $J>0$). This gives an exponentially large degeneracy if $h_t=h_l=0$. A straightforward exercise in degenerate perturbation theory for small transverse field $|h_t| \ll |J|$ for the monomer fluctuations \eqref{eq:Xmono} gives:
\begin{equation}
H = J \sum_\alpha \sum_{\langle i,j\rangle_\alpha} \mathcal Z_i^\alpha \mathcal Z_j^\alpha - \frac{63}{256} \frac{h_t^6}{|J|^5} \sum_p W_p + O\left( \frac{h_t^{10}}{J^9} \right), \label{eq:perturbation}
\end{equation}
where $W_p = \prod_{i=1}^6 \mathcal X_{p_i}^{\alpha_i}$ is the plaquette operator defined in Eq.~\eqref{eq:Wp} (i.e., each $\alpha_i$ corresponds to the bond which is sticking out of the plaquette). Note that this is a stabilizer\cref{footnote:stabilizer} Hamiltonian! It is instructive to depict it visually:
\begin{equation}
H_\textrm{stabilizer} = J \sum_v \raisebox{-8pt}{\includegraphics[scale=0.4]{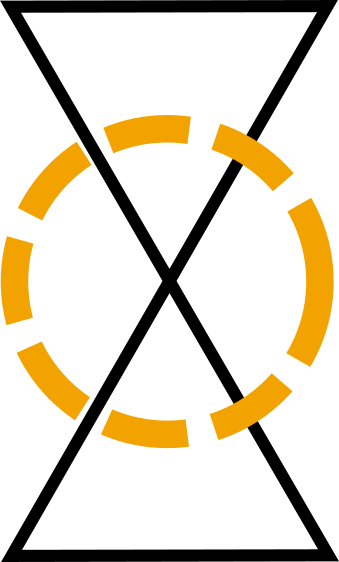}} - K \sum_p \raisebox{-18pt}{\includegraphics[scale=0.4]{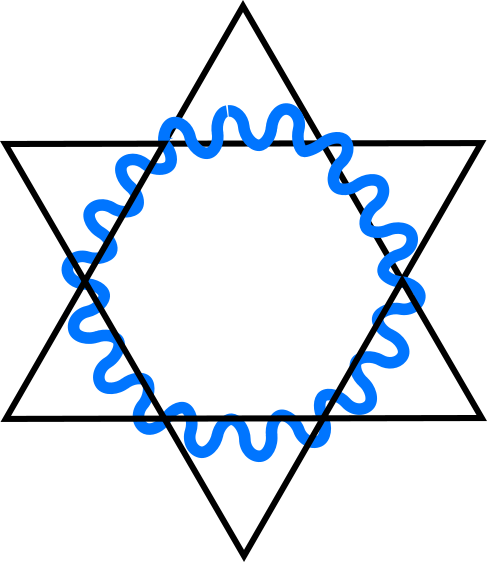}}, \label{eq:stabilizer}
\end{equation}
corresponding to a two-site vertex and six-site plaquette operator; we have $K = \frac{63}{256} \frac{h_t^6}{|J|^5}$ as the effective description of $H_\e$ for small $h_t$. The ground state is thus characterized by the stabilizers $G_v=-1$ \eqref{eq:gauss} and $W_p=1$.

The terms in Eq.~\eqref{eq:stabilizer} commute and square to unity. The first term is diagonal and enforces the dimer constraint (if $J>0$) and the second resonates all possible 32 dimer configurations on a kagom\'e lattice star. Hence, if we consider the dimer constraint as a hard constraint in the Hilbert space, these 32 dimer moves are the ones proposed by Misguich, Serban and Pasquier to give rise to a solvable dimer liquid \cite{Misguich02}. There are several benefits to rewriting their model in this spin-$3/2$ Hilbert space: (1) since the dimer constraint is no longer hardwired into the Hilbert space, we can define local off-diagonal operators such that the resonance term is just a product operator; (2) we do not require any arbitrary choice of reference dimer configuration (in contrast to the pseudospin representation of Ref.~\onlinecite{Misguich02}); (3) it is more natural when we interpolate to models where the dimer constraint is lightly violated (such as the ruby lattice model; see Sec.~\ref{subsec:connectruby}).

\subsection{Arbitrary field: duality to spin-1/2 transverse field Ising model on the kagom\'e lattice \label{subsec:monomerduality}}

For all values of the field $h_t$, the plaquette operator $W_p$ is a symmetry of $H_\e$ (see Eq.~\eqref{eq:Wpsym}). Using this, we can define a duality mapping of $H_\e$ to the transverse-field Ising model on the kagom\'e lattice. To do so, we define the spin-1/2 Pauli operator on a vertex $v$ of the kagom\'e lattice as $\sigma^x_v = \mathcal Z^\alpha_i \mathcal Z^\alpha_j$, where $\langle i,j\rangle$ is the corresponding bond of the honeycomb lattice and $\alpha \in \{x,y,z\}$ its orientation (see Fig.~\ref{fig:fromkagometohoneycomb}(a)). Moreover, for two neighboring vertices $v,v'$ of the kagom\'e lattice, we define $\sigma^z_{v} \sigma^z_{v'} = \mathcal X^\alpha_i$, where  $\alpha$ again corresponds to the orientation picked out by $\langle v,v'\rangle$ and $i$ is the site of the spin-3/2 honeycomb lattice corresponding to the kagom\'e triangle. (Equivalently, we can define a single $\sigma^z_v$ as a nonlocal $\mathcal X$-string; see Appendix~\ref{app:duality} for details.) We thus obtain the dual Hamiltonian on the kagom\'e lattice:
\begin{equation}
\tilde H_\e = J \sum_v \sigma^x_v + h_t \sum_{\langle v,v'\rangle} \sigma^z_v \sigma^z_{v'}. \label{eq:dual}
\end{equation}

In stating Eq.~\eqref{eq:dual}, we have implicitly assumed that the ground state of $H_\e$ is in the sector $W_p = 1$. If for a certain plaquette $W_p=-1$, the dual model \eqref{eq:dual} would obtain a frustrated bond around that hexagonal plaquette. Such frustration is unlikely to lower the energy, suggesting\footnote{We are not aware of any rigorous argument for this; especially in the regime $h_t <0 $ where triangles are frustrated, one might wonder whether introducing frustration along the hexagonal plaquettes could help. This is why we performed independent numerical checks.} that the ground state should indeed be in the $W_p=1$ sector. Note that we analytically derived this property in Eq.~\eqref{eq:perturbation} for small field. For more generic values of $h_t$, we have numerically tested and confirmed that $W_p=1$; we discuss the numerical set-up in more detail in the next subsection.

The phase diagram of this dual model \eqref{eq:dual} is known. First, let us observe that setting $h_t \to 0$ recovers the (stabilizer) Hamiltonian for the trivial paramagnet, consistent with the original theory being a \2 quantum spin liquid \eqref{eq:perturbation}. For large negative $h_t$, we expect Eq.~\eqref{eq:dual} to go into a symmetry-breaking ferromagnetic phase (dual to a trivial phase in our original model). This is indeed known to occur at $h_t \approx -0.34|J|$ \cite{Bloete02}. However, $h_t>0$ gives an Ising model with frustrated triangles.
Various approaches---including Monte Carlo simulations---have shown that this leads to a featureless paramagnet for all values of the transverse field \cite{Moessner00,Moessner01b,Nikolic05,Powalski13}; this paramagnet is adiabatically connected to the product state limit $h_t \to 0$. This implies that in our original spin-3/2 model, the \2 spin liquid is robust for all $h_t \gtrapprox -0.34|J|$, as shown in Fig.~\ref{fig:phasediagram}.

\subsection{Strong-field limit: spin-1 quadrupolar Kitaev model \label{subsec:strongmonomer}}

\begin{figure}
    \centering
\begin{tikzpicture}
\node at (2.2,3) {\includegraphics[scale=0.65]{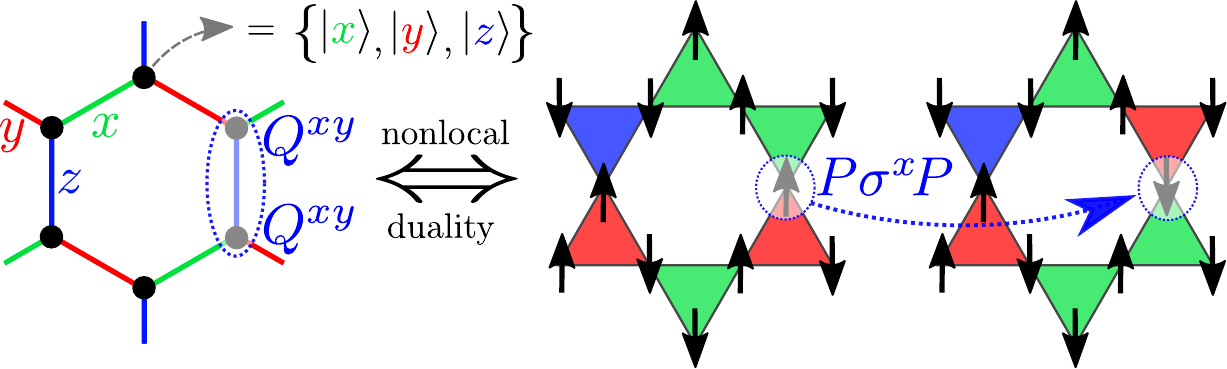}};
\node at (0,0) {\includegraphics[scale=0.25]{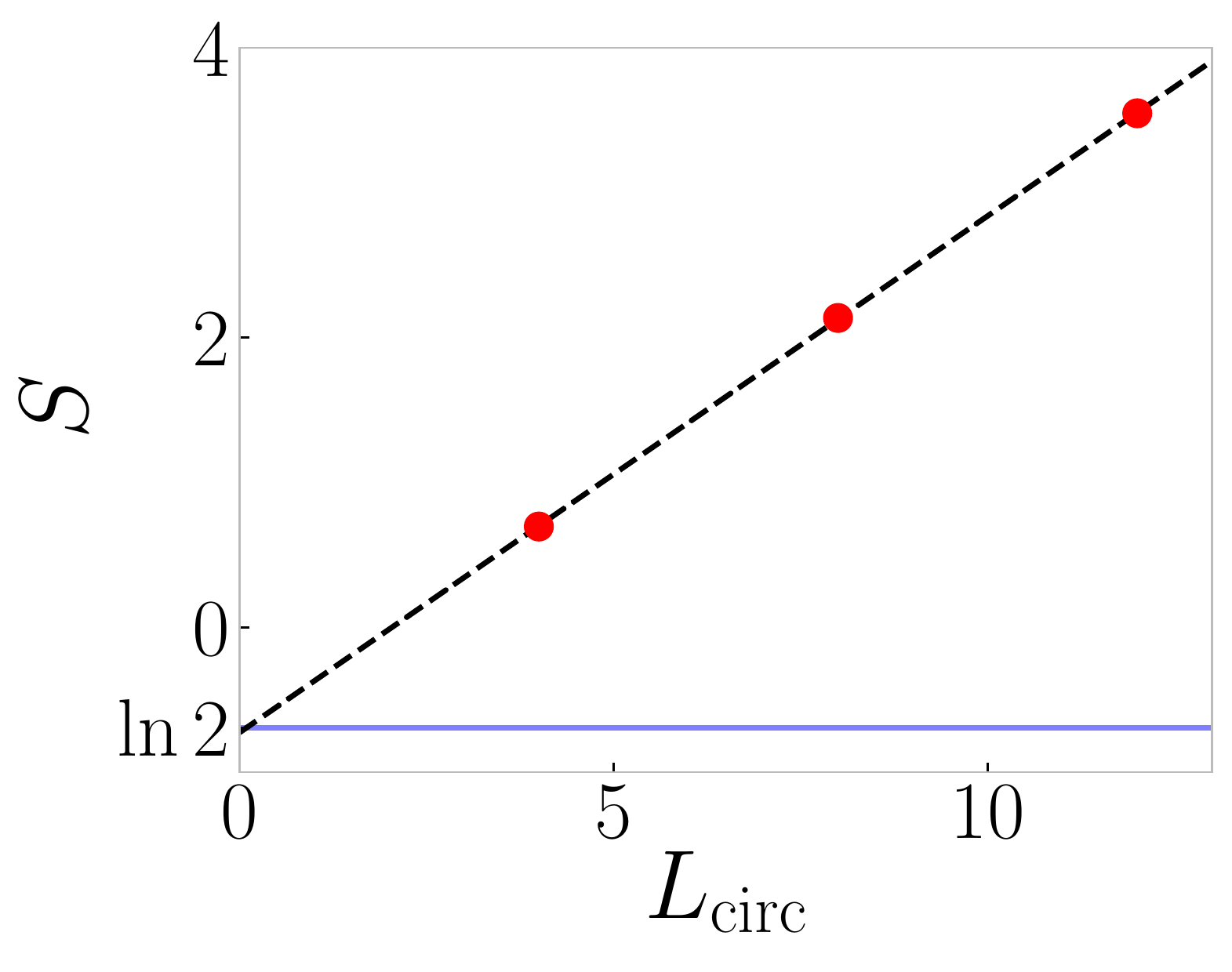}};
\node at (4.2,0) {\includegraphics[scale=0.25]{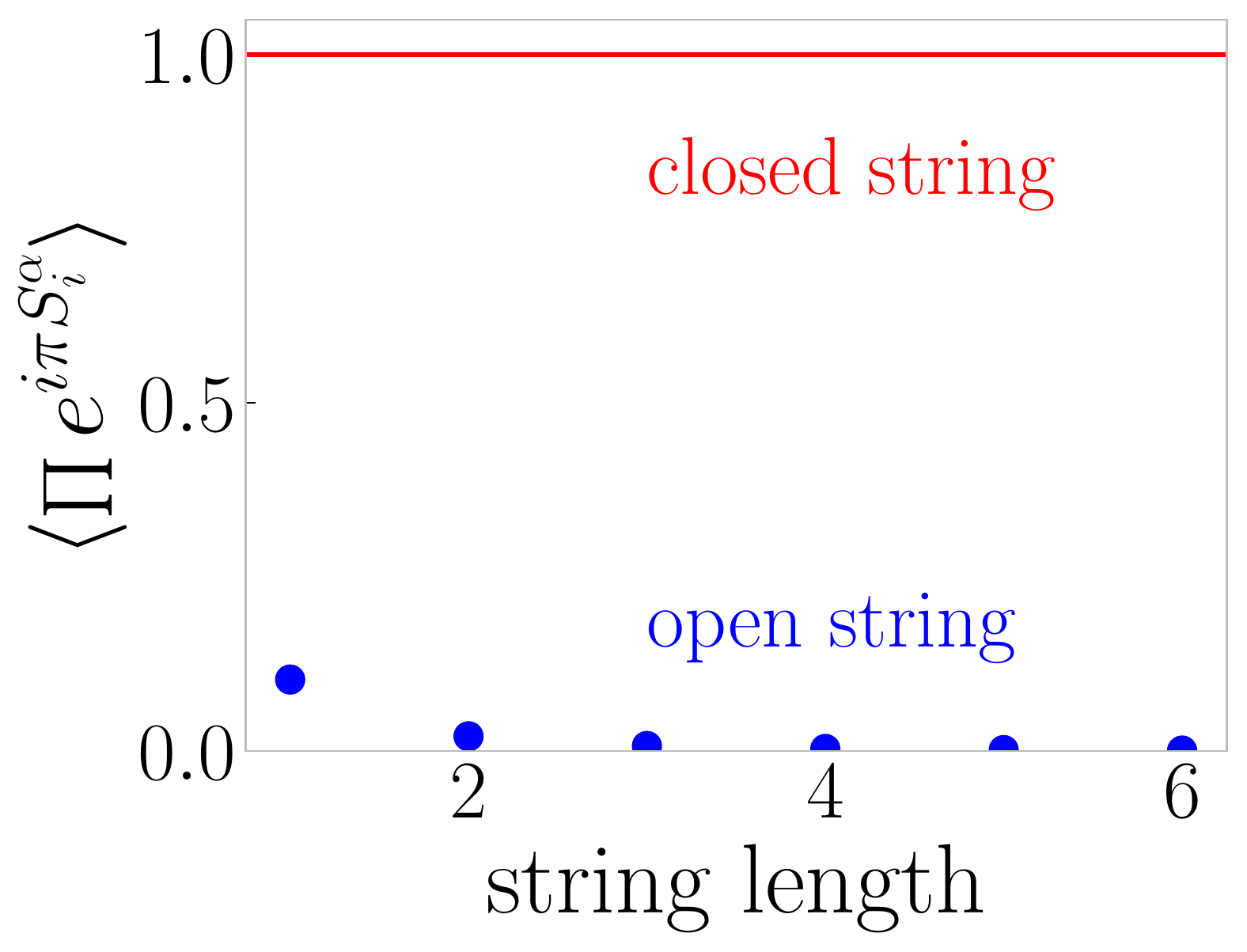}};
\node at (-2+0.1,1.5) {(b)};
\node at (-2+0.1,4) {(a)};
\end{tikzpicture}
    \caption{\textbf{Spin-1 quadrupolar Kitaev liquid and its dual frustrated PXP kagom\'e model.} If we consider our spin-3/2 parent model $H_\e$ of an emergent dimer model with $e$-anyon fluctuations, then in the large field/fluctuation limit, we obtain an effective spin-1 model on the honeycomb lattice (see Sec.~\ref{subsec:strongmonomer}). (a) Its interactions are Kitaev-like, except with quadrupole operators $Q^{\alpha \beta}$ rather than spin operators $S^\alpha$. This is dual to the frustrated Ising model on the kagom\'e lattice, in the weak-field limit where we project the transverse field $\sigma^x$ into a $PXP$-type term. (b) Since the latter is known to be disordered \cite{Moessner00,Moessner01b,Nikolic05,Powalski13}, we learn that the dual spin-1 quadrupolar Kitaev model forms a \2 topological liquid. We independently confirm this by finding a topological entanglement entropy $\gamma \approx \ln 2$, and by showing that while closed loops of the Wilson operator are unity in the ground state, open strings decay exponentially fast; the latter is a sign of 1-form symmetry breaking which defines topological order \cite{Gaiotto_2015,McGreevy22}, and can be interpreted as a limiting case of the Fredenhagen-Marcu string order parameter \cite{Fredenhagen83,Fredenhagen86,Marcu86,Gregor11}, detecting that $e$-anyons have \emph{not} condensed.}
    \label{fig:spin1quadrupolar}
\end{figure}

We thus know that even infinitely-strong $e$-anyon fluctuations $h_t \to +\infty$ give rise to a \2 spin liquid, which is moreover adiabatically connected to the emergent dimer liquid at $h_t \to 0$. The effective description of this large-field limit is rather interesting. For large $h_t>0$, $\sum_{\alpha=x,y,z} \mathcal X^\alpha$ wants to be as negative as possible, which is frustrated since $\mathcal X^x \mathcal X^y \mathcal X^z = 1$. Hence, the best we can do for each spin-3/2 site of the honeycomb lattice is to have one $\mathcal X^\alpha=1$ and the other two being $-1$. This gives rise to an effective qutrit whose basis states can be labeled by $\{ |\tilde x\rangle,|\tilde y\rangle, |\tilde z \rangle \}$ defined by $\mathcal X^\alpha |\tilde \beta\rangle = -(-1)^{\delta_{\alpha \beta}} |\tilde \beta\rangle$. If we equate this with the $p$-orbital basis of a spin-1, $\{|x\rangle,|y\rangle,|z\rangle\rangle\}$\footnote{To wit \cite{toththesis}, $|x\rangle = \frac{i}{\sqrt{2}}\left(\ket{\uparrow}- \ket{\downarrow}\right)$, $|y\rangle= \frac{1}{\sqrt{2}} \left( \ket{\uparrow} + \ket{\downarrow} \right)$ and $|z\rangle = -i|0\rangle$.}, then $\mathcal X^\alpha = e^{i \pi S^\alpha}$. Hence, we already see that our effective spin-1 model will have a conserved quantity for every plaquette: $W_p =\prod \mathcal X^\alpha = \prod e^{i \pi S^\alpha}$; this is the same local integral of motion of the spin-1 Kitaev model \cite{Baskaran08,Zhu20,Hickey20,Dong20,Koga20,Khait21,Lee21,Chen22,Bradley22}!

To obtain the effective spin-1 Hamiltonian induced by the $J$-term, we need only project $\mathcal Z^\alpha$ into this qutrit\footnote{The simplest way to obtain this is by first performing a Hadamard transformation $\mathcal Z^\alpha \leftrightarrow \mathcal X^\alpha$ (see Sec.~\ref{subsec:hadamard}) after which our qutrit corresponds to the basis in Fig.~\ref{fig:fromkagometohoneycomb}(a) where we project out $|0\rangle$; then the projected operator can be read off from Eq.~\eqref{eq:Xdef}.}:
\begin{equation}
\begin{array}{ccl}
P \mathcal Z^xP &= &\ket{\tilde y}\bra{\tilde z} + \ket{\tilde z}\bra{\tilde y}, \\ 
P \mathcal Z^y P &= &\ket{\tilde z}\bra{\tilde x} + \ket{\tilde x}\bra{\tilde z},\\
P \mathcal Z^z P &= &\ket{\tilde x}\bra{\tilde y} + \ket{\tilde y}\bra{\tilde x}.
\end{array}
\end{equation}
In terms of spin-1 operators, this action coincides with that of the quadrupole operator\footnote{In contrast, the spin operators act as $S^\alpha \ket{\beta} = i \varepsilon_{\alpha \beta \gamma} \ket{\gamma}$ in the $p$-orbital basis. We can thus interpret the quadrupole operators as a time-reversal invariant version of the spin operators. Indeed, $-Q^{xy} = \{ Q^{yz},Q^{xz} \}$ compared to $iS^z = [S^x,S^y]$.} $Q^{\alpha \beta} = \{ S^\alpha, S^\beta \}$:
\begin{equation}
P\mathcal Z^x P = -Q^{yz},  \; P \mathcal Z^y P = -Q^{xz}, \; P \mathcal Z^z P = -Q^{xy}.
\end{equation}
In conclusion, the $h_t \to +\infty$ limit of $H_\e$ gives a spin-1 honeycomb model with
\begin{equation}
H_\textrm{eff} = J_x \sum_{\langle i,j\rangle_x} Q^{yz}_i Q^{yz}_i + 
J_y \sum_{\langle i,j\rangle_y} Q^{xz}_i Q^{xz}_i +
J_z \sum_{\langle i,j\rangle_z} Q^{xy}_i Q^{xy}_i \label{eq:Hquadrupolar}
\end{equation}
where in our case $J=J_x=J_y=J_z$.
This is the spin-1 Kitaev model, except with the spin operators replaced by quadrupole operators. Indeed, there is arguably not a unique natural choice of what `spin-1 Kitaev model' should mean, unlike for the spin-1/2 case where quadrupole operators vanish ($\{ \sigma^x, \sigma^y\} = 0$).

Remarkably, such a `spin-1 quadrupolar Kitaev model' has not yet appeared in the literature. (However, recently a spin-3/2 version was studied \cite{Farias20}, which requires perturbations to lift extensive degeneracies.) The analysis in Sec.~\ref{subsec:monomerduality} shows that it is a spin liquid. In fact, one can repeat the nonlocal duality mapping (to a spin-1/2 kagom\'e Ising model) for the spin-1 model: the qutrit encodes which of the three bonds of the dual kagom\'e lattice has a frustrated Ising interaction (see Fig.~\ref{fig:spin1quadrupolar}(a)). Each quadrupolar Kitaev interaction then simply maps to a $P\sigma^x P$ spin-flip in the constrained manifold of frustrated Ising states\footnote{The Hilbert space dimension of this frustrated space is $\sim 4.50636^T$ where $T$ is the number of kagom\'e triangles \cite{Kano53}, which is remarkably well-approximated by a naive Pauling estimate for the spin-1 model: $(3^2/2)^T$, based on having two spin-1's per unit cell and one $W_p=1$ plaquette condition.}. The trivial disordered phase of this dual model \cite{Moessner00,Moessner01b,Nikolic05,Powalski13} shows that Eq.~\eqref{eq:Hquadrupolar} is a \2 spin liquid.

The above mapping relies on the ground state being in the $W_p=1$ sector. We have numerically confirmed that this is true using density matrix renormalization group (DMRG) \cite{White92,White93} simulations on infinitely-long cylinders \cite{Stoudenmire12} for the XC-$n$ geometry\footnote{This refers to a lattice which is periodic in the $y$-direction as plotted in Fig.~\ref{fig:fromkagometohoneycomb}(a) with $n/2$ unit cells.} with $n=4,8,12$. The simulations were performed using the TeNPy library \cite{Hauschild18}. We find a remarkably small correlation length $\xi \approx a$ (with $a$ being the lattice spacing of the honeycomb lattice); correspondingly, a bond dimension $\chi = 2000$ was sufficient to converge our quantities of interest, even for the rather large XC-12 cylinder. In addition to confirming $W_p=1$, we have also extracted\footnote{We consider a bipartition of the infinite cylinder \cite{Jiang12,Jiang13}. For each circumference in Fig.~\ref{fig:spin1quadrupolar}(b), we plot the entanglement entropy averaged over all four distinct topological ground states to minimize finite-size effects. The dashed line is a fit to $a L_\textrm{circ}-\gamma$.} topological entanglement entropy \cite{Kitaev06b,Levin06} (see Fig.~\ref{fig:spin1quadrupolar}(b)), which is close to the expected value \cite{Hamma05}. Moreover, we find that although closed strings of $\mathcal X^\alpha$ (or $e^{i \pi S^\alpha}$ in the spin-1 notation) leave the ground state invariant, \emph{open} strings have correlation functions which decay quickly to zero. This is sign of 1-form symmetry breaking which defines topological order \cite{Gaiotto_2015,McGreevy22}; it is a special case of the Fredenhagen-Marcu string order parameter where the denominator is unity \cite{Fredenhagen83,Fredenhagen86,Marcu86,Gregor11}. The vanishing of this topological string operator is dual to the corresponding Ising model not having long-range order, confirming our above analysis.

In conclusion, we have found that in the presence of strongly-frustrated $e$-anyon fluctuations, we obtain an effective spin-1 model where the dimer-enforcing interactions have morphed into a novel spin-1 Kitaev-type model. Its ground state is a remarkably robust \2 spin liquid, with a correlation length of roughly a lattice spacing. In that sense, it seems like a more ideal spin liquid compared to the previously-studied spin-1 Kitaev model (with spin rather than quadrupole operators) \cite{Baskaran08,Zhu20,Hickey20,Dong20,Koga20,Khait21,Lee21,Chen22,Bradley22}, where it remains unclear whether it is a \2 or gapless spin liquid. In fact, in Sec.~\ref{subsec:kitaevpath} we discuss a natural path of Hamiltonians interpolating from our novel model to the conventional spin-1 Kitaev model, which could potentially be used to provide new insight into the latter.

\section{Quantum liquids from $f$-anyon fluctuations \label{sec:spinon}}

Here we study the emergent dimer model in the presence of fermionic $f$-anyon fluctuations, i.e., $H_\f$ in Eq.~\eqref{eq:model} with the transverse field $\mathcal X_\f$ in Eq.~\eqref{eq:Xspinon}. Similar to the case with $e$-anyon fluctuations, we find that weak fluctuations give rise to an exactly solvable stabilizer Hamiltonian for a dimer liquid. However, the two different choices of (a)chirality in Eq.~\eqref{eq:Xspinon} lead to \emph{distinct} symmetry-enriched versions  (Sec.~\ref{subsec:weakspinon}). More generally, we find that $H_\f$ is exactly-solvable for all values of $h_t$, leading to non-Abelian and gapless topological phases for larger values of $h_t$ (Sec.~\ref{subsec:intermediatespinon}). Finally, similar to the case studied in the previous section, we find that the large-field limit $h_t \to \infty$ gives an effective Kitaev model, although now for $S=1/2$ rather than $S=1$ (Sec.~\ref{subsec:strongspinon}).

\subsection{Weak-field limit: distinct dimer liquid SETs \label{subsec:weakspinon}}

If we repeat the degenerate perturbation theory exercise from Sec.~\ref{subsec:weakmonomer} for $H_\f$, then to leading-order in the transverse field, we obtain:
\begin{equation}
H = J \sum_\alpha \sum_{\langle i,j\rangle_\alpha} \mathcal Z_i^\alpha \mathcal Z_j^\alpha \pm \frac{63}{256} \frac{h_t^6}{|J|^5} \sum_p W_p' + O\left( \frac{h_t^{10}}{J^9} \right), \label{eq:perturbationchiral}
\end{equation}
where the sign $\pm$ is the same as in Eq.~\eqref{eq:Xspinon}; to wit, $+$ is the achiral model and $-$ is the chiral model. Here $W_p'$ is the plaquette operator defined in Eq.~\eqref{eq:Wpprime}, which is a conserved quantity for $H_\f$ for arbitrary values of $h_t$ and $J$ (with $h_l = 0$). We can interpret $W_p'$ as the worldline of an $f$-anyon around the plaquette. Eq.~\eqref{eq:perturbationchiral} implies that $W_p' = \mp 1$ in the ground state, i.e., $W_p' = -1$ for the achiral model and $W_p' = 1$ for the chiral model.

The different eigenvalues for the plaquette operator tell us that the fixed-point dimer liquids of the chiral and achiral models are distinct symmetry-enriched topological (SET) phases in the presence of translation symmetry. Moreover, the achiral model is in the same phase as the dimer liquid we found for the fluctuating monomers in Eq.~\eqref{eq:perturbation}. Indeed, in the dimer limit (where we have the Gauss law in Eq.~\eqref{eq:gauss}) we can equate $W_p' = -W_p$; this follows from the observation that the parity loop in $W_p'$ encloses an odd number of vertices of the kagom\'e lattice. Since the achiral model has $W_p'=-1$, it satisfies $W_p=+1$ in the dimer limit, agreeing with Eq.~\eqref{eq:perturbation}. In contrast, the chiral model has $W_p=-1$ and is thus a distinct SET. We might call this a `double odd' $\mathbb Z_2$ gauge theory: it has one $e$-anyon per vertex and one $m$-anyon per plaquette! In other words, whereas a usual dimer model with an odd number of dimers touching a unit cell (as in the kagom\'e dimer model)   has one `background' $e$-anyon per unit cell, the chiral model leads to a `background' $f$-anyon.

\subsection{Exact solution for arbitrary field: Majorana Fermi surfaces and chiral non-Abelian Ising topological order \label{subsec:intermediatespinon}}

The model with fermionic fluctuations is exactly solvable for arbitrary values of $h_t$ and $J$ (with $h_l=0$). Our strategy closely follows the famous solution of the spin-1/2 Kitaev model \cite{Kitaev06}, which involves introducing four Majorana operators satisfying a parity constraint $b^x b^y b^z c = 1$, such that the spin-1/2 Pauli operators can be written as $\sigma^\alpha_n = i b^\alpha_n c_n$. Rewriting the Kitaev model with those operators transforms it into a free-fermion problem. Here we similarly solve our model by using the Majorana representation of a spin-3/2 \cite{Wang09,Yao09,Yao11,Chua11,Whitsitt12,Natori16,Natori17,deCarvalho18,Natori18,Seifert20,Farias20,Natori20,Chulliparambil20,Ray21,Jin21,Chulliparambil21}.

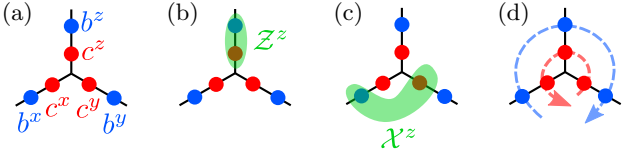
\begin{figure}
    \centering
    \begin{tikzpicture}
    \node at (0,0) { \includegraphics[scale=0.5]{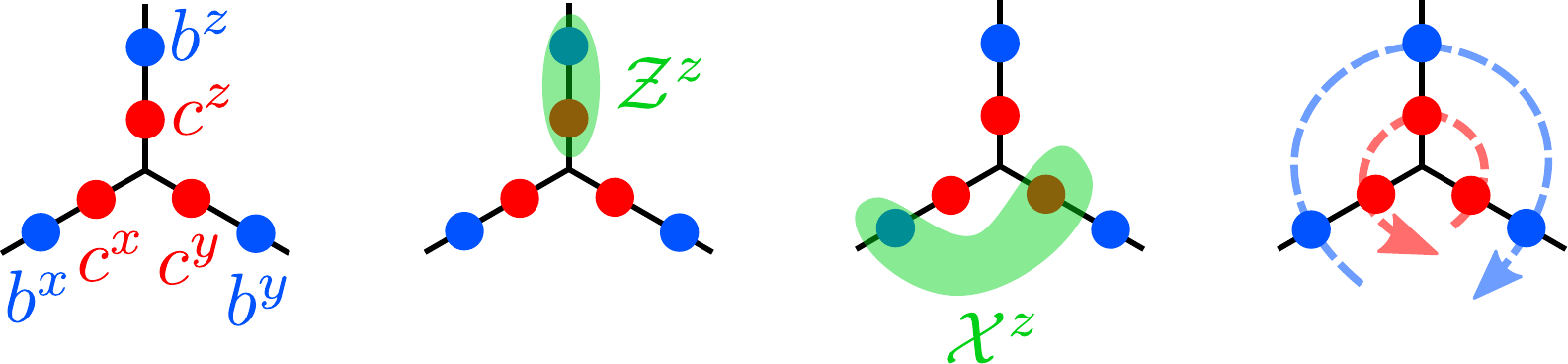}};
    \node at (-4,0.85) {(a)};
    \node at (-4+2.2,0.85) {(b)};
    \node at (-4+2*2.2,0.85) {(c)};
    \node at (-4+3*2.2,0.85) {(d)};
    \end{tikzpicture}
    \caption{\textbf{Majorana representation of a spin-3/2.} (a) Our 4-state system (encoding the dimers on a triangle of the kagom\'e lattice) can be represented by six Majorana operators with a parity constraint $i b^x b^y b^z c^x c^y c^z = 1$. (b) The diagonal string operator $\mathcal Z^\alpha$ introduced in Eq.~\eqref{eq:Zdef} (which measures the parity of dimers) is represented as $i b^\alpha c^\alpha$ (shown for $\alpha= z$). (c) The off-diagonal string operator $\mathcal X^\alpha$ introduced in Eq.~\eqref{eq:Xdef} (which shuffles dimers around) is represented as $i b^{\alpha+1} c^{\alpha-1}$ on the A sublattice (shown for $\alpha = z$). (d) By permuting the Majoranas, e.g., $c^x \to c^y \to c^z \to c^x$ and $b^x \to b^z \to b^y \to b^x$, the off-diagonal string operator $\mathcal X^\alpha$ is transformed into the diagonal operator $\mathcal Z^\alpha$ \cite{Verresen21} thereby making it a measurable observable in Rydberg atom arrays \cite{Semeghini21}. }
    \label{fig:majoranas}
\end{figure}

\emph{Spin-3/2 in the Majorana Representation.} To every site of the honeycomb lattice, we associate six Majorana operators\footnote{To wit, these are Hermitian fermionic operators which square to the identity.} satisfying the parity constraint $i b^x b^y b^z c^x c^y c^z = 1$. Note that this has the correct dimension ($\sqrt{2}^6/2=4$) to represent a 4-level system. We orient the operators on the lattice as shown in Fig.~\ref{fig:majoranas}, with $\alpha=x,y,z$ following the orientation convention of Fig.~\ref{fig:fromkagometohoneycomb}. We call $b^\alpha$ the bond Majoranas, and $c^\alpha$ the matter Majoranas. If we equate $\mathcal Z^\alpha = i b^\alpha c^\alpha$ (Fig.~\ref{fig:majoranas}(b)), then the parity constraint correctly implies $\mathcal Z^x \mathcal Z^y \mathcal Z^z=1$. For $\mathcal X^\alpha$ it will be convenient to have a different pairing for the two sublattices:
\begin{align}
\textrm{on A:} \quad \mathcal X^x &= i b^y c^z, \quad \mathcal X^y =ib^z c^x, \quad \mathcal X^z = i b^x c^y, \\
\textrm{on B:} \quad \mathcal X^x &= i b^z c^y, \quad \mathcal X^y =ib^x c^z, \quad \mathcal X^z = i b^y c^x,
\end{align}
where A (B) refers to the odd (even) sites of the honeycomb lattice (i.e., upward- and downward-pointing triangles of the kagom\'e lattice respectively). These also satisfy $ \mathcal X^x \mathcal X^y \mathcal X^z = 1$.

Let us now rewrite $H_\f$ in this notation. First, the $J$-term, which energetically enforces the dimer constraint, becomes:
\begin{equation}
J\mathcal Z^\alpha_m \mathcal Z^\alpha_n = -J\left( i b^\alpha_m b^\alpha_n \right) \left( i c^\alpha_m c^\alpha_n \right) = - i J \hat u_{m,n} c^\alpha_m c^\alpha_n.
\end{equation}
While is this is a four-body term, we find that---similar to the Kitaev model solution \cite{Kitaev06}---the bond pairing $\hat u_{m,n} = \pm 1$ is a local integral of motion. Indeed, the on-site fermionic fluctuations $\mathcal X_\f$ \eqref{eq:Xspinon} do not involve the bond Majoranas:
\begin{align}
\textrm{on A:} \quad \mathcal X_\f &= i \mathcal X^x \mathcal Z^y + i \mathcal X^y \mathcal Z^z + i \mathcal X^z \mathcal Z^x \\
&= -i\left( c^y c^z + c^z c^x + c^x c^y \right), \label{eq:XA} \\
\textrm{on B:} \quad \mathcal X_\f &= \pm \left(i \mathcal X^x \mathcal Z^z + i\mathcal X^y \mathcal Z^x + i \mathcal X^z \mathcal Z^y\right) \\
&= \pm i \left( c^y c^z + c^z c^x + c^x c^y \right). \label{eq:XB}
\end{align}
Note that the \emph{chiral} model corresponds to choosing the minus sign, in which case we see that the Majorana fermions hop with the same chirality on both triangles; in contrast, the \emph{achiral} model does not have net chirality per unit cell.

To determine the ground state value of $\hat u_{m,n} = \pm 1$, we need to relate it to the conserved plaquette operator $W_p'$ \eqref{eq:Wpprime}.
A simple calculation shows that:
\begin{equation}
W_p' = \prod_{(m,n) \in \partial p} \hat u_{m,n},
\end{equation}
where we now fix our ordering such that $\hat u_{m,n}$ has $m$ on an A sublattice and $n$ on a B sublattice. In the previous subsection on the limit of weak transverse field $|h_t| \ll J$, we derived that $W_p' = 1$ ($W_p'=-1$) for the chiral (achiral) model. Hence, for the chiral model we can choose the gauge \cite{Kitaev06} where $u_{m,n} = 1$ on all bonds, giving a translation-invariant free-fermion problem with a six-site unit cell---two triangles per unit cell with three matter Majoranas each. However, for the achiral model we have to work with a unit cell of 12 Majorana fermions, where we make $u_{m,n} = -1$ on one bond to ensure this background flux condition.

We have thus reduced $H_\f$ to a free-fermion problem, from which we can easily extract the phase diagram. In fact, for our chiral model this happens to coincide with that of the Yao-Kivelson model \cite{Yao07} (see also Sec.~\ref{subsec:star}). In particular, at $|h_t| = |J|/\sqrt{3}$, our $\mathbb Z_2$ dimer liquid undergoes a transition into a gapped non-Abelian phase with chiral Ising topological order. For the Yao-Kivelson model, and hence for our chiral model, it is known that the ground state remains in the flux-free sector for all values of the tuning parameters \cite{Yao07}. However, for our achiral model, we find a flux transition: whereas $W_p' = -1$ for $|h_t| < |J|$, it changes to $W_p' =1$ for $|h_t|>|J|$. The resulting phase diagram is shown in Fig.~\ref{fig:phasediagram}: the achiral model also has a transition at $|h_t| = |J|/\sqrt{3}$, but now into a gapless phase supporting a Majorana Fermi surface. The aforementioned flux transition occurs within this phase, leading to a discontinuous change of the Fermi surface. A similar phenomenology was studied in Refs.~\onlinecite{Gazit_2017,Heqiu22}.

\subsection{Strong-field limit: spin-$1/2$ Kitaev model \label{subsec:strongspinon}}

In Sec.~\ref{subsec:strongmonomer} we saw that $H_\e$ effectively became a spin-1 Kitaev-type model in the large-field limit. Here we obtain a similar result: $H_\f$ reduces to the (gapless isotropic) spin-1/2 Kitaev model. More precisely,
\begin{equation}
\lim_{|h_t| \to \infty} H = \mp \frac{J}{3} \sum_{\alpha=x,y,z} \sum_{\langle i,j\rangle_\alpha} \sigma^\alpha_i \sigma^\alpha_j,
\label{eq:Hspinhalf}
\end{equation}
where the sign implies that the achiral (chiral) model gives rise to the FM (AFM) Kitaev model.

One way of seeing this is by utilizing the Majorana description of the previous subsection. Let us consider $\mathcal X_\f$ in Eq.~\eqref{eq:XA}. Note that this commutes with
\begin{equation}
c = \frac{c^x+c^y+c^z}{\sqrt{3}}. \label{eq:c}
\end{equation}
For large $h_t \mathcal X_\f$, we will pin two of the three matter Majoranas, leaving Eq.~\eqref{eq:c} as the remaining free matter Majorana. In other words, we can think of the large $h_t$ field as \emph{fusing} the three Majoranas into one. In this limit we obtain the effective parity condition $b^x b^y b^z c = 1$ (see Appendix~\ref{app:spinthreehalf}). We thus recover the Majorana description of a single spin-1/2 \cite{Kitaev06}. To map our operators, we can simply replace $c^\alpha \to c/\sqrt{3}$. In particular, $\mathcal Z^\alpha = i b^\alpha c^\alpha \to i b^\alpha c/\sqrt{3}$. Hence, projecting $\mathcal Z^\alpha$ into this constrained space gives us the Pauli operator $\frac{\sigma^\alpha}{\sqrt{3}}$.

In the achiral case (where Eq.~\eqref{eq:XB} has a different sign), repeating the above exercise leads to the substitution $\mathcal Z^\alpha \to - \frac{\sigma^\alpha}{\sqrt{3}}$ on the B sublattice. This is the reason that the chiral and achiral models lead to a different sign in Eq.~\eqref{eq:Hspinhalf}.

We emphasize that this provides a striking connection between dimer models and Kitaev physics: the $J$-term started its life as a dimer-constraint-enforcing condition, but in the presence of large (fermionic) dimer-defect fluctuations, it naturally morphed into the Kitaev interaction.

\section{Applications \label{sec:applications}}

In this section, we utilize the insights gained from the above model---and more generally the spin-3/2 Hilbert space and its operators. In particular, in Sec.~\ref{subsec:connectruby} we adiabatically connect the solvable dimer liquid we found in Eq.~\eqref{eq:perturbation} to the previously-studied ruby lattice model \cite{Verresen21}, thereby confirming that the latter is indeed a spin liquid. Sec.~\ref{subsec:hadamard} uses the Majorana representation to elucidate the Hadamard transformation proposed in Ref.~\onlinecite{Verresen21} and used in Ref.~\onlinecite{Semeghini21} to measure the off-diagonal string operators in a Rydberg atom array. We then revisit our newly discovered spin-1 quadrupolar Kitaev model in Sec.~\ref{subsec:kitaevpath} and show that there is a natural relation to the spin-1 Kitaev model. Sec.~\ref{subsec:toric} discusses the connection between the solvable dimer liquid \eqref{eq:perturbation} and the toric code model. Lastly, Sec.~\ref{subsec:XYZ} uses the spin-3/2 framework to demonstrate how one can construct and analyze generalizations of our parent model, focusing on a `$XYZ$'-type model for illustrative purposes, where we again find robust spin liquids due to anyon fluctuations. 

\subsection{Adiabatically connecting ruby lattice spin liquid to solvable stabilizer Hamiltonian \label{subsec:connectruby}}

Let us denote the model in Eq.~\eqref{eq:model} as $H_a(J,h_t,h_l) $. We can then consider interpolating the ruby lattice model to the solvable stabilizer Hamiltonian found in Eq.~\eqref{eq:stabilizer}:
\begin{equation}
H(\lambda) = (1-\lambda) \; H_\textrm{ruby}(J,h_t,h_l) + \lambda \; H_\textrm{stabilizer}(J,J). \label{eq:Hlambda}
\end{equation}
This particular choice of interpolation keeps the diagonal Ising interaction constant throughout, and we have set $K=J$ in Eq.~\eqref{eq:stabilizer}.
For concreteness, we consider $h_l = 0.9 J$ and $h_t = 0.06 J$, i.e., $\frac{h_t}{J-h_l} = 0.6$, such that for $\lambda=0$ we start in the middle of the claimed $\mathbb Z_2$ spin liquid in Fig.~\ref{fig:phasediagram}, as reported in Ref.~\onlinecite{Verresen21}.

As explained in Sec.~\ref{sec:monomer}, we use cylinder DMRG to obtain the ground state of this two-dimensional system by wrapping it on infinitely-long cylinders with variable circumference. As we tune $\lambda: 0 \to 1$, we find a finite correlation length along this path, which moreover is converged in system size---as shown in Fig.~\ref{fig:rubyinterpolation}(a), the result barely changes upon doubling the circumference. Similar to Sec.~\ref{subsec:strongspinon}, we extract the topological entanglement entropy as plotted in Fig.~\ref{fig:rubyinterpolation}(b), which is roughly constant along the whole path. Finally, the expectation values of the two stabilizers, i.e. $-G_v$ for a vertex (see Eq.~\eqref{eq:gauss}) and $W_p$ for a plaquette (see Eq.~\eqref{eq:Wp}) monotonically increase towards unity as we approach the exactly solvable limit. This confirms that the claimed $\mathbb Z_2$ spin liquid \cite{Verresen21} connects to the fixed-point dimer liquid. Moreover, we have already analytically demonstrated that this same fixed-point limit connects to the $H_\e$ and (achiral) $H_\f$ models studied in Secs.~\ref{sec:monomer} and \ref{sec:spinon}, respectively.


\begin{widetext}

\begin{center}
\begin{figure*}[h]
\begin{tikzpicture}
\node at (0,0.1) {\includegraphics[scale=0.36]{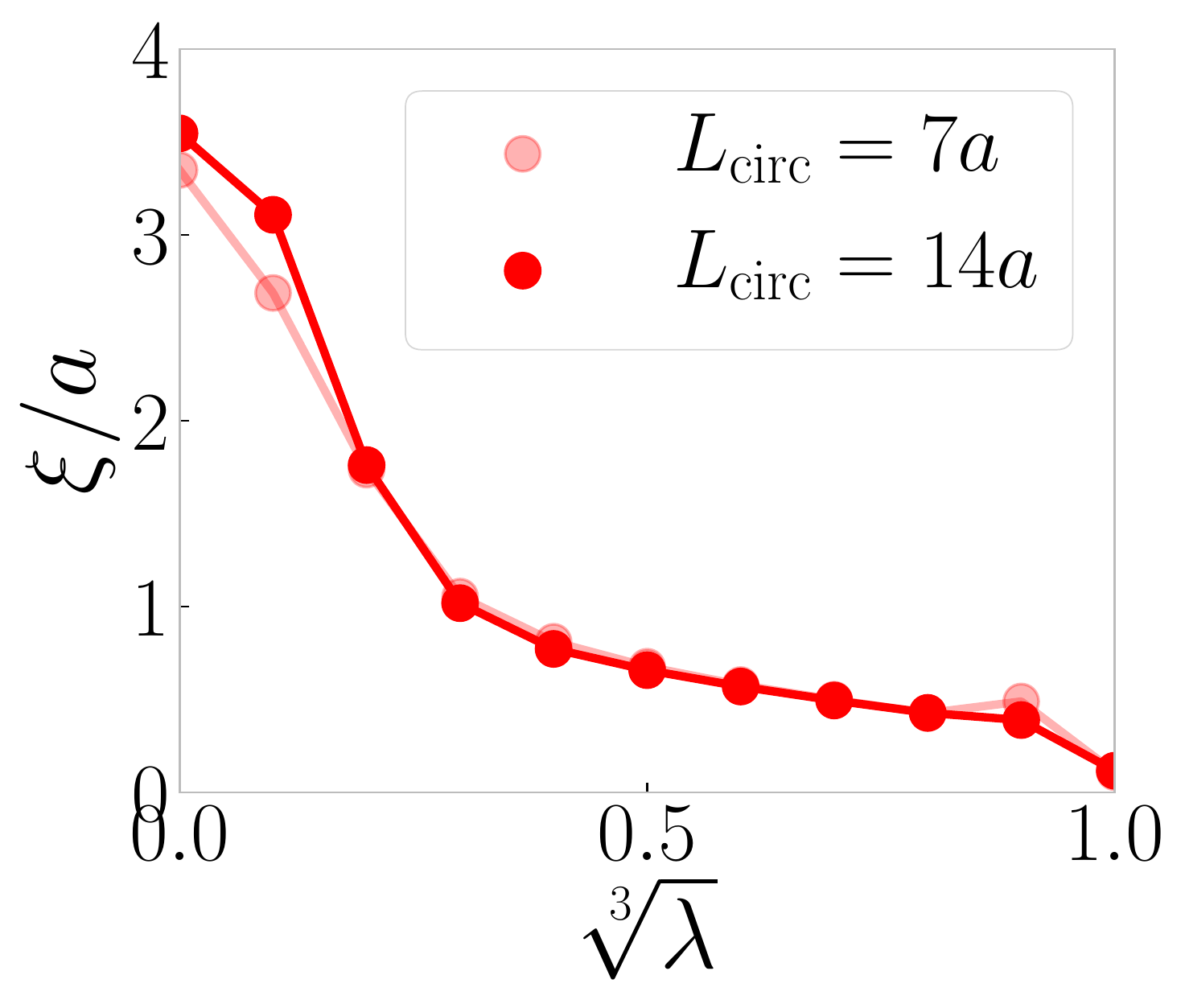}};
\node at (5.5,0.1)
{\includegraphics[scale=0.36]{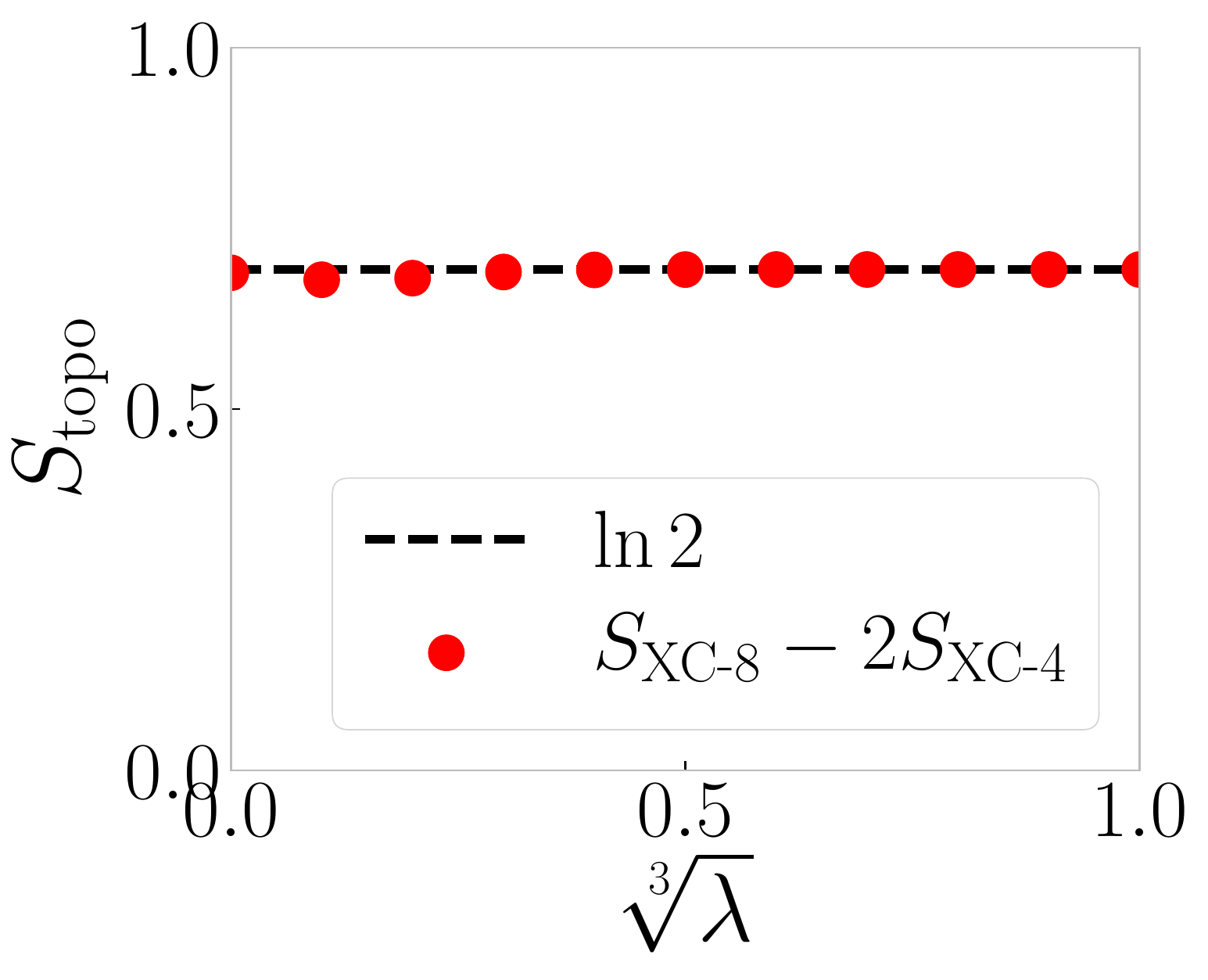}};
\node at (11.5,0.1)
{\includegraphics[scale=0.36]{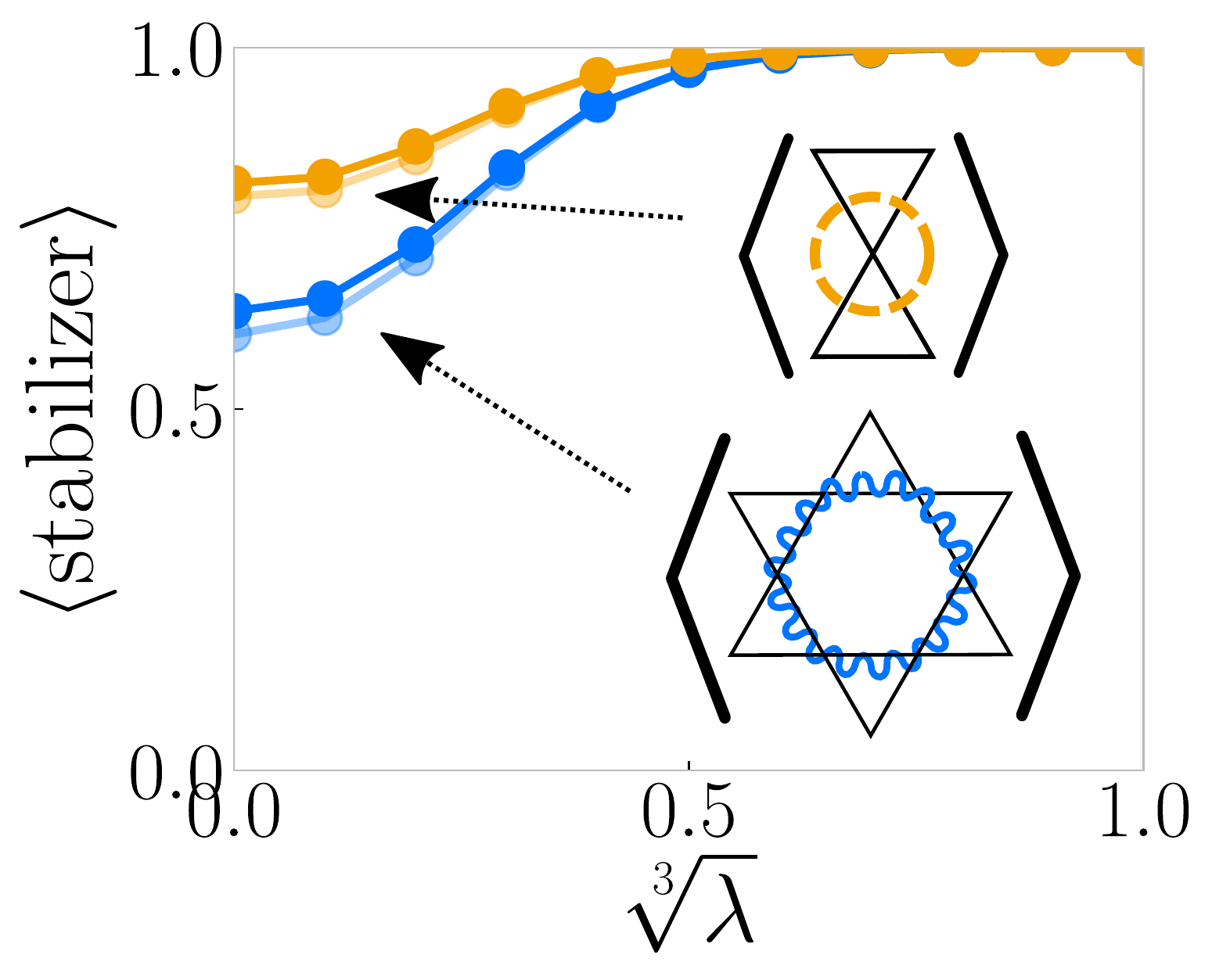}};
\draw[->] (-1.87,-2) -- (-1.87,-1.7);
\draw[->] (2.37,-2) -- (2.37,-1.7);
\node at (-1.8,-2.2) {ruby lattice SL \cite{Verresen21}};
\node at (2.5,-2.2) {solvable liquid \cite{Misguich02}};
\node at (-2.65,2.1) {(a)};
\node at (2.9,2.1) {(b)};
\node at (8.8,2.1) {(c)};
\end{tikzpicture}
\caption{\textbf{Interpolating from spin-$1/2$ ruby lattice Ising model to solvable dimer liquid.} Eq.~\eqref{eq:Hlambda} interpolates from an emergent dimer model at $\lambda=0$ (Eq.~\eqref{eq:model} with Eq.~\eqref{eq:Xruby}, which encodes the ruby lattice model \cite{Verresen21}; see Sec.~\ref{subsec:ruby}) to the stabilizer Hamiltonian \eqref{eq:stabilizer} for a dimer liquid at $\lambda=1$ (see Sec.~\ref{subsec:weakmonomer}). (a) We compute the ground state correlation length (in units of the lattice spacing of the ruby lattice) using iDMRG and we find that it decreases as we tune toward the solvable point; in particular, we see no sign of a transition, confirming that the ruby lattice Ising model is indeed a spin liquid. The fainter data is for XC-4, which has $L_\textrm{circ}/a = \sqrt{3} \times 4 \approx 7$; the solid data is for XC-8 with $L_\textrm{circ}/a \approx 14$. (b) The topological entanglement entropy is roughly constant throughout, close to the expected value of $\ln 2$. (c) The fixed-point dimer liquid is defined by both types of stabilizer ($-G_v$ in Eq.~\eqref{eq:gauss} and $W_p$ in Eq.~\eqref{eq:Wp}) being unity; we track their expectation value along the interpolation, with the fainter (solid) data being for XC-4 (XC-8) cylinders.}
    \label{fig:rubyinterpolation}
\end{figure*}
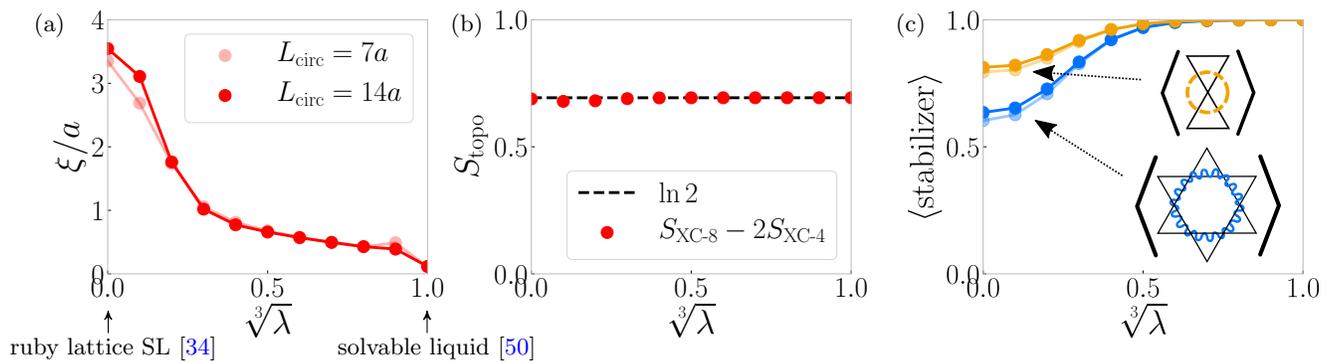
\end{center}

\end{widetext}

\subsection{The Hadamard transformation for a dimer model \label{subsec:hadamard}}

When we introduced the $\mathcal Z^\alpha$ and $\mathcal X^\alpha$ operators in Sec.~\ref{sec:modelhilb}, we commented on the fact that the existence of the latter is rather non-trivial. In particular, for dimer models on more generic lattices, one usually cannot define an operator on the Hilbert space which shuffles dimers around. Here we show that an even stronger property holds: there are natural on-site rotations which transform the diagonal and off-diagonal operators into one another. The first instance of this was described in Ref.~\onlinecite{Verresen21}, which reported a $\mathbb Z_3$ transformation interchanging $\mathcal X^\alpha$ and $\mathcal Z^\alpha$. This `change of basis' was essential to the experimental implementation of the ruby lattice proposal \cite{Semeghini21}, as it allowed to effectively measure off-diagonal string operators---key to arguing that the state is a coherent superposition rather than a mixture. Here we add to this in two ways: Sec.~\ref{subsec:Z2Had} introduces a $\mathbb Z_2$ Hadamard transformation, and Sec.~\ref{subsec:Z3Had} uses the Majorana representation to give insight into the previously reported $\mathbb Z_3$ transformation, which in turn also clarifies how to implement the novel $\mathbb Z_2$ Hadamard transformation in the existing Rydberg atom array platform.

In this section we consider on-site transformations (with respect to each triangle which is effectively a 4-state system), so we will drop/omit spatial indices for simplicity---all operators are to be understood as acting on the same site.

\subsubsection{$\mathbb Z_2$ Hadamard \label{subsec:Z2Had}}

For convenience, let us use the short-hand $\mathcal Z = \mathcal Z^x + \mathcal Z^y + \mathcal Z^z $ and $\mathcal X = \mathcal X^x + \mathcal X^y + \mathcal X^z $. We define:
\begin{equation}
U = e^{i \frac{\pi}{4} \mathcal Z} e^{i \frac{\pi}{4} \mathcal X} e^{i \frac{\pi}{4} \mathcal Z}. \label{eq:UHadamard}
\end{equation}
(In fact, one can insert arbitrary signs into any of the three exponents without changing the following properties.) Using the fact that
\begin{equation}
\left( \mathcal X^\alpha \right)^2 = \mathcal X^x \mathcal X^y \mathcal X^z = 1 = \mathcal Z^x \mathcal Z^y \mathcal Z^z = \left( \mathcal Z^\alpha \right)^2,
\end{equation}
we see that $U^2 \propto 1$, i.e., this is a $\mathbb Z_2$ transformation. Moreover, using the (anti)commutation relations for $\mathcal X^\alpha$ and $\mathcal Z^\beta$ in Eq.~\eqref{eq:semionic}, one obtains that:
\begin{equation}
U \mathcal X^\alpha U = \mathcal Z^\alpha \quad \textrm{and} \quad U \mathcal Z^\alpha U = \mathcal X^\alpha. \label{eq:hadamard}
\end{equation}
This thus constitutes what we might call a Hadamard transformation, in analogy to the qubit mapping which exchanges Pauli-$Z$ and Pauli-$X$ operators. 


If one is able to implement $H_\e$, one can straightforwardly realize the transformation in Eq.~\eqref{eq:UHadamard}. Indeed, note that $\mathcal X = \mathcal X_\e$ defined in Eq.~\eqref{eq:Xmono}, whereas $\mathcal Z$ is the longitudinal field in Eq.~\eqref{eq:model}. However, what if one only has access to $\mathcal X_\f$ or $\mathcal X_\textrm{ruby}$? We turn to this now.

\subsubsection{$\mathbb Z_3$ isomorphism through the Majorana notation \label{subsec:Z3Had}}

The Majorana notation introduced in Sec.~\ref{subsec:intermediatespinon} is not just useful for solving $H_\f$. More generally, it gives insight into the structure of our spin-3/2 Hilbert space and its operators. In particular, let us recall that we can express:
\begin{equation}
\mathcal Z^\alpha = i b^\alpha c^\alpha \quad \textrm{and} \quad \mathcal X^\alpha = i b^{\alpha+1} c^{\alpha-1},
\end{equation}
where we define `$x+1$' as meaning `$y$', `$x-1$' as meaning `$z$', and so on. In this representation, it is clear that if we cycle the matter Majoranas $c^x \to c^y \to c^z \to c^x$, then we can convert $\mathcal X^\alpha$ into $\mathcal Z^\alpha$ and vice versa. The generator of this cycling is exactly $\frac{1}{2\sqrt{3}} \mathcal X_\f$, where the $\sqrt{3}$ appears due to the $\mathbb Z_3$ nature of the cycling. More precisely, in Appendix~\ref{app:iso} we derive how the unitary rotation $\exp\left({\frac{2\pi i}{3} \frac{\mathcal X_\f}{2 \sqrt{3}} }\right)$ implements the shift $c^x \to c^y \to c^z \to c^x$, whilst leaving $b^\alpha$ untouched.

Hence, if we have control over $\mathcal X_\f$, we can generate rotations between the off-diagonal and diagonal string operators. However, this is a chiral operator which we might not have access to, like in the $H_\textrm{ruby}$ model, which is of particular experimental relevance. However, there is a natural non-chiral transformation to consider, namely cycling $c^x \to c^y \to c^z \to c^x$ at the same time as $b^x \to b^z \to b^y \to b^x$; see Fig.~\ref{fig:majoranas}(d). Note that this transforms $\mathcal X^\alpha = i b^{\alpha+1} c^{\alpha-1} \to i b^{\alpha} c^{\alpha} = \mathcal Z^\alpha$. In Appendix~\ref{app:spinthreehalf}, we show how this simple Majorana rotation is generated by $e^{i \frac{\pi}{4} \mathcal Z} \mathcal X_\textrm{ruby} e^{-i \frac{\pi}{4} \mathcal Z}$. This thus reproduces the proposal of Ref.~\onlinecite{Verresen21} from a new perspective.

Finally, let us note that this perspective also clarifies how to implement the $\mathbb Z_2$ Hadamard \eqref{eq:hadamard} using the same resources (which are available, say, in the ruby lattice model). Observe that $\mathcal Z = \sum_\alpha i b^\alpha c^\alpha$ generates a rotation between the bond and matter Majoranas. Clearly, if we combine this with the above $\mathbb Z_3$ cycling, we obtain a $\mathbb Z_2$ transformation\footnote{This is perhaps seen most clearly by tracing the paths in the transition graph: {\scriptsize $\begin{array}{ccccccc} b^x & \leftarrow & b^y & \leftarrow & b^z & \leftarrow & b^x\\
\updownarrow & & \updownarrow & & \updownarrow & & \updownarrow \\
 c^x & \to & c^y & \to & c^z & \to & c^x
\end{array}$}} which interchanges $\mathcal X^\alpha \leftrightarrow \mathcal Z^\alpha$. Details---including the exact representation in terms of the Rydberg atom Hamiltonians---can be found in Appendix~\ref{app:spinthreehalf}. It would be interesting to use this proposal in future realizations of the ruby lattice spin liquid, and to see whether it gives better or worse results compared to the $\mathbb Z_3$ proposal which has been implemented \cite{Semeghini21}.

\subsection{A natural family of spin-1 Kitaev models \label{subsec:kitaevpath}}

In Sec.~\ref{subsec:strongmonomer}, we saw how strong and frustrated $e$-anyon fluctuations effectively reduced $H_\e$ to a novel spin-1 quadrupolar Kitaev model \eqref{eq:Hquadrupolar} on the honeycomb lattice. We noted it has the same conserved plaquette operator as the usual spin-1 Kitaev model. Here we further explore the connection between these two models.

Let us define the following interpolation\footnote{It is amusing to note that these obey a type of $q$-deformed Lie algebra: $[S^x(\varphi),S^y(\varphi)]_q = i S^z (\varphi)$ where $[A,B]_q = q AB - q^{-1} BA$ with $q=e^{i\varphi/2}$. For $\varphi=0 \mod 2\pi$ this gives a commutator, and for $\varphi=\pi \mod 2\pi$ an anti-commutator.} of the spin and quadrupole operators:
\begin{align}
S^x(\varphi) &= \cos\left(\frac{\varphi}{6} \right) S^x + \sin\left(\frac{\varphi}{6} \right) Q^{yz}, \\
S^y(\varphi) &= \cos\left(\frac{\varphi}{6} \right) S^y + \sin\left(\frac{\varphi}{6} \right) Q^{zx}, \\
S^z(\varphi) &= \cos\left(\frac{\varphi}{6} \right) S^z + \sin\left(\frac{\varphi}{6} \right) Q^{xy}.
\end{align}
We define a corresponding Kitaev-type model:
\begin{equation}
H(\varphi) = J \sum_{\alpha=x,y,z} \sum_{\langle i ,j\rangle_\alpha} S^\alpha_i(\varphi) S^\alpha_j(\varphi). \label{eq:spin1interpolation}
\end{equation}
In Appendix~\ref{app:quadruopole} we show that up to unitary transformation, $H(\varphi)$ only depends on $\varphi \mod 2 \pi$. In particular, for $\varphi = 0 \mod 2\pi$ we have usual spin-1 Kitaev model \cite{Baskaran08,Zhu20,Hickey20,Dong20,Koga20,Khait21,Lee21,Chen22,Bradley22}, whereas $\varphi = \pi \mod 2\pi$ is unitarily equivalent to the spin-1 quadrupolar Kitaev model \eqref{eq:Hquadrupolar}, since one can unitarily relate $(S^x(\pi),S^y(\pi),S^z(\pi)) = - \left(  Q^{yz} , Q^{xz} , Q^{xy} \right)$.

The spin-1 Hamiltonian \eqref{eq:spin1interpolation} has a conserved plaquette operator $W_p = \prod e^{i \pi S^\alpha(\varphi)} = \prod e^{i \pi S^\alpha(0)}$ for all values of $\varphi$. This follows from the simple fact\footnote{In fact, there is a simple formula for more general rotations: $
\footnotesize e^{i \theta S^\alpha(\varphi)} S^\beta(\varphi) e^{-i \theta S^\alpha(\varphi)} = \cos(\theta) S^\beta(\varphi) + i \sin(\theta)[S^\alpha(\varphi),S^\beta(\varphi)]$.} that
\begin{equation}
e^{i \pi S^\alpha} S^\beta(\varphi) e^{- i \pi S^\alpha} = - (-1)^{\delta_{\alpha \beta}} S^\beta(\varphi),
\end{equation}
which in turn follows from the above definition of $S^\beta(\varphi)$, remembering that $Q^{ab} = \{ S^a, S^b \}$. Moreover, note that $W_p$ is independent of $\varphi$.

The fact $H(\varphi)$ commutes with $W_p$ implies we can perform the duality transformation (encountered in Sec.~\ref{sec:monomer}) to a frustrated Ising model on the kagom\'e lattice. In Sec.~\ref{subsec:strongmonomer} we found that in the case of the quadrupolar model (i.e., $\varphi = \pi \mod 2\pi$), the dual Hamiltonian is\footnote{Note that a global unitary transformation $\prod \sigma^z$ makes this dual model sign-problem-free.} $H = J \sum_v P\sigma^x P$, where $P$ enforces the condition that the two adjacent triangles are and remain frustrated; in other words, we can flip a spin only if it is part of frustrated bonds on \emph{both} adjacent triangles (see Fig.~\ref{fig:spin1quadrupolar}(a)). The only modification for $\varphi \neq \pi \mod 2 \pi$ is that this term acquires a phase factor: in particular, if we attempt to flip a kagom\'e spin on a $\alpha$-bond of the corresponding honeycomb lattice, then the $P\sigma^x P$ term has a $-e^{i \varphi/3}$ ($-e^{-i \varphi/3}$) phase factor if both frustrated bonds lie along the $\alpha+1$ ($\alpha-1$) direction (e.g., if $\alpha=z$, then $\alpha+1=x$ and $\alpha-1=y$). By absorbing phase factors into the basis states, one can show that only $\varphi \mod 2\pi$ matters; see Appendix~\ref{app:quadruopole}. If the two frustrated bonds do \emph{not} lie in the same direction (as in Fig.~\ref{fig:spin1quadrupolar}(a)), there is no additional phase factor.

The above suggests a way of gaining new insight into the spin-1 Kitaev model---after all, in the latter model there is still some disagreement about whether the ground state is a \2 spin liquid or not \cite{Baskaran08,Zhu20,Hickey20,Dong20,Koga20,Khait21,Lee21,Chen22,Bradley22}. Firstly, Eq.~\eqref{eq:spin1interpolation} can be investigated numerically; if one finds that this gives rise to an adiabatic path, then this would confirm the \2 spin liquid of the spin-1 Kitaev model (since Sec.~\ref{subsec:strongmonomer} already established that $\varphi = \pi \mod 2 \pi$ is a \2 spin liquid). Secondly, the dual representation gives a slightly more economical encoding of the model, since instead of the $3^2$-dimensional unit cell (consisting of two spin-1's on the honeycomb lattice), we have a frustrated Ising model with a $2^3$-dimensional unit cell.
We leave such numerical explorations for future work.

\subsection{Kagom\'e dimer model as honeycomb toric code \label{subsec:toric}}

In Secs.~\ref{sec:monomer}, \ref{sec:spinon} and \ref{subsec:connectruby}, we saw that the parent model \eqref{eq:model} with the three choices of transverse fields ($\mathcal X_\e$, $\mathcal X_\f$ and $\mathcal X_\textrm{ruby}$) are all adiabatically connected to the fixed-point dimer liquid on the kagom\'e lattice \eqref{eq:stabilizer}, with the caveat that for \emph{chiral} $f$-anyon fluctuations, we found that the plaquette stabilizer is $-W_p$ rather than $W_p$. We note that this fixed-point model is a stabilizer Hamiltonian: the entire spectrum is solvable. This makes it distinct from the Rokhsar-Kivelson model \cite{RK} which is a commuting projector Hamiltonian such that only the ground state is known; in fact, the kagom\'e dimer model seems more akin to the toric code model in this respect \cite{Kitaev_2003}. Here we show that this is not a coincidence: one can locally rewrite the kagom\'e dimer liquid as a toric code model on the honeycomb lattice.

In fact, there is a well-known connection between kagom\'e dimer models and honeycomb loop models. This originates with the work by Elser and Zeng \cite{Elser93} showing that dimer coverings on the kagom\'e lattice can be encoded in a particular `arrow representation'. Misguich, Serban and Pasquier used this to construct their exactly solvable dimer liquid \cite{Misguich02,Misguich03}. In particular, if one fixes such an arrow representation, then any other state defines a loop state on the honeycomb lattice, with the solvable model corresponding to a toric code Hamiltonian \cite{Buerschaper14,Iqbal14,Iqbal20}. Here we describe a different mapping which dispenses with the choice of a reference state by introducing `odd' loop states.

We focus on the case where the dimer constraint is exact, meaning we only consider states where $G_v=-1$ (see Eq.~\eqref{eq:gauss}); the fixed-point model is then $H_\textrm{MSP} = - \sum_p W_p$, which was introduced by Misguich, Serban and Pasquier. (Our representation in Eq.~\eqref{eq:stabilizer} embeds this in a larger Hilbert space where the dimer constraint is only energetically enforced, which has the benefit of making $W_p$ in Eq.~\eqref{eq:Wp} a product operator over six `sites'; see Sec.~\ref{subsec:star} for its connection to a toric code model on a \emph{decorated} honeycomb lattice \cite{Schuch12}.).

\begin{figure}
    \centering
    \begin{tikzpicture}
    \node at (0,0) {\includegraphics[scale=0.5]{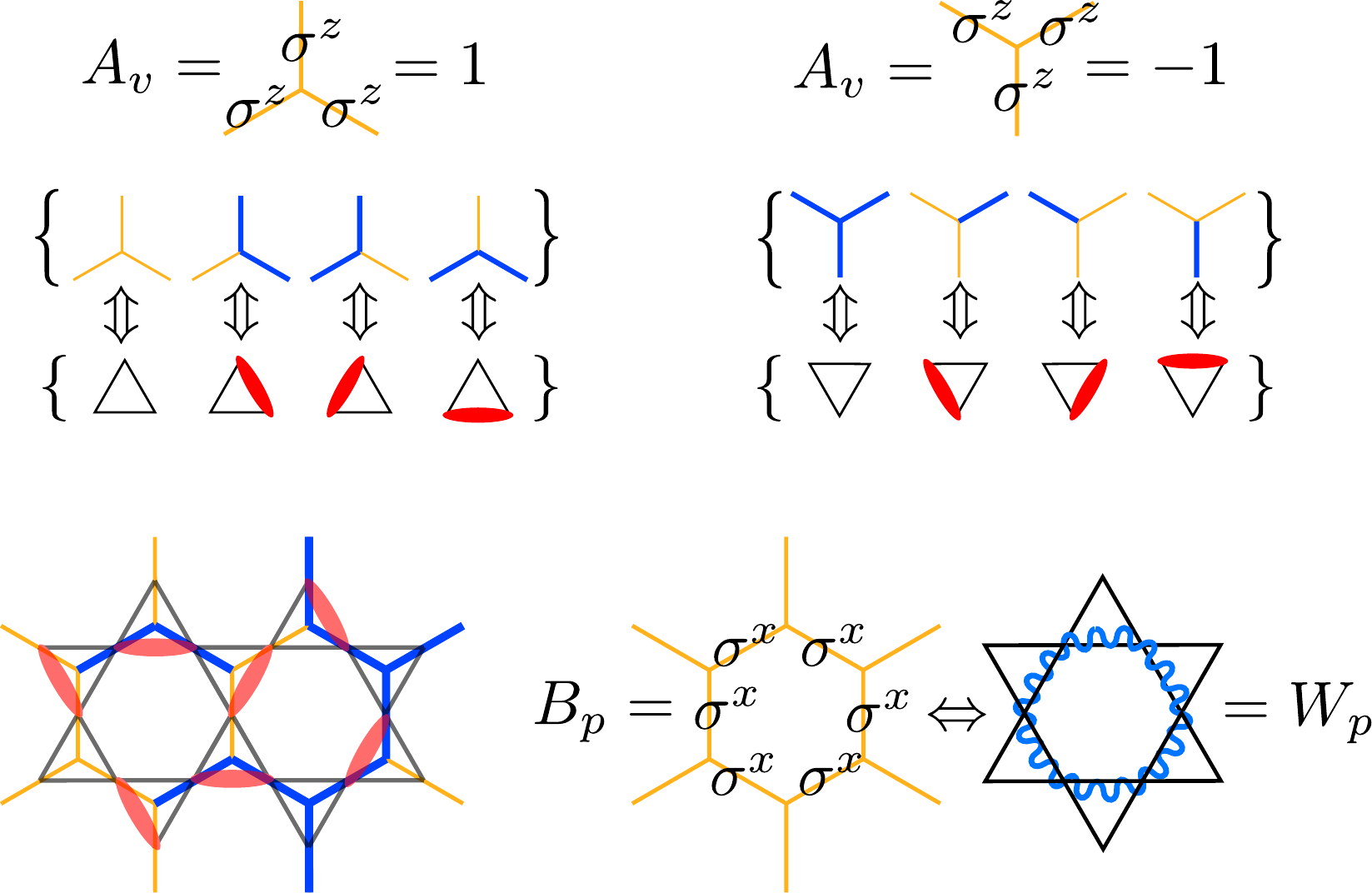}};
    \node at (-4,2.6) {(a)};
    \node at (0,2.58) {(b)};
    \node at (-4,-0.8) {(c)};
    \node at (-0.9,-0.8) {(d)};
    \end{tikzpicture}
    \caption{\textbf{Equivalence between kagom\'e dimer model and honeycomb toric code.} There is a local map between the Hilbert space of dimer coverings of the kagom\'e lattice and `odd loop' states on the honeycomb lattice, by which we mean that (a) on the A sublattice we enforce $A_v=1$ such that we have loop states (with $\sigma^z=-1$ signifying a loop, shown in solid blue) but (b) on the B sublattice $A_v=-1$ such that an odd number of strings terminate there. (c) An example of the local equivalence between a dimer state and such an `odd loop' state. (d) Under this local mapping, the usual plaquette operator $B_p$ of the toric code is equivalent to the dimer resonance $W_p$ defined in Eq.~\eqref{eq:Wp}.}
    \label{fig:toric}
\end{figure}
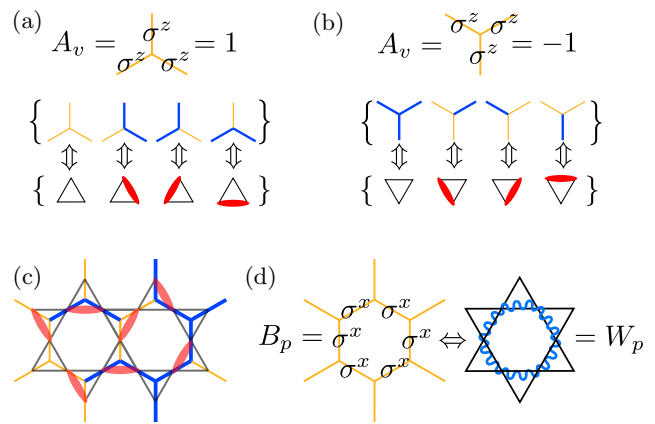

Let us briefly recall the toric code model for the particular case of the honeycomb lattice. We place qubits on the links of the lattice, and for each vertex we have the three-body stabilizer $A_v = \prod_{l \in v} \sigma^z_l$ (where $l \in v$ signifies that the qubit lives on a link $l$ emanating from said vertex $v$) and for each plaquette the six-body stabilizer $B_p = \prod_{l \in p} \sigma^x_l$. We observe that $A_v$ and $B_p$ commute and square to the identity. In fact, up to subtleties which depend on boundary conditions, $A_v = \pm 1$ and $B_p = \pm 1$ label all the states in the Hilbert space\footnote{Note that the counting works out: there are three qubits per unit cell (since the bonds of the honeycomb lattice form the kagom\'e lattice), and indeed there are two $A_v$'s and one $B_p$ per unit cell.}. The usual toric code Hamiltonian \cite{Kitaev_2003} is then $H_\textrm{TC} = -\sum_v A_v - \sum_p B_p$, i.e., the ground state satisfies $A_v = 1 = B_p$.

We have already discussed how a dimer model on the kagom\'e lattice is an odd \2 gauge theory, i.e., there is a background $e$-anyon per unit cell. The usual toric code has $A_v = 1$, meaning that there is no $e$-anyon in the ground state. However, one can modify the toric code Hamiltonian by toggling the sign of certain $A_v$ terms, such that the ground state satisfies $A_v = -1$. In the absence of translation symmetry, this is unitarily equivalent\footnote{Despite mapping between two translation-invariant models, the unitary operator will itself not be translation-invariant. This is similar to having to choose a reference arrow representation for mapping kagom\'e dimer coverings to honeycomb loop states \cite{Elser93,Buerschaper14,Iqbal14}.} to the original model, however it can define a distinct SET protected by translation symmetry. We can thus only hope to relate the kagom\'e dimer model to a honeycomb toric code model if we ensure that they are in the same SET phase. To obtain a toric code with an $e$-anyon per unit cell, we consider the modified Hamiltonian $H_\textrm{TC}' =  -\sum_{v \in A} A_v + \sum_{v \in B} A_v - \sum_p B_p$, where we have toggled $A_v$ on one of the two sublattices of the honeycomb lattice.

We now locally relate the kagom\'e dimer and honeycomb toric code models. We already mentioned that we will only consider exact dimer configurations (i.e., $G_v=-1$). Similarly, we will enforce the $A_v$ terms exactly---$A_v=1$ ($A_v=-1$) on the $A$ ($B$) sublattice. In this limit, the toric code model is also called a pure gauge theory \cite{Wegner71,Wilson74,FradkinShenker}, since there is no dynamical matter: only magnetic fluxes can have dynamics, with e.g. a perturbed Hamiltonian $H=-\sum B_p + g \sum_l \sigma^z_l$. For these restricted Hilbert spaces, there is a local identification as shown in Fig.~\ref{fig:toric}(a) and (b), with an example in Fig.~\ref{fig:toric}(c); we note that this is closely related to the notion of an arrow representation \cite{Elser93}, although here we leverage the bipartite structure of the honeycomb lattice. It is then easy to see how Hamiltonian terms map: $B_p$ corresponds to $W_p$ in Eq.~\eqref{eq:Wp}, whereas $\sigma^z_l$ on a bond in the $\alpha \in \{x,y,z\}$ direction (see Fig.~\ref{fig:fromkagometohoneycomb}(a) for labeling) maps to $\mathcal Z^\alpha$ on the upward-pointing triangle of the kagom\'e lattice (or equivalently $-\mathcal Z^\alpha$ on the downward triangle).

\subsection{Emergent dimer liquids in a generalized $XYZ$ model \label{subsec:XYZ}}

The framework introduced in Sec.~\ref{sec:model} (and its associated graphical notation) can be used more broadly to define and analyze interesting models. Here we consider an exemplary case-study, where the anyon fluctuations of the emergent dimer model are not introduced by single-site terms, but rather two-body interactions. In particular, we will still take the dimer-enforcing interaction, $\sum_{\langle i,j\rangle_\alpha} \mathcal Z^\alpha_i \mathcal Z^\alpha_j$, as our starting point. A natural two-body off-diagonal interaction to consider is to simply replace $\mathcal Z^\alpha \to \mathcal X^\alpha$. Indeed, we will see that $H = \sum_{\langle i,j\rangle_\alpha} \left( J_x \mathcal X^\alpha_i \mathcal X^\alpha_j + J_z \mathcal Z^\alpha_i \mathcal Z^\alpha_j \right)$ leads to two robust $\mathbb Z_2$ spin liquids, with a transition at $|J_x|=|J_z|$. In fact, we will directly analyze a broader parameter space, inspired by there being a natural isomorphism $\mathcal X^\alpha \to \mathcal Y^\alpha \to \mathcal Z^\alpha \to \mathcal X^\alpha$, where $\mathcal Y^\alpha := - \mathcal X^\alpha \mathcal Z^\alpha$ (see Appendix~\ref{app:spinthreehalf}). Hence, we consider the following spin-3/2 honeycomb model:
\begin{equation}
H = \sum_{\alpha=x,y,z} \sum_{\langle i,j\rangle_\alpha} \left( J_x \mathcal X^\alpha_i \mathcal X^\alpha_j  + J_y \mathcal Y^\alpha_i \mathcal Y^\alpha_j  + J_z \mathcal Z^\alpha_i \mathcal Z^\alpha_j \right), \label{eq:HXYZ}
\end{equation}
where we are using the $x,y,z$ labeling of the bonds in Fig.~\ref{fig:fromkagometohoneycomb}. Using our graphical notation:
\begin{equation}
H = \sum_v \Big( J_x \raisebox{-9.6pt}{\includegraphics[scale=0.25]{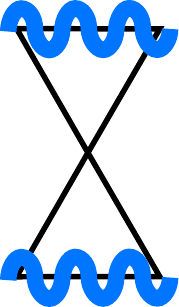}} +
J_y \raisebox{-9.6pt}{\includegraphics[scale=0.25]{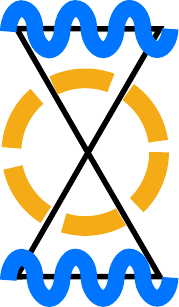}} +
J_z \raisebox{-7pt}{\includegraphics[scale=0.25]{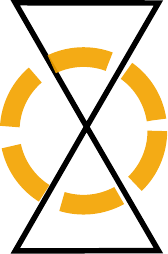}}
\Big). \label{eq:HXYZgraphical}
\end{equation}
It is worth pointing out that unitary transformations can arbitrarily permute $J_x,J_y,J_z$ as well as toggle their signs---up to the sign of the product $J_xJ_yJ_z$, which is immutable.

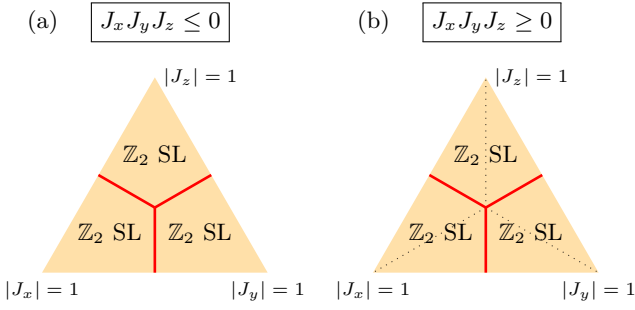
\begin{figure}
\centering
\begin{tikzpicture}
\node at (0,0){
    \begin{tikzpicture}
    \coordinate (a) at (0,0);
    \coordinate (b) at (1.5,{1.5*sqrt(3)});
    \coordinate (c) at (3,0);
    \fill[fill=neworange!60,opacity=1]  (barycentric cs:a=1,b=0,c=0) --  (barycentric cs:a=0,b=1,c=0) --  (barycentric cs:a=0,b=0,c=1);
    \node[below] at (barycentric cs:a=1,b=0,c=0) {\scriptsize $|J_x|=1$};
    \node[below] at (barycentric cs:a=0,b=0,c=1) {\scriptsize $|J_y|=1$};
    \node[right] at (barycentric cs:a=0,b=1,c=0) {\scriptsize $|J_z|=1$};
    \draw[line width=1,red] (barycentric cs:c=1,b=0,a=1) -- (barycentric cs:c=1,b=1,a=1);
    \draw[line width=1,red] (barycentric cs:c=1,b=1,a=0) -- (barycentric cs:c=1,b=1,a=1);
    \draw[line width=1,red] (barycentric cs:c=0,b=1,a=1) -- (barycentric cs:c=1,b=1,a=1);
    \node at (barycentric cs:c=3,b=1,a=1) {$\mathbb Z_2$ SL};
    \node at (barycentric cs:c=1,b=3,a=1) {$\mathbb Z_2$ SL};
    \node at (barycentric cs:c=1,b=1,a=3) {$\mathbb Z_2$ SL};
    \node[above,yshift=10pt,xshift=-10pt] at (barycentric cs:a=0,b=1,c=0) {(a) $\quad \boxed{J_xJ_yJ_z \leq 0}$};
    \end{tikzpicture}
};
\node at (4.4,0){
    \begin{tikzpicture}
    \coordinate (a) at (0,0);
    \coordinate (b) at (1.5,{1.5*sqrt(3)});
    \coordinate (c) at (3,0);
    \fill[fill=neworange!60,opacity=1]  (barycentric cs:a=1,b=0,c=0) --  (barycentric cs:a=0,b=1,c=0) --  (barycentric cs:a=0,b=0,c=1);
    \draw[-,dotted,opacity=0.8] (barycentric cs:a=1,b=1,c=1) -- (barycentric cs:a=1,b=0,c=0);
    \draw[-,dotted,opacity=0.8] (barycentric cs:a=1,b=1,c=1) -- (barycentric cs:a=0,b=1,c=0);
    \draw[-,dotted,opacity=0.8] (barycentric cs:a=1,b=1,c=1) -- (barycentric cs:a=0,b=0,c=1);
    \node[below] at (barycentric cs:a=1,b=0,c=0) {\scriptsize $|J_x|=1$};
    \node[below] at (barycentric cs:a=0,b=0,c=1) {\scriptsize $|J_y|=1$};
    \node[right] at (barycentric cs:a=0,b=1,c=0) {\scriptsize $|J_z|=1$};
    \draw[line width=1,red] (barycentric cs:c=1,b=0,a=1) -- (barycentric cs:c=1,b=1,a=1);
    \draw[line width=1,red] (barycentric cs:c=1,b=1,a=0) -- (barycentric cs:c=1,b=1,a=1);
    \draw[line width=1,red] (barycentric cs:c=0,b=1,a=1) -- (barycentric cs:c=1,b=1,a=1);
    \node at (barycentric cs:c=3,b=1,a=1) {$\mathbb Z_2$ SL};
    \node at (barycentric cs:c=1,b=3,a=1) {$\mathbb Z_2$ SL};
    \node at (barycentric cs:c=1,b=1,a=3) {$\mathbb Z_2$ SL};
    \node[above,yshift=10pt,xshift=-10pt] at (barycentric cs:a=0,b=1,c=0) {(b) $\quad \boxed{J_xJ_yJ_z \geq 0}$};
    \end{tikzpicture}
};
\end{tikzpicture}
    \caption{\textbf{Emergent dimer liquid in a spin-3/2 model from two-body off-diagonal terms.} Complementary to the model in Eq.~\eqref{eq:model} where deconfinement was stabilized by single-site anyon fluctuations in the spin-3/2 honeycomb model, here we consider the case with two-body fluctuations \eqref{eq:HXYZ}. The conservation of two plaquette operators allows for a nonlocal duality mapping \eqref{eq:Z2Z2dictionary} to the spin-1/2 XYZ honeycomb magnet, resulting in the above phase diagrams with extended spin liquids, using barycentric coordinates where $|J_x|+|J_y|+|J_z| = 1$. (a) If $J_xJ_yJ_z \leq 0$, the ground state always satisfies $W_p = 1 = \tilde G_p$; the three Ising phase of the spin-1/2 XYZ magnet map to three spin liquids. (b) The phase diagram is very similar for $J_xJ_yJ_z \geq 0$, except along the black dotted lines, where there is an extensive ground state degeneracy of the aforementioned plaquette operators.
    }
    \label{fig:XYZphasediagram}
\end{figure}

We will now obtain the phase diagram in Fig.~\ref{fig:XYZphasediagram}. As a first step, let us gain some intuition in the perturbative regimes, where one of the $J_\alpha$ dominates. Without loss of generality (i.e., up to an on-site unitary transformation), we can presume that we then have large $J_z>0$. In the dimer regime (where the Gauss law $G_v = -1$), we see that in Eq.~\eqref{eq:HXYZ}, the $J_x$ term is effectively renormalized to $J_x-J_y$; let us consider the generic case $J_x \neq J_y$. Clearly this term brings us out of the dimer manifold, but it is simple to see that at sixth order in perturbation theory, we effectively generate the fixed-point dimer Hamiltonian \eqref{eq:stabilizer}, with the dimer resonance $W_p = 1$ in the ground state. In this limit we thus recover the same emergent dimer liquid as we found for our $e$-anyon fluctuating model $H_\e$ in Sec.~\ref{subsec:weakmonomer}. In this case, the deconfined phase is stabilized by $e$-anyon fluctuations which come in groups of \emph{four}, rather than \emph{two}.

Another similarity to our $H_\e$ model is that $W_p$ is a conserved plaquette operator for the whole parameter range of the model. However, a striking difference is that Eq.~\eqref{eq:HXYZ} also conserves the product of Gauss laws around a plaquette:
\begin{equation}
\tilde G_p = \prod_{v \in p} G_v = 
\raisebox{-15pt}{\includegraphics[scale=0.25]{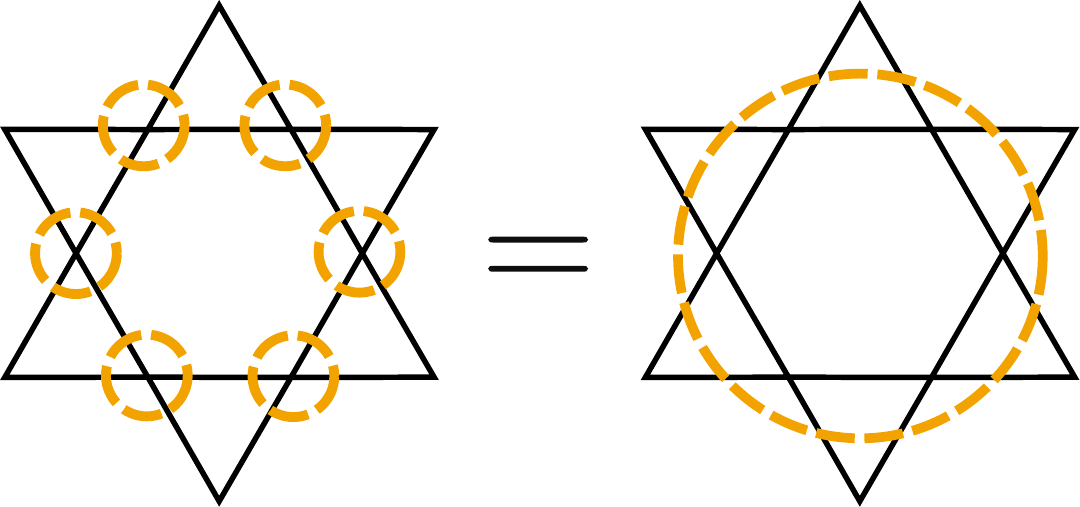}}.
\end{equation}
Indeed, one can straightforwardly confirm that $[\tilde G_p,H] = 0$, which is simplest to see using the graphical notation (Eq.~\eqref{eq:HXYZgraphical}). In the aforementioned perturbative regime, we find that both of these conserved quantities are in the flux-free sector: $W_p = 1 = \tilde G_p$. It turns out that this is true for almost \emph{all} values $J_x,J_y,J_z$, which we have checked with DMRG. The only exception is along the black dotted lines in Fig.~\ref{fig:XYZphasediagram}(b), where we find that although $W_p = 1 = \tilde G_p$ still belongs to the ground state manifold, it is degenerate with other sectors. This is in line with our above analysis where we found that if $J_z>0$ and $J_x = J_y$, then perturbation theory could not generate the $W_p$ term. Let us henceforth focus on the generic case, where we set aside this measure zero case.

Let us recall that in Sec.~\ref{subsec:monomerduality}, we used $W_p=1$ to perform a nonlocal change of variables, mapping our spin-3/2 honeycomb model $H_\e$ into a spin-1/2 kagom\'e model. Since Eq.~\eqref{eq:HXYZ} also conserves $W_p$, we could employ the same mapping; however, in this case it does not reduce to a previously-studied model. Fortunately, by using both $W_p=1$ and $\tilde G_p=1$, we can devise a different duality, which maps to effective spin-1/2 operators on the honeycomb lattice. Indeed, the conservation laws allow us to effectively reduce the on-site Hilbert space dimension from four to two. We provide more details about this novel mapping in Appendix~\ref{app:duality}, which can be interpreted as a $\mathbb Z_2 \times \mathbb Z_2$ Kramers-Wannier (or `gauging') map. Let us here summarize the resulting dictionary for these spin-1/2 operators, for nearest-neighbors $\langle i,j \rangle$ on a bond of type $\alpha \in \{x,y,z\}$ on the honeycomb lattice:
\begin{equation}
\sigma^x_i \sigma^x_j = -\mathcal X^\alpha_i \mathcal X^\alpha_j, \;
\sigma^y_i \sigma^y_j = -\mathcal Y^\alpha_i \mathcal Y^\alpha_j, \;
\sigma^z_i \sigma^z_j = -\mathcal Z^\alpha_i \mathcal Z^\alpha_j. \label{eq:Z2Z2dictionary}
\end{equation}
One can indeed confirm that the Pauli algebra is satisfied. Thus in the ground state sector $W_p = 1 = \tilde G_p$, we can rewrite $H$ in Eq.~\eqref{eq:HXYZ} as:
\begin{equation}
H_\textrm{eff} = - \sum_{\langle i,j\rangle} \left( J_x \sigma^x_i \sigma^x_j + J_y \sigma^y_i \sigma^y_j + J_z \sigma^z_i \sigma^z_j \right).
\end{equation}

Remarkably, we obtain the spin-$1/2$ XYZ model on the honeycomb lattice! This has a well-known phase diagram, containing a gapless symmetry-breaking phase extending from the solvable ferromagnetic Heisenberg point ($J_x=J_y=J_z$ with $J_xJ_yJ_z>0$) across the $XY$ magnet ($J_z=0$ and $J_x=J_y$ and permutations thereof) \cite{Varney11} to the Heisenberg anti-ferromagnet ($J_x=J_y=J_z$ with $J_xJ_yJ_z<0$) \cite{Fouet01}; for other values of parameters the model is gapped out into Ising phases. This leads to the phase diagram in Fig.~\ref{fig:XYZphasediagram}, where the three different Ising orders of the spin-$1/2$ model correspond to $\mathbb Z_2$ spin liquids in our original model, due to the nonlocal change of variables. The model in Eq.~\eqref{eq:HXYZ} thus constitutes another example of how fluctuating anyon defects can stabilize deconfinement.

\section{Local rewriting of spin-$3/2$ model as two-body spin-$1/2$ `grandparent' models \label{sec:equivalence}}

Thus far, we have mainly discussed the above model(s) as (emergent) dimer models on the kagom\'e lattice or as spin-$3/2$ Ising models on the honeycomb lattice. In this section, we show that they can be \emph{locally} rewritten as spin-$1/2$ models. We emphasize that the local aspect is key, since it implies that the physics is preserved. (In contrast, it is well-known that nonlocal mappings, such as the Kramers-Wannier transformation, can radically change the physical interpretation of a model. An example of such a nonlocal rewriting of $H_\e$ was discussed in Sec.~\ref{subsec:monomerduality}, where a \2 spin liquid was mapped to a trivial paramagnet,  or in Sec.~\ref{subsec:XYZ} where a different nonlocal mapping related \2 spin liquids to Ising symmetry-breaking phases.)

\subsection{Ruby lattice \label{subsec:ruby}}

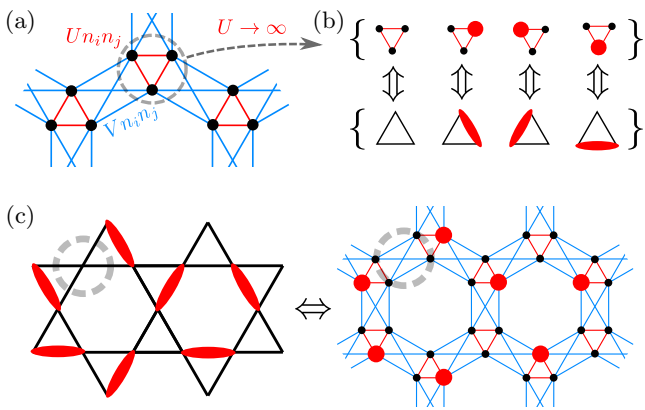
\begin{figure}
    \centering
    \begin{tikzpicture}
    \node at (0,0) {\includegraphics[scale=0.45]{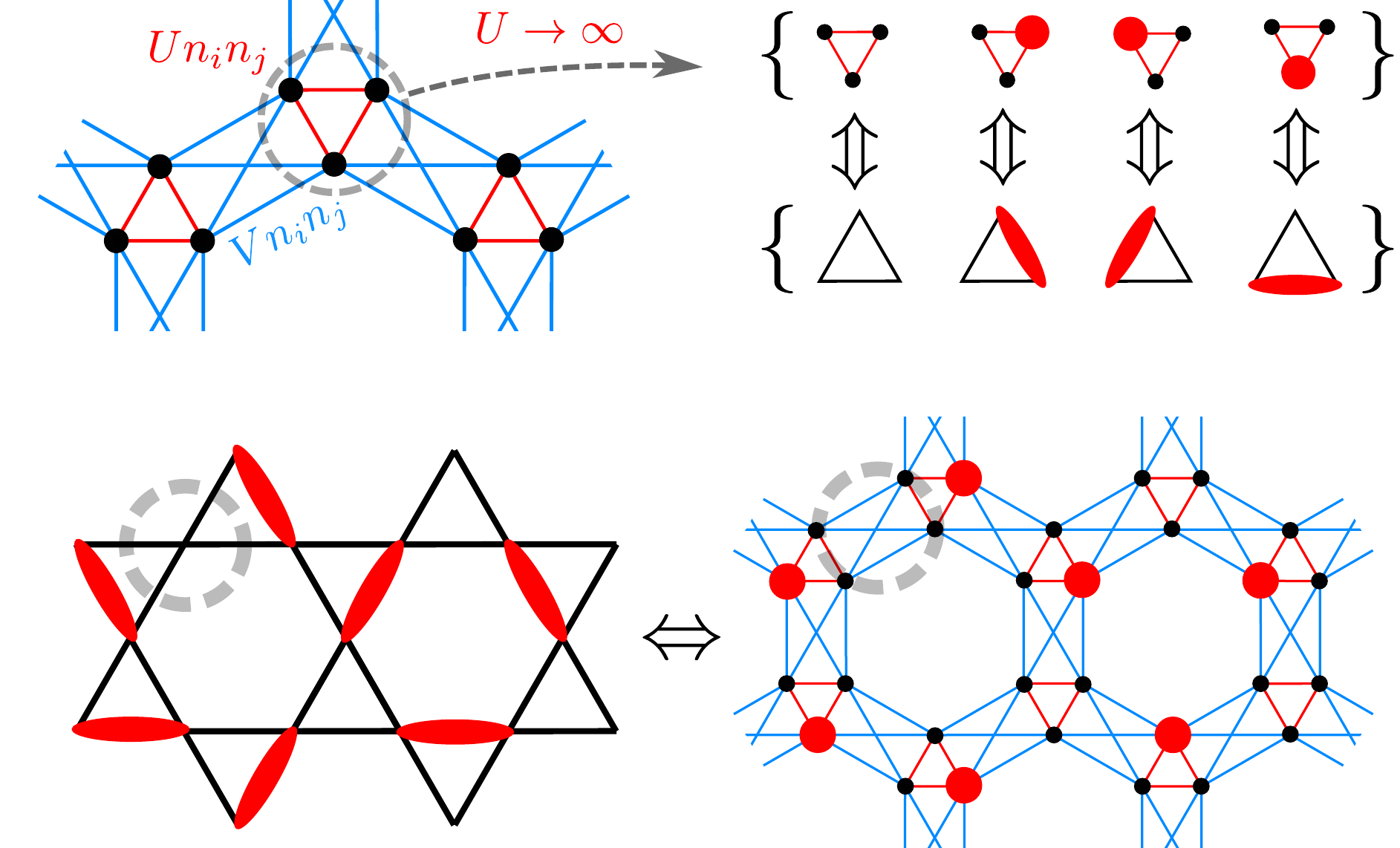}};
    \node at (-4,2.5) {(a)};
    \node at (0.1,2.5) {(b)};
    \node at (-4,-0.1) {(c)};
    \end{tikzpicture}
    \caption{\textbf{Spin-1/2 model on ruby lattice gives effective spin-3/2 model on honeycomb lattice.} (a) The black dots denote spin-1/2's on the ruby lattice. We have density-density repulsion ($n = \frac{1+\sigma^z}{2}$) with strength $U n_i n_j$ on the red bonds, and $V n_i n_j$ on the blue bonds. (b) If $U \to +\infty$, each triangle of the ruby lattice becomes an effective 4-state system. We show how this corresponds to an emergent dimer model Hilbert space on the kagom\'e lattice, i.e., the spin-3/2 Hilbert space on the honeycomb lattice introduced in Sec.~\ref{sec:model}. In the main text we discuss how the spin-1/2 Hamiltonian leads to our parent model in Eq.~\eqref{eq:model}. (c) Example of how a kagom\'e dimer state is mapped to a spin-1/2 (or hardcore boson) configuration on the ruby lattice. Violations of the dimer constraint correspond to rectangles with no (or two) hardcore boson(s).}
    \label{fig:ruby}
\end{figure}

We consider spin-1/2's on the ruby lattice as shown in Fig.~\ref{fig:ruby}(a). We envision it as a system of hardcore bosons with a Hamiltonian $H = H_\textrm{diag} + H_\textrm{off-diag}$, where the diagonal piece encodes repulsive interactions $\sim U$ on the red triangles and $\sim V$ on the blue bonds of the rectangles:
\begin{equation}
H_\textrm{diag} = U \sum_{\langle i,j\rangle_\textrm{red}} n_i n_j + V \sum_{\langle i,j \rangle_\textrm{\color{blue}blue}} n_i n_j - \delta \sum_i n_i.
\end{equation}

If we consider the blockaded limit $U \to \infty$, then each triangle becomes an effective 4-state system: either it is empty, or a hardcore boson occupies one of the three corners. As shown in Fig.~\ref{fig:ruby}(b), this is in a natural 1-to-1 correspondence with the emergent dimer Hilbert space we introduced in Sec.~\ref{sec:model}, i.e., spin-3/2's on the honeycomb lattice. Under this correspondence, for a given rectangle of the ruby lattice we can identify $\sum_{\langle i,j \rangle_\textrm{\color{blue}blue}} n_i n_j = \frac{1-\mathcal Z^\alpha_i}{2} \frac{1-\mathcal Z^\alpha_j}{2}$ where $\alpha\in \{x,y,z\}$ corresponds to the orientation of the rectangle. Similarly, up to a constant, we have that the chemical potential on a triangle of the ruby lattice $\sum_{i \in \triangle} n_i = -\frac{1}{4} \left( \mathcal Z^x + \mathcal Z^y + \mathcal Z^z \right)$.

In conclusion:
\begin{equation}
\lim_{U \to \infty} H_\textrm{diag} = \frac{V}{4} \sum_{\alpha=x,y,z} \sum_{\langle i,j\rangle_\alpha} \mathcal Z^\alpha_i \mathcal Z^\alpha_j + \frac{\delta-V}{4} \sum_i \mathcal Z_i.
\end{equation}
Hence, \emph{the blockaded limit of this spin-1/2 `grandparent' model on the ruby lattice leads to our spin-3/2 parent model} \eqref{eq:model} with $J = \frac{V}{4}$ and $h_l = \frac{V-\delta}{4}$, or equivalently, $\delta = 4(J-h_l)$. The three choices of transverse field, $\mathcal X_\textrm{ruby}$, $\mathcal X_\e$ and $\mathcal X_\f$, correspond to choosing a particular $H_\textrm{off-diag}$ on the ruby lattice, as we discuss now.

\subsubsection{$H_\textrm{ruby}$ as a spin-1/2 model on the ruby lattice}

Choosing $H_\textrm{off-diag} = \frac{\Omega}{2} \sum_i \sigma^x_i$ means we can only create/destroy dimers; i.e., there are no direct dimer hoppings. This corresponds to $\frac{\Omega}{2}\sum_{\alpha=x,y,z} \frac{1+\mathcal Z^\alpha}{2} \mathcal X^\alpha$, which is exactly $\mathcal X_\textrm{ruby}$ in Eq.~\eqref{eq:Xruby}. Hence, choosing the transverse field in the spin-1/2 ruby lattice model (in the blockade limit) leads to $H_\textrm{ruby}$ with $h_t = \frac{\Omega}{4}$.

This model is of particular interest since it can be approximately realized in Rydberg atom arrays \cite{Verresen21,Semeghini21}.
Various instances were explored in Ref.~\onlinecite{Verresen21}, including models with longer-range interactions. However, one particularly simple instance is where we also take the blockade limit for the blue bonds: $V\to \infty$ (note that this requires taking $J,h_l \to \infty$ such that $\delta=4(J-h_l)$ remains finite). In this case, we obtain the so-called PXP model on the ruby lattice. Using various probes based on entanglement, correlations and degeneracies, Ref.~\onlinecite{Verresen21} reported a \2 spin liquid for $0.5 \lessapprox |\Omega|/\delta \lessapprox 0.7$ \cite{Verresen21}. Using the above dictionary, this leads to the spin liquid phase indicated in Fig.~\ref{fig:phasediagram}. In Sec.~\ref{subsec:connectruby}, we have confirmed this reported spin liquid by adiabatically connecting it to the fixed-point dimer liquid.

\subsubsection{$H_\e$ as a spin-1/2 model on the ruby lattice}

If we add spin-flop terms which allow for dimer hoppings within the triangle, we obtain our $e$-anyon fluctuating model. More precisely, in a given triangle, we can equate $H_\textrm{off-diag} = \sum_{i \in \triangle} \sigma^x_i + \sum_{\langle i,j\rangle_\triangle} \sigma^x_i \sigma^x_j = \mathcal X_\e$, where this is to be understood as being projected into the blockaded Hilbert space; for this reason, we can equally well write $\mathcal X_\e = \sum_{i \in \triangle} \sigma^x_i + \sum_{\langle i,j\rangle_\triangle} \left( \sigma^+_i \sigma^-_j +h.c. \right)$.

To summarize, the $e$-anyon fluctuating model $H_\e$ can be written as a spin-1/2 model on the ruby lattice:
\begin{align}
H_\e &= 4J\sum_{\langle i,j\rangle_\textrm{\color{blue}blue}} \bigg(n_i- \frac{1}{4}\bigg)\bigg(n_j -\frac{1}{4}\bigg) \\
&\quad + h_t \sum_{\langle i,j\rangle_\textrm{red}} \left(\sigma^+_i \sigma^-_j +h.c.\right)+ h_t \sum_i \sigma^x_i, \label{eq:XXZ}
\end{align}
where we enforce the blockade condition of not having more than one particle per triangle (i.e., we actually have an additional term $U \sum_{\langle i,j\rangle_\textrm{red}} n_i n_j$ with large $U$). This is simply an XXZ model in a tilted field, with different values\footnote{Ref.~\onlinecite{Buerschaper14} considered an XXZ model with a longitudinal field where the couplings are the same on both types of bonds; numerical arguments were used to argue the emergence of a dimer liquid. In contrast, our model \eqref{eq:XXZ} has an exactly solvable limit.} of the anisotropy on the blue and red bonds. It is remarkable that this relatively simple Hamiltonian has an exactly-solvable dimer liquid for $h_t \to 0$, whose spin liquid phase extends all the way to $h_t \to + \infty$ where we obtain an effective Kitaev model \eqref{eq:Hquadrupolar} (see Sec.~\ref{subsec:strongmonomer}).

\subsubsection{$H_\f$ as a spin-1/2 model on the ruby lattice}

We can similarly obtain the model $H_\f$ with fluctuating $f$-anyons as a spin-1/2 model on the ruby lattice. In fact, it is the same\footnote{Note that one does not need to replace $\sigma^x_i$ by $\sigma^y_i$ since these can interchanged by a rotation generated by $\sigma^z_i$.} as Eq.~\eqref{eq:XXZ} with the only change that we insert a phase factor $i$ in front of the spin-flop term: $\sigma^+_i \sigma^-_j \to i \sigma^+_i \sigma^-_j$. The hardcore boson thus effectively hops in a magnetic field. This endows this term with a directionality. If we take it to be, say, clockwise on all triangles, we obtain the chiral $H_\f$ model, whereas choosing alternating orientations leads to the achiral model.

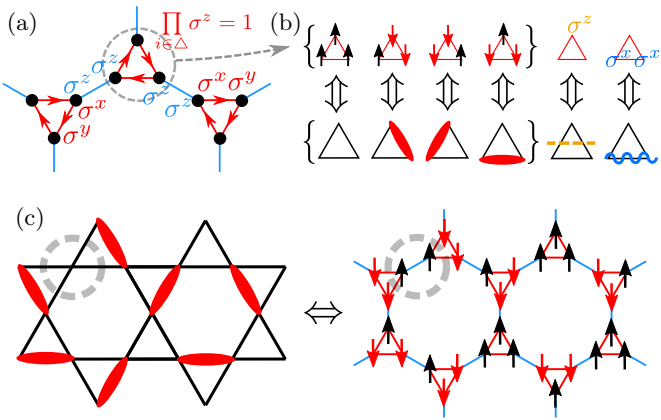
\begin{figure}
    \centering
    \begin{tikzpicture}
    \node at (0,0) {\includegraphics[scale=0.48]{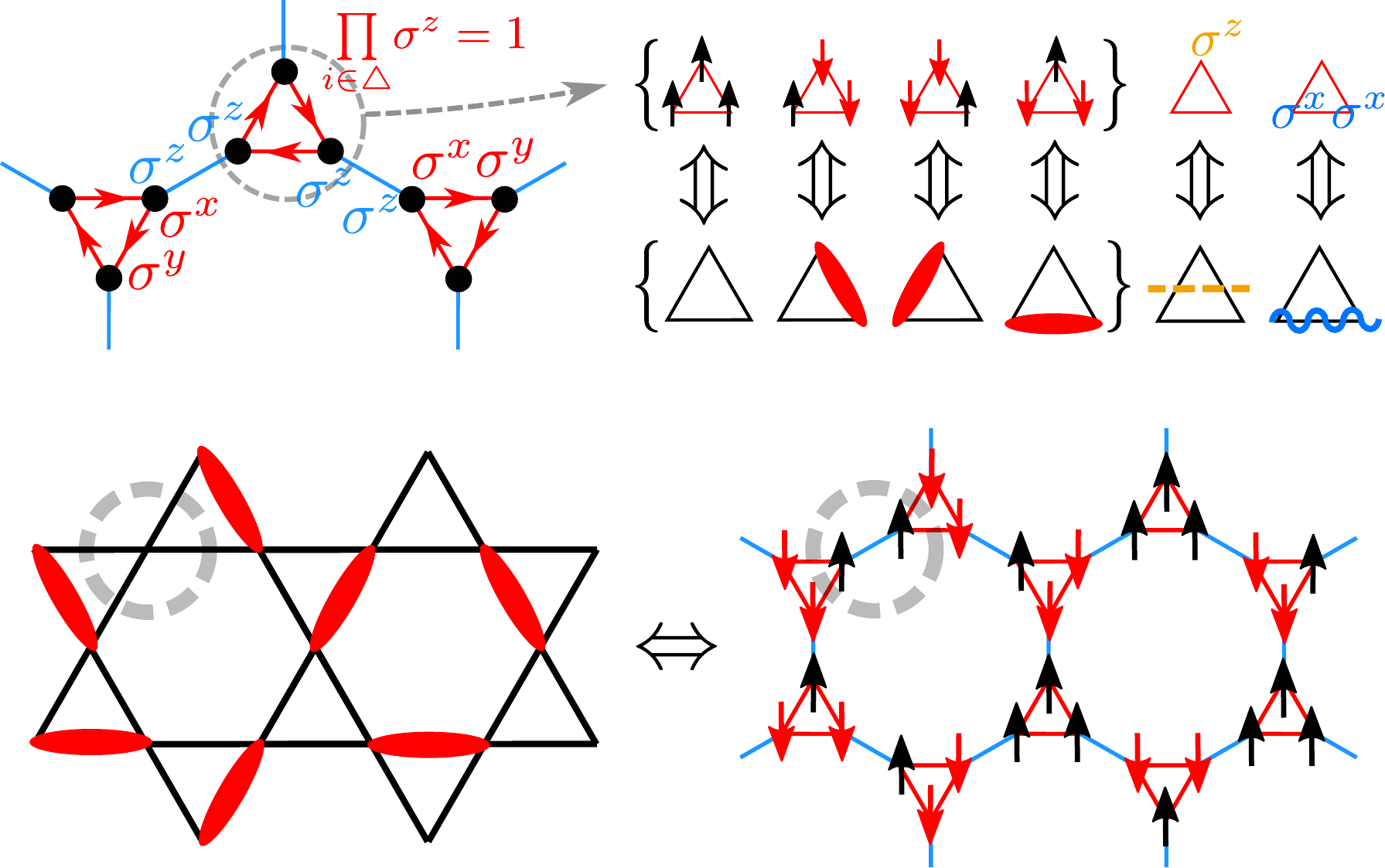}};
    \node at (-4.1,2.5) {(a)};
    \node at (-0.6,2.5) {(b)};
    \node at (-4,-0.1) {(c)};
    \end{tikzpicture}
    \caption{\textbf{Spin-1/2 model on star lattice gives effective spin-3/2 model on honeycomb lattice.} (a) The black dots form the star lattice (also known as the Fisher lattice). We only consider Hamiltonians which commute with $\prod_{i \in \triangle} \sigma^z_i$ on each triangle; e.g., $H_\f$ corresponds to the Yao-Kivelson model \cite{Yao07} with $\sigma^x \sigma^y$ couplings along the red bonds, and $\sigma^z \sigma^z$ on the blue bonds. (b) If we fix $\prod_{i \in \triangle} \sigma^z_i= 1$, we obtain an effective spin-3/2 model on the honeycomb lattice. We also show how our $\mathcal Z^\alpha$ \eqref{eq:Zdef} and $\mathcal X^\alpha$ \eqref{eq:Xdef} operators act on this spin-1/2 model. (c) Example of how a kagom\'e dimer state is mapped to a spin-1/2 model on the star lattice. Violations of the dimer constraint correspond to ferromagnetic blue bonds.}
    \label{fig:star}
\end{figure}

\subsection{Star lattice \label{subsec:star}}

In the previous subsection, the presence or absence of a dimer on the kagom\'e lattice was encoded in the presence or absence of a hardcore boson at the midpoint of the bond (forming the vertices of the ruby lattice). Another option for a local encoding is to represent the dimer by two hardcore bosons at the \emph{endpoints} of the dimer. We explore this here.

Consider the star lattice in Fig.~\ref{fig:star}(a). We will study (two-body) Hamiltonians where the ground state satisfies three-body operator condition $\prod_{i \in \triangle} \sigma^z_i = 1$ on \emph{each} triangle. In this sector, a triangle becomes an effective 4-state system, which we can identify with our spin-3/2 Hilbert space as shown in Fig.~\ref{fig:star}(b). Under this correspondence, the $\mathcal Z^\alpha$ and $\mathcal X^\alpha$ operators also have a simple form, as shown. This implies that the coupling $\mathcal Z^\alpha_i \mathcal Z^\alpha_j$ in the honeycomb model in Eq.~\eqref{eq:model} becomes an Ising coupling $\sigma^z_i \sigma^z_j$ for the corresponding bond of the star lattice. Similarly, it tells us that $\mathcal X_\e$ \eqref{eq:Xmono} is simply a sum of $\sigma^x_i \sigma^x_j$ around the triangle.

The $e$-anyon fluctuating model $H_\e$ is thus realized as follows in a spin-1/2 model on the star lattice:
\begin{equation}
H_\e = J \sum_{\langle i,j\rangle_\textrm{\color{blue}blue}} \sigma^z_i \sigma^z_j + h_t \sum_{\langle i,j\rangle_\textrm{red}} \sigma^x_i \sigma^x_j. \label{eq:starmono}
\end{equation}
Note that this indeed commutes with $\prod_{i \in \triangle} \sigma^z_i$. However, Eq.~\eqref{eq:starmono} is degenerate for all choices of $\prod_{i \in \triangle} \sigma^z_i = \pm 1$, which can be seen by noting that Eq.~\eqref{eq:starmono} commutes with $\sigma^x_i \sigma^x_j$ on blue bonds. One option is to simply fix the value of the sector to be unity for each triangle; for instance, this is a valid option if one prepares the ground state of Eq.~\eqref{eq:starmono} by adiabatic state preparation where one can start in a product state $\ket{\uparrow}^{\otimes N}$. Another option is, of course, to add $-\sum_\triangle \prod_{i \in \triangle} \sigma^z_i$ to the Hamiltonian, but this is a three-body operator and thus not physically plausible. Instead, it is more natural to perturb Eq.~\eqref{eq:starmono} with a small field $-\varepsilon \sum_{i} \sigma^z_i$. Since $H_\e$ is gapped, small values of $\varepsilon$ will not disturb the phase diagram, but third-order perturbation theory selects $\prod_{i \in \triangle} \sigma^z_i = 1$ (if $0<\varepsilon \ll |h_t|$).

Encoding the model $H_\f$ with fluctuating $f$-anyons corresponds to replacing $\sigma^x_i \sigma^x_j$ in Eq.~\eqref{eq:starmono} by $\sigma^x_i \sigma^y_j$. This model has been studied before: it is a known rewriting \cite{Dusuel08} of the Yao-Kivelson (YK) model \cite{Yao07}. In the YK model, one takes these $\sigma^x_i \sigma^y_j$ couplings to, say, run clockwise around each triangle; in Fig.~\ref{fig:star}(a) the arrow on the bond represents $\sigma^x \sigma^y$ with the first (second) Pauli matrix at the tail (head) or the arrow. Note that the Hamiltonian itself is not chiral: it is invariant under spinful time-reversal symmetry, and moreover one can exchange $\sigma^x \leftrightarrow \sigma^y$ by an on-site unitary transformation, such as conjugating the model by $e^{i \pi \sigma^z/4} \sigma^x$ on every site. However, the ground state is spontaneously chiral, and one of the two ground states fixes $\prod_{n \in \triangle} \sigma^z_n = 1$ everywhere \cite{Yao07}. Hence, in this case we do not need to add extra terms. Ref.~\onlinecite{Yao07} showed that when the red bonds are dominant, the ground state is in a non-Abelian spin liquid with chiral edge modes; when the blue bonds are dominant, one obtains a $\mathbb Z_2$ spin liquid. This follows from the free-fermion solution (Sec.~\ref{subsec:intermediatespinon}), where these are the weak- and strong-pairing phases, respectively. To the best of our knowledge, it had not yet been pointed out that the weak-field limit gives a solvable dimer liquid (Sec.~\ref{subsec:weakspinon}), and the strong-field limit reduces to the spin-1/2 Kitaev model (Sec.~\ref{subsec:strongspinon}).

Lastly, let us remark that the weak-field limit of Eq.~\eqref{eq:starmono} generates a plaquette term $\prod \sigma^x$ on the star lattice. Together with the Ising interaction $\sigma^z_i \sigma^z_j$ on the blue bonds as well as the triangle condition $\prod_{i \in \triangle} \sigma^z_i= 1$, these are exactly the three stabilizers one would write down for a toric code model on a \emph{decorated} honeycomb lattice, where we dress every bond with an additional vertex. In fact, this toric code model has been discussed before \cite{Schuch12}, where it was obtained through a tensor network representation of the Misguich-Serban-Pasquier dimer liquid \cite{Misguich02}. If we take this model and drive the bond terms to be infinitely strong ($J \to \infty$), we recover the honeycomb toric code model of Sec.~\ref{subsec:toric}.

\section{Conclusions and future directions \label{sec:conclusions}}
In summary, we have highlighted an extensive connection between kagom\'e dimer models, Kitaev honeycomb magnets, and the ruby lattice spin liquid. A central property we leverage is that loop operators generating dimer resonances can be defined for the kagom\'e dimer model without having to specify the underlying configuration. This property is special to lattices with corner sharing triangles and does not generalize to dimer models on the square or triangular lattices. In fact, this property persists when we admit violations of the dimer constraint.

We defined a parent Hamiltonian based on a 4-state model on each kagom\'e triangle (i.e., carrying one dimer or being empty). This set-up, which can equivalently be thought of as a spin-3/2 model on the honeycomb lattice, allows for a unified description of different models. Interestingly, this spin-3/2 model has only Ising-like (i.e., diagonal) interactions, and is perturbed by single-site transverse fields (with a generalization in Sec.~\ref{subsec:XYZ}). The choice of transverse field leads to different known and novel models which we explored in detail. (Moreover, we explicitly showed how these models can emerge from two-body spin-1/2 models on the ruby or star lattices.) The implications of this are manifold, of which we recount only a few. First, this sheds new light on the ruby Rydberg spin liquid, highlighting its connections to two exactly soluble spin liquids, and an adiabatic path connecting the exact dimer liquid to the ruby model further confirms the spin liquid phase in the latter. Second, we have gained new understanding of how local anyon fluctuations can stabilize deconfined phases of an emergent gauge theory---this viewpoint is sufficiently fundamental to naturally link together seemingly unrelated models. Third, we found a (nonlocal) duality of the transverse-field Ising model on the kagom\'e lattice to the generalized kagom\'e dimer model ($H_\e$). This duality is distinct from earlier approaches \cite{Nikolic05} and captures both the ferromagnetic and the frustrated antiferromagnetic Ising coupling. The latter case is known to be disordered even for infinitesimal transverse fields \cite{Moessner00}, which indicates a spin liquid state in the dual model. Interestingly this corresponds to a new $S=1$ quadrupolar Kitaev model that has not previously been discussed, which is in a \2 spin liquid. Fourth, we more generally observe the relation between emergent dimer constraints and an effective Kitaev interaction, including the $S=1/2$ Kitaev model for strong $f$-anyon fluctuations, confirming that this transmutation reflects a deep connection.

\begin{widetext}

\begin{table}[h]
\begin{center}
\begin{tabular}{c|c}
$d_\textrm{unit cell}$ & Models \\ \hline
$8^2$ & `grandparent' models on \hyperref[subsec:ruby]{ruby} and \hyperref[subsec:star]{star} lattices; \hyperref[subsec:star]{Yao-Kivelson model} \cite{Yao07}; \hyperref[subsec:star]{toric code} (decorated honeycomb) \cite{Schuch12} \\
$4^2$ & \hyperref[sec:modelham]{spin-3/2 parent model} \eqref{eq:model} (honeycomb lattice) = \hyperref[subsec:ruby]{ruby lattice with Rydberg blockaded triangles}\\
$3^2$ & \hyperref[subsec:strongmonomer]{spin-1 quadrupolar Kitaev honeycomb model} \eqref{eq:Hquadrupolar} \\
$\sqrt{2}^6=2^3$ & \hyperref[subsec:intermediatespinon]{free-fermion solution} of $H_\f$; \hyperref[subsec:monomerduality]{Ising model on kagom\'e lattice} \eqref{eq:dual} as dual of $H_\e$ \\ 
$2^2$ & \hyperref[subsec:strongspinon]{spin-1/2 Kitaev honeycomb model} \cite{Kitaev06} \\ $2$ & \hyperref[subsec:weakmonomer]{hard dimer model on kagom\'e lattice} \cite{Misguich02}; \hyperref[subsec:toric]{toric code gauge theory on honeycomb lattice}
\end{tabular}
\caption{\textbf{Overview of unified models.} The first column indicates the size of the Hilbert space per unit cell. } 
\end{center}
\end{table}

\end{widetext}

Let us briefly mention several promising future directions opened up by this work.

\emph{Extended phase diagram.} It will be worth exploring the rich global phase diagram as sketched schematically in Fig.~\ref{fig:schematic}, where transitions between different topological orders and symmetry enrichment thereof would be interesting for future work. Thus far, we have established all three axes  and the fact that they are all connected to the same fixed-point dimer liquid (with the caveat that the chiral model has the opposite sign of the plaquette stabilizer). Completing this picture should reveal a rich landscape of phases and transitions. 

\emph{3D generalizations.} Extensions to 3D lattices would be of great interest, given the existing 3D Kitaev models and materials \cite{Kimchi_Review,Analytis,Takayama_2015,Surendran} as well as frustrated 3D lattices. In fact, the hyperkagom\'e lattice \cite{Okamoto_2007} is a lattice of corner-sharing triangles, such that much of our framework directly carries over. For instance, we can define $H_\e$ on this 3D lattice and the perturbation theory calculations in Sec.~\ref{subsec:weakmonomer} give us an emergent fixed-point dimer liquid on the hyperkagom\'e lattice. Also our duality mapping carries through, as well as our effective $S=1$ quadrupolar Kitaev model in the large-field limit, which now lives on the hyperhoneycomb lattice.

\emph{Duality of frustrated Ising models.} We saw that the known paramagnetic phase of the transverse-field Ising model on the kagom\'e model \cite{Moessner00,Moessner01b,Nikolic05,Powalski13} gave rise to novel spin liquid model which only has two-body interactions. This raises the question of which other known paramagnets are hiding interesting new spin liquid models.  (In addition, the other duality mapping discussed in Appendix~\ref{app:duality} is able to map symmetry-breaking phases to spin liquids.)  Moreover, related to the above suggestion of 3D generalizations, it would be interesting to know whether the frustrated Ising model on the hyperkagom\'e lattice is also disordered. Perhaps an interesting first step would be to explore the same question on the `infinite-dimensional' Husimi cactus lattice \cite{Husimi50}.

\emph{Other topological orders.} Can our framework be expanded to include other types of spin liquids? Here a natural starting point is the double-semion (DS) topological order. For instance, Refs.~\onlinecite{Buerschaper14,Iqbal14,Qi15} showed that introducing certain phase factors in the dimer resonances of Ref.~\onlinecite{Misguich02} gives a dimer model for DS order; can we generate this effective resonance using local anyon fluctuations? Moreover, recent work \cite{Ellison22} has shown that using a 4-state system one can even write down a stabilizer model for DS order (rather than just a commuting projector model \cite{Buerschaper14,Iqbal14,Qi15}); is this also possible using our formalism?

\emph{Quadrupolar models.} While spin-1/2 models cannot form quadrupoles, spin-1 models can. Remarkably, in Sec.~\ref{subsec:strongmonomer} we found that a quadrupolar analogue of the spin-1 Kitaev model gives a more robust spin liquid. Perhaps such `quadrupolar brother models' could be interesting to explore more generally. Here it is worth recalling that we found a natural interpolation between spin and quadrupole operators (Sec.~\ref{subsec:kitaevpath}), and it would be very interesting to see whether this gives an adiabatic path to the spin-1 Kitaev model \cite{Baskaran08,Zhu20,Hickey20,Dong20,Koga20,Khait21,Lee21,Chen22,Bradley22}. Moreover, it has been reported that the spin-1 Kitaev model in a field gives rise to a gapless liquid \cite{Zhu20,Hickey20}; it is unclear whether this property also holds for the quadrupolar model.

\emph{Connection between geometric and exchange frustration.} A key finding of our work is the connection between dimer models and Kitaev interactions. In particular, one is morphed into the other by tuning a field. Can this be extended to other cases? Here a natural starting point is the study of other types of constrained models and their related deconfined phases \cite{Pankov07,Devakul18,Jandura20,You21}.

\emph{Solid-state realizations.} We have seen how various models with simple Ising interactions and transverse fields can have complex phase diagrams and exotic ground states. Future work understanding the lattices, magnetic ions and interactions that could result in solid state realizations of some of these models is a promising future direction \cite{Inosov_2018,wolf2000ising}. Although historically much attention has been focused on $SU(2)$ spin symmetric spin liquid candidate, our work here makes the case for studying low symmetry spin systems. While spin-ice and Kitaev magnets also fit into this larger direction, here we emphasize that the ability to tune magnetic fields along different directions, which correspond to distinct perturbations in the low symmetry limit, can be used to guide the material towards small islands of spin liquid phases in parameter space.  A second requirement is that the  magnetic exchange couplings be not much larger than the magnetic fields  one can  apply in the lab. These twin requirements should motivate a broader study of candidate magnetic materials with rare-earth ions \cite{wolf2000ising}. Interestingly, in certain instances where the magnetic moments have a non-Kramers origin, mechanical strain can play the role of magnetic field, introducing a new route to applying effective magnetic fields on quantum magnets \cite{Balents,FisherKivelson,Mourigal}. This is a promising mechanism to systematically explore in future work. A different direction is to investigate pathways to the physical realization \cite{Jackeli09} of the quadrupolar $S=1$ Kitaev model which is predicted here to be a spin liquid. Note, related theoretical proposals are either not in the $S=1$ space \cite{PhysRevB.102.075110}, or the proposed $S=1$ models employ spin rather than quadrupolar couplings \cite{PhysRevLett.123.037203}.

\emph{Cold-atom implementation.} The strong diagonal interactions in Rydberg atom arrays \cite{Browaeys_2020}, combined with the versatility of local off-diagonal fields and potentially hopping terms, provide an enticing direction for realizing some of our models and perturbations thereof. Indeed, $H_\textrm{ruby}$ is naturally realized in Rydberg atom arrays \cite{Verresen21,Semeghini21}; in Sec.~\ref{subsec:hadamard} we identified a novel way of measuring its off-diagonal string operators in future experimental realizations. Moreover, we established a closely-related model, $H_\e$, which has an exactly solvable limit and displays a very robust spin liquid; it would thus be interesting to implement this model as well. In fact, in Sec.~\ref{subsec:ruby} we explicitly rewrote $H_\e$ as a spin-1/2 XXZ model (in a tilted field) on the ruby lattice. Various cold-atom platforms are able to realize XXZ interactions \cite{Whitlock17,Circular18,Scholl21,Signoles21}, making this an exciting direction for further exploration.  Lastly, it would be worthwhile to explore to what extent the spin-3/2 parent models can be experimentally realized by encoding the 4-state qudit into Rydberg excited states of an atom \cite{Kruckenhauser22}. 

\vspace{7pt}

\emph{Note added.} When the present manuscript was being prepared for the arXiv, a pre-print appeared (arXiv:2205.13000v1 \cite{Tarabunga22}) which adiabatically connects the ruby lattice spin liquid proposed in Ref. \onlinecite{Verresen21} to a toric code model (involving the study of a model which is akin to our $H_\e$ for the case of unfrustrated $e$-anyon fluctuations, i.e., $h_t<0$). In addition, we have presented a complementary parent Hamiltonian approach which not only establishes such an adiabatic path (Sec.~\ref{subsec:connectruby}) but also connects to various other models (including $H_\f$, the frustrated $H_\e$, and their effective Kitaev model descriptions).

\begin{acknowledgements}
R.V. would like to thank Urban Seifert for enlightening discussions about the quadrupole model, Marcus Bintz, Nick G. Jones, Norbert Schuch and Nat Tantivasadakarn for insightful comments on the manuscript, and Steve Kivelson for the same in addition to encouragement at early stages of this project. DMRG simulations were performed using the TeNPy Library \cite{Hauschild18}.
R.V. is supported by the Harvard Quantum Initiative Postdoctoral Fellowship in Science and Engineering.
R.V. and A.V. are supported by the Simons Collaboration on Ultra-Quantum Matter, which is a grant from the Simons Foundation (651440, A.V.). 
\end{acknowledgements}

\bibliography{main_resubmitv2.bbl}

\onecolumngrid
\appendix

\begin{figure}[h]
    \centering
    \begin{tikzpicture}
    \node at (0,0) {\includegraphics[scale=0.6]{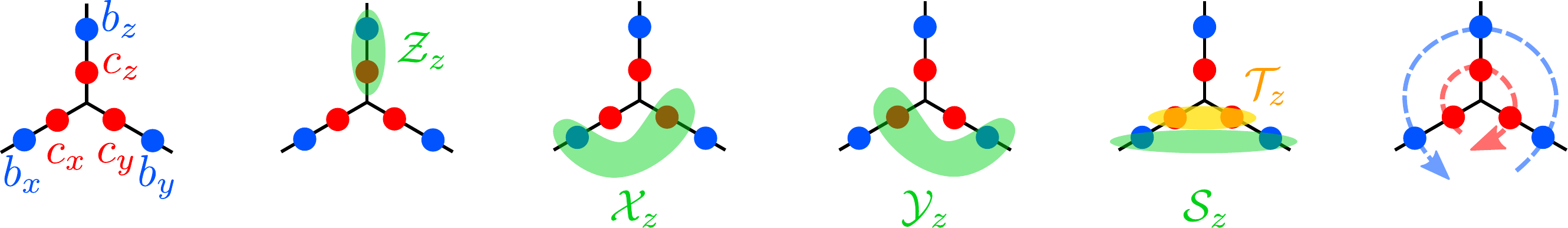}};
    \node at (-8,1) {(a)};
    \node at (-8+2.8,1) {(b)};
    \node at (-8+2*2.8,1) {(c)};
    \node at (-8+3*2.8+0.2,1) {(d)};
    \node at (-8+4*2.8+0.3,1) {(e)};
    \node at (-8+5*2.8+0.3,1) {(f)};
    \end{tikzpicture}
    \caption{(a) Six Majoranas with even parity condition $i b^x b^y b^z c^x c^y c^z = 1$ represent a 4-state system. (b) $\mathcal Z^z$ operator. (c) $\mathcal X^z$ operator. (d) $\mathcal Y^z$ operator. (e) Chiral $\mathcal S^z$ and $\mathcal T^z$ operators. (f) Using the $\mathcal S^\alpha$ and $\mathcal T^\alpha$ operators, one can generate the $\mathbb Z_3$ clocks which permute the operators; note that this transforms the operator $\mathcal Z^z$ in panel (a) to the $\mathcal X^z$ operator in panel (b)!}
    \label{fig:MajoranaHilbertspace}
\end{figure}
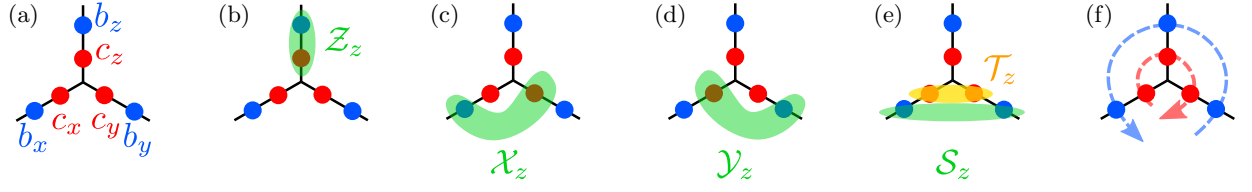

\section{Spin-3/2 Hilbert space \label{app:spinthreehalf}}

This appendix summarizes some of the key properties of our on-site Hilbert space. For completeness and ease of reference, it has some overlap with the main text.

\subsection{Algebra: Majoranas and bosonic operators}

We consider six Majorana operators with the condition $i b^x b^y b^z c^x c^y c^z = 1$. We define:
\begin{equation}
\boxed{\mathcal Z^\alpha = i b^\alpha c^\alpha, \quad
\mathcal X^x = i b^y c^z, \quad
\mathcal X^y = i b^z c^x, \quad
\mathcal X^z = i b^x c^y} \; .
\end{equation}
These six are the essential operators in the sense that all other operators we need to consider can be defined as products of these six. (In fact, we only need $\mathcal Z^x$, from which the other operators can be obtained by applying the two $\mathbb Z_3$ isomorphisms of Section~\ref{app:iso}.)

It is also useful to define
\begin{equation}
\boxed{\mathcal Y^\alpha = -\mathcal Z^\alpha \mathcal X^\alpha, \textrm{ i.e., } \mathcal Y^x = i b^z c^y, \quad \mathcal Y^y = i b^x c^z, \quad \mathcal Y^z = i b^y c^x} \; .
\end{equation}

Some useful relations:
\begin{equation}
\left( \mathcal X^\alpha \right)^2 = \left( \mathcal Y^\alpha \right)^2 = \left( \mathcal Z^\alpha \right)^2 = 1 = \mathcal X^x \mathcal X^y \mathcal X^z = \mathcal Y^x \mathcal Y^y \mathcal Y^z = \mathcal Z^x \mathcal Z^y \mathcal Z^z
\end{equation}
\begin{equation}
[\mathcal X^\alpha,\mathcal X^\beta] = [\mathcal Y^\alpha,\mathcal Y^\beta] = [\mathcal Z^\alpha,\mathcal Z^\beta] = 0
\end{equation}
\begin{equation}
\{ \mathcal X^\alpha, \mathcal Y^\beta \}_{(-1)^{\delta_{\alpha,\beta}}}
= \{ \mathcal Y^\alpha, \mathcal Z^\beta \}_{(-1)^{\delta_{\alpha,\beta}}}
= \{ \mathcal Z^\alpha, \mathcal X^\beta \}_{(-1)^{\delta_{\alpha,\beta}}} = 0
\end{equation}

Finally, we also introduce the $b$-pairing and $c$-pairing operators:
\begin{equation}
\boxed{\mathcal S^x = -i b^y b^z = i \mathcal Z^z \mathcal X^x, \quad
\mathcal S^y = - i b^z b^x = i \mathcal Z^x \mathcal X^y  , \quad
\mathcal S^z = - i b^x b^y = i \mathcal Z^y \mathcal X^z}
\end{equation}
and
\begin{equation}
\boxed{\mathcal T^x = -i c^y c^z = i \mathcal X^x \mathcal Z^y, \quad
\mathcal T^y = - i c^z c^x = i \mathcal X^y \mathcal Z^z , \quad
\mathcal T^z = - i c^x c^y = i \mathcal X^z  \mathcal Z^x } \; .
\end{equation}
These operators have different algebraic properties than the ones defined above. Indeed, they are effectively Pauli operators:
\begin{equation}
[\mathcal S^\alpha,\mathcal S^\beta] = 2i \varepsilon_{\alpha \beta \gamma} \mathcal S^\gamma \qquad \textrm{and} \qquad \{ \mathcal S^\alpha, \mathcal S^\beta \} = 2 \delta_{\alpha \beta} .
\end{equation}
Similarly for $\mathcal T^\alpha$. We observe that $[\mathcal S^\alpha,\mathcal T^\beta]=0$.

Expressing the kinetic terms/transverse fields encountered in the main text in terms of the above operators, we have:
\begin{align}
\mathcal X_\e &= \mathcal X^x + \mathcal X^y + \mathcal X^z \\
\mathcal X_\textrm{ruby} &= \mathcal X^x + \mathcal X^y + \mathcal X^z - \mathcal Y^x - \mathcal Y^y - \mathcal Y^z \\
\mathcal X_\f &= \left\{ \begin{array}{rl}
\mathcal T^x+\mathcal T^y+\mathcal T^z & \textrm{ on A sublattice, i.e., up-triangles;} \\
\mp \left( \mathcal S^x + \mathcal S^y + \mathcal S^z \right) & \textrm{ on B sublattice, i.e., down-triangles.}
\end{array} \right.
\end{align}
More precisely, the last line equals $\mathcal S^x + \mathcal S^y + \mathcal S^z$ for the chiral model and $-\left(\mathcal S^x + \mathcal S^y + \mathcal S^z\right)$ for the achiral model; see also Eqs.~\eqref{eq:XA} and \eqref{eq:XB}.

\subsection{$\mathbb Z_3$ isomorphisms \label{app:iso}}

It is instructive to consider the $\mathbb Z_3$ operators $R_b$ and $R_c$ which cycle through the Majoranas, i.e., $R_b: b^\alpha \to b^{\alpha+1}$ and $R_c: c^\alpha \to c^{\alpha+1}$. These are generated by $\frac{1}{2} \sum_\alpha \frac{\mathcal S^\alpha}{\sqrt{3}}$ and $\frac{1}{2} \sum_\alpha \frac{\mathcal T^\alpha}{\sqrt{3}}$, respectively. More explicitly, let
\begin{equation}
R_c = e^{\frac{2\pi i}{3} \frac{ \mathcal T^x + \mathcal T^y + \mathcal T^z }{2\sqrt{3}} } = \frac{1+c^y c^z + c^z c^x + c^x c^y}{2}.
\end{equation}
Then indeed:
\begin{align}
R_c^\dagger c^x R_c
&= \frac{c^x}{4} \left( 1 - c^y c^z + c^z c^x + c^x c^y \right) \left( 1+c^y c^z + c^z c^x + c^x c^y \right) \\
&= \frac{c^x}{4} \left( 2 c^z c^x + 2 c^x c^y - 2 c^y c^z \left( c^z c^x + c^x c^y \right)  \right) = \frac{c^x}{4} \left(4 c^x c^y\right) = c^y.
\end{align}
The effect of $R_c$ is thus to cycle:
\begin{equation}
\mathcal Z^\alpha = i b^\alpha c^\alpha \to  \mathcal X^{\alpha-1} \to \mathcal Y^{\alpha+1} \to \mathcal Z^\alpha,
\end{equation}
where $x+1=y$ et cetera; e.g., $\mathcal Z^z \to \mathcal X^y \to \mathcal Y^x \to \mathcal Z^z$.

The net effect of $R_b R_c$ is then to cycle:
\begin{equation}
\mathcal X^x \to \mathcal X^y \to \mathcal X^z \to \mathcal X^x
\end{equation}
and similarly for $\mathcal Y^\alpha, \mathcal Z^\alpha, \mathcal S^\alpha$ and $\mathcal T^\alpha$.

Finally, $R_b R_c^\dagger$ maps:
\begin{equation}
\mathcal Z^\alpha \to \mathcal X^\alpha \to \mathcal Y^\alpha \to \mathcal Z^\alpha.
\end{equation}

\subsection{\2 Hadamard \label{app:hadamard}}

We can also directly define a $\mathbb Z_2$ isomorphism which exchanges, say, $\mathcal Z^\alpha \leftrightarrow \mathcal X^\alpha$. This is generated by $\mathcal Y^\alpha$. Indeed, let us consider $U_H = e^{ i \frac{\pi}{4} \left(1 + \mathcal Y^x + \mathcal Y^y + \mathcal Y^z \right) }$ (observe that $U^2 = 1$). Then:
\begin{equation}
U_H \mathcal Z^\alpha U_H = \mathcal Z^\alpha e^{i\frac{\pi}{2} \left(1+\mathcal Y^\alpha\right)} = - \mathcal Z^\alpha \mathcal Y^\alpha = \mathcal X^\alpha,
\end{equation}
and similarly $U_H \mathcal X^\alpha U_H = \mathcal Z^\alpha$. Similarly one can of course swap $\mathcal X^\alpha \leftrightarrow \mathcal Y^\alpha$ by using $\mathcal Z^\alpha$ as a generator.

\subsection{Reduction to spin-$1/2$}

Suppose that for each $S=3/2$ site, we add a field in the Hamiltonian \begin{equation}
H_\textrm{field} = \lambda \sum_\alpha \mathcal T^\alpha = -i \lambda \left( c^x c^y + c^y c^z + c^z c^x \right).
\end{equation}
For large $\lambda$, this will pin two of the three modes. The one that remains is the one that commutes with $H_\textrm{field}$ (i.e., the `zero mode'). If we define $c := \frac{c^x + c^y + c^z}{\sqrt{3}}$, we see that
\begin{equation}
[c,H_\textrm{field}] = 0.
\end{equation}
To be more explicit, we can define the new basis $\{ c,c',c'' \}$ with $c := \frac{c^x + c^y + c^z}{\sqrt{3}}$, $c' = \frac{2 c^x - c^y - c^z}{\sqrt{6}}$ and $c'' = \frac{c^y - c^z}{\sqrt{2}}$. Then
\begin{equation}
c' c'' = \frac{1}{\sqrt{12}} \left( 2 c^x - c^y - c^z \right) \left( c^y - c^z \right) = \frac{1}{\sqrt{12}} \left( 2 c^x c^y + 2 c^z c^x + 2 c^y c^z \right),
\end{equation}
hence:
\begin{equation}
H_\textrm{field} = -i\lambda \sqrt{3} c' c''.
\end{equation}
For large $\lambda$, we pin $i c'c'' = 1$, but $c \propto c^x + c^y + c^z$ is unaffected! Moreover, one can confirm that $c^x c^y c^z = cc'c''$, such that the total parity condition becomes:
\begin{equation}
1 = i b^x b^y b^z c^x c^y c^z = i b^x b^y b^z c \underbrace{ c' c''}_{=-i} = b^x b^y b^z c.
\end{equation}

We have thus effectively collapsed the three `$c$-type' Majoranas into one, i.e., we now effectively have a total of four Majoranas ($b^x,b^y,b^z,c$) instead of six, recovering Kitaev's description of a spin-$1/2$! In particular,
\begin{equation}
\mathcal Z^\alpha \to \frac{ i b^\alpha c }{\sqrt{3}},
\end{equation}
such that we see that for this large field, the three commuting operators $\mathcal Z^\alpha$ can be identified with the (unnormalized) Pauli operators $\frac{ \sigma^\alpha }{\sqrt{3}}$. The prefactor is due to the fact that we first have to decompose $c^x = \frac{1}{\sqrt{3}} c + \sqrt{\frac{2}{3}} c'$ and then we project out $c'$.

We also see that in this limit, we can identify $\mathcal X^\alpha \to \mathcal Z^{\alpha+1}$ and $\mathcal Y^\alpha \to \mathcal Z^{\alpha-1}$. This is consistent with the $\mathbb Z_3$ isomorphism $R_c$ acting trivially on $c$. Hence, we see that e.g. $\mathcal X^x,\mathcal Y^x, \mathcal Z^x$ generate the Pauli algebra in the large-field limit. We can now also identify $\mathcal S^\alpha \to \sqrt{3}\mathcal Z^\alpha = \sigma^\alpha$.

\subsection{Matrix representation}

We set
\begin{equation}
\mathcal Z^x = \left( \begin{array}{cccc}
1 & 0 & 0 & 0 \\
0 & 1 & 0 & 0 \\
0 & 0 & -1 & 0 \\
0 & 0 & 0 & -1
\end{array} \right), \; \;
\mathcal Z^y = \left( \begin{array}{cccc}
1 & 0 & 0 & 0 \\
0 & -1 & 0 & 0 \\
0 & 0 & 1 & 0 \\
0 & 0 & 0 & -1
\end{array} \right), \; \; \mathcal Z^z = \left( \begin{array}{cccc}
1 & 0 & 0 & 0 \\
0 & -1 & 0 & 0 \\
0 & 0 & -1 & 0 \\
0 & 0 & 0 & 1
\end{array} \right)
\end{equation}
and
\begin{equation}
\mathcal X^x = \left( \begin{array}{cccc}
0 & 1 & 0 & 0 \\
1 & 0 & 0 & 0 \\
0 & 0 & 0 & 1 \\
0 & 0 & 1 & 0
\end{array} \right), \; \;
\mathcal X^y = \left( \begin{array}{cccc}
0 & 0 & 1 & 0 \\
0 & 0 & 0 & 1 \\
1 & 0 & 0 & 0 \\
0 & 1 & 0 & 0
\end{array} \right), \; \; \mathcal X^z = \left( \begin{array}{cccc}
0 & 0 & 0 & 1 \\
0 & 0 & 1 & 0 \\
0 & 1 & 0 & 0 \\
1 & 0 & 0 & 0
\end{array} \right),
\end{equation}
which indeed obey the above algebra.

From this we derive:
\begin{equation}
\mathcal Y^x = \left( \begin{array}{cccc}
0 & -1 & 0 & 0 \\
-1 & 0 & 0 & 0 \\
0 & 0 & 0 & 1 \\
0 & 0 & 1 & 0
\end{array} \right), \; \;
\mathcal Y^y = \left( \begin{array}{cccc}
0 & 0 & -1 & 0 \\
0 & 0 & 0 & 1 \\
-1 & 0 & 0 & 0 \\
0 & 1 & 0 & 0
\end{array} \right), \; \; \mathcal Y^z = \left( \begin{array}{cccc}
0 & 0 & 0 & -1 \\
0 & 0 & 1 & 0 \\
0 & 1 & 0 & 0 \\
-1 & 0 & 0 & 0
\end{array} \right)
\end{equation}
and
\begin{equation}
\mathcal S^x = \left( \begin{array}{cccc}
0 & i & 0 & 0 \\
-i & 0 & 0 & 0 \\
0 & 0 & 0 & -i \\
0 & 0 & i & 0
\end{array} \right), \; \;
\mathcal S^y = \left( \begin{array}{cccc}
0 & 0 & i & 0 \\
0 & 0 & 0 & i \\
-i & 0 & 0 & 0 \\
0 & -i & 0 & 0
\end{array} \right), \; \; \mathcal S^z = \left( \begin{array}{cccc}
0 & 0 & 0 & i \\
0 & 0 & -i & 0 \\
0 & i & 0 & 0 \\
-i & 0 & 0 & 0
\end{array} \right)
\end{equation}
and
\begin{equation}
\mathcal T^x = \left( \begin{array}{cccc}
0 & -i & 0 & 0 \\
i & 0 & 0 & 0 \\
0 & 0 & 0 & -i \\
0 & 0 & i & 0
\end{array} \right), \; \;
\mathcal T^y = \left( \begin{array}{cccc}
0 & 0 & -i & 0 \\
0 & 0 & 0 & i \\
i & 0 & 0 & 0 \\
0 & -i & 0 & 0
\end{array} \right), \; \; \mathcal T^z = \left( \begin{array}{cccc}
0 & 0 & 0 & -i \\
0 & 0 & -i & 0 \\
0 & i & 0 & 0 \\
i & 0 & 0 & 0
\end{array} \right)
\end{equation}

\subsection{Connection to Rydberg atom array realization}

There are only three operators which are available in the Rydberg atom array context, corresponding to the density and the projected Pauli-$X$ and Pauli-$Y$ operators:
\begin{align}
\frac{1}{4} \sum_\alpha \left( 1 - \mathcal Z^\alpha\right)
&= \left( \begin{array}{cccc}
0 & 0 & 0 & 0 \\
0 & 1 & 0 & 0 \\
0 & 0 & 1 & 0 \\
0 & 0 & 0 & 1
\end{array} \right) = `n'
\\
\frac{1}{2} \sum_\alpha \left( \mathcal X^\alpha - \mathcal Y^\alpha \right) &= \left( \begin{array}{cccc}
0 & 1 & 1 & 1 \\
1 & 0 & 0 & 0 \\
1 & 0 & 0 & 0 \\
1 & 0 & 0 & 0
\end{array} \right) = `PXP'
\\
\frac{1}{2}\sum_\alpha \left( \mathcal T^\alpha - \mathcal S^\alpha \right) &= \left( \begin{array}{cccc}
0 & -i & -i & -i \\
i & 0 & 0 & 0 \\
i & 0 & 0 & 0 \\
i & 0 & 0 & 0
\end{array} \right) = `PYP' = U^\dagger \left(PXP\right) U \textrm{ with } U = e^{-i \frac{\pi}{2} n}
\end{align}
Fortunately, the latter is enough to generate the $\mathbb Z_3$ isomorphism, such that we can rotate the offdiagonal $\mathcal X^\alpha$ operators into diagonals $\mathcal Z^\alpha$. In fact, using $n$ we can also generate the $\mathbb Z_2$ isomorphism which swaps $\mathcal X^\alpha \leftrightarrow \mathcal Y^\alpha$. Sandwiching this with the $\mathbb Z_3$ isomorphism thus allows us to implement the $\mathbb Z_2$ Hadamard $\mathcal Z^\alpha \leftrightarrow \mathcal X^\alpha$.

We can be more explicit: define
\begin{equation}
R = e^{- \frac{2\pi i}{3} \frac{PYP}{\sqrt{3}} } = e^{i \frac{\pi}{2} n} e^{- \frac{2\pi i}{3} \frac{PXP}{\sqrt{3}} } e^{-i \frac{\pi}{2} n} = -\frac{1}{2} \left( \begin{array}{rrrr}
1 & 1 & 1 & 1 \\
-1 & -1 & 1 & 1 \\
-1 & 1 & -1 & 1 \\
-1 & 1 & 1 & -1
\end{array} \right),
\end{equation}
which nicely agrees with Eq.~(D4) in the appendix of Ref.~\onlinecite{Verresen21}.
Note that $R^3 = 1$. Moreover:
\begin{equation}
R^\dagger \mathcal Z^\alpha R = \mathcal X^\alpha, \qquad R^\dagger \mathcal X^\alpha R = \mathcal Y^\alpha, \qquad R^\dagger \mathcal Y^\alpha R = \mathcal Z^\alpha.
\end{equation}
The $\mathbb Z_2$ operator which toggles $\mathcal X^\alpha \leftrightarrow \mathcal Y^\alpha$ is given by
\begin{equation}
S = e^{-i\pi n} = \left( \begin{array}{cccc}
1 & 0 & 0 & 0 \\
0 & -1 & 0 & 0 \\
0 & 0 & -1 & 0 \\
0 & 0 & 0 & -1
\end{array} \right).
\end{equation}
Finally, then, we obtain the Hadamard transformation:
\begin{equation}
\boxed{ U_H } = SR = \boxed{ e^{-i \frac{\pi}{2} n} e^{- \frac{2\pi i}{3} \frac{PXP}{\sqrt{3}} } e^{-i \frac{\pi}{2} n} } \; = -\frac{1}{2} \left( \begin{array}{rrrr}
1 & 1 & 1 & 1 \\
1 & 1 & -1 & -1 \\
1 & -1 & 1 & -1 \\
1 & -1 & -1 & 1
\end{array} \right) .
\end{equation}
Indeed:
\begin{align}
\boxed{ U_H^\dagger \mathcal Z^\alpha U_H } &= R^\dagger (S\mathcal Z^\alpha S) R = R^\dagger \mathcal Z^\alpha R = \boxed{ \mathcal X^\alpha } \;, \\
\boxed{ U_H^\dagger \mathcal X^\alpha U_H } &= R^\dagger( S\mathcal X^\alpha S)R = R^\dagger \mathcal Y^\alpha R = \boxed{ \mathcal Z^\alpha} \; .
\end{align}

\section{Kramers-Wannier dual descriptions of the spin-3/2 Hilbert space \label{app:duality}}

Here we explicitly state two types of dualities that were used in the main text to solve certain models. We note that these dualities are general maps which can be applied to other models of interest. Both dualities can effectively be interpreted as gauging certain symmetries, as we discuss.

\subsection{Gauging a $\mathbb Z_2$ symmetry in a spin-1/2 kagom\'e lattice model}

Suppose one has a spin-3/2 model, as in the main text, with the following conserved plaquette operator:
\begin{equation}
W_p = \prod_{\alpha \in p} \mathcal X^\alpha = \raisebox{-18pt}{\includegraphics[scale=0.6]{Wp.pdf}}. 
\end{equation}
If one is in the sector with $W_p = 1$ for all plaquettes, then one can define effective spin-1/2 operators for a Hilbert space of qubits on the vertices of the kagom\'e lattice as follows (where we use the graphical notation introduced in Sec.~\ref{sec:model} to denote $\mathcal Z^\alpha$ and $\mathcal X^\alpha$):
\begin{equation}
\raisebox{-25pt}{\includegraphics[scale=0.25]{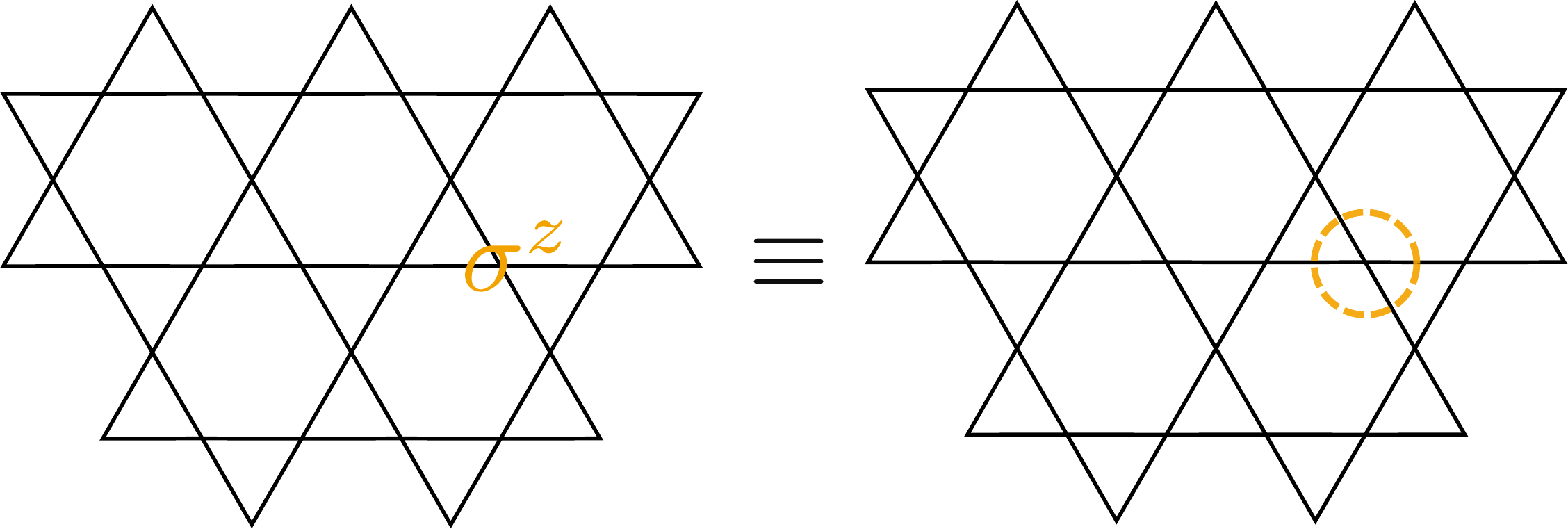}}
\quad \textrm{and} \quad \raisebox{-25pt}{\includegraphics[scale=0.25]{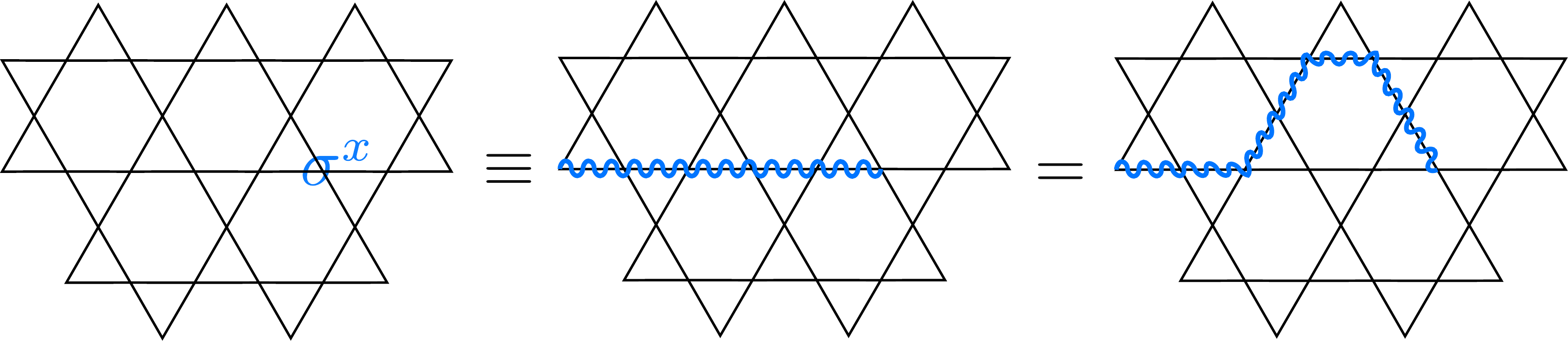}} .
\end{equation}
Here the $\sigma^x$ operator is defined by a string operator in the original variables; we take it to trail off to infinity, ignoring boundary effects (as is customary for nonlocal duality mappings). Note that $\sigma^x$ is independent of the location of the string operator defining it, since it can be freely moved using the $W_p=1$ condition; only the endpoint is physical. It is straightforward to see that the above $\sigma^z$ and $\sigma^x$ operators satisfy the desired Pauli algebra.

Using this mapping, a local spin-3/2 model with the local condition $W_p=1$ can be (nonlocally) rewritten as a local spin-1/2 model on the kagom\'e lattice. In reverse, any spin-1/2 model on the kagom\'e lattice with global Ising symmetry ($\prod \sigma^z$) can be dualized using our spin-3/2 variables. One can interpret this as a convenient way of gauging a global Ising symmetry on the kagom\'e lattice. Correspondingly, a paramagnetic phase of the spin-1/2 kagom\'e model becomes a $\mathbb Z_2$ spin liquid in the spin-3/2 variables. This is the mapping we used in Sec.~\ref{subsec:monomerduality} when we rewrote $H_\e$ as a spin-1/2 transverse-field Ising model on the kagom\'e lattice (where we furthermore performed a Hadamard transformation $\sigma^x \leftrightarrow \sigma^z$ such that the Ising interaction was in the diagonal basis).

We note that a remarkable property of the above duality is that whereas usually gauging an Ising symmetry leads to multi-body interactions (such as the famous toric code Hamiltonian \cite{Kitaev_2003}), the above spin-3/2 variables can lead to just two-body interactions, as was the case for $H_\e$ in the main text.

\subsection{Gauging a $\mathbb Z_2 \times \mathbb Z_2$ symmetry in a spin-1/2 honeycomb lattice model}

Here we consider the case where every plaquette has \emph{two} conserved plaquette operators:
\begin{equation}
W_p = \prod_{\alpha \in p} \mathcal X^\alpha = \raisebox{-18pt}{\includegraphics[scale=0.6]{Wp.pdf}} \qquad \textrm{and} \qquad \tilde G_p = \prod_{v \in p} G_v = \raisebox{-20pt}{\includegraphics[scale=0.3]{newplaquettev2.pdf}}.
\end{equation}
This occurs, for instance, in the `$XYZ$'-type model discussed in Sec.~\ref{subsec:XYZ} of the main text. If we are in the sector $W_p = 1 = \tilde G_p$, then these local conserved quantities allow us to define the following effective spin-$1/2$ operators on the honeycomb lattice:
\begin{align}
&\raisebox{-40pt}{\includegraphics[scale=0.3]{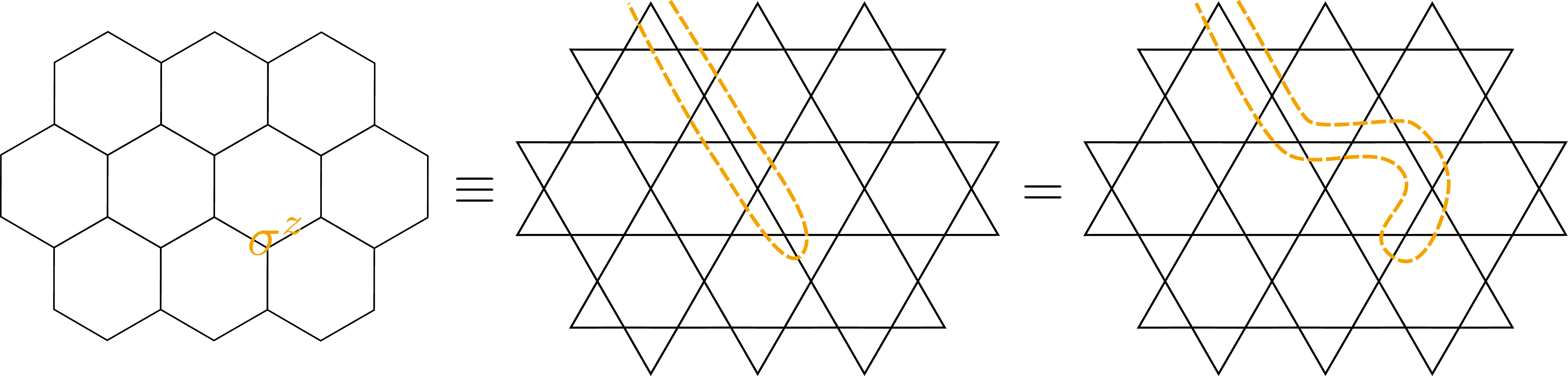}} \\
&\raisebox{-40pt}{\includegraphics[scale=0.3]{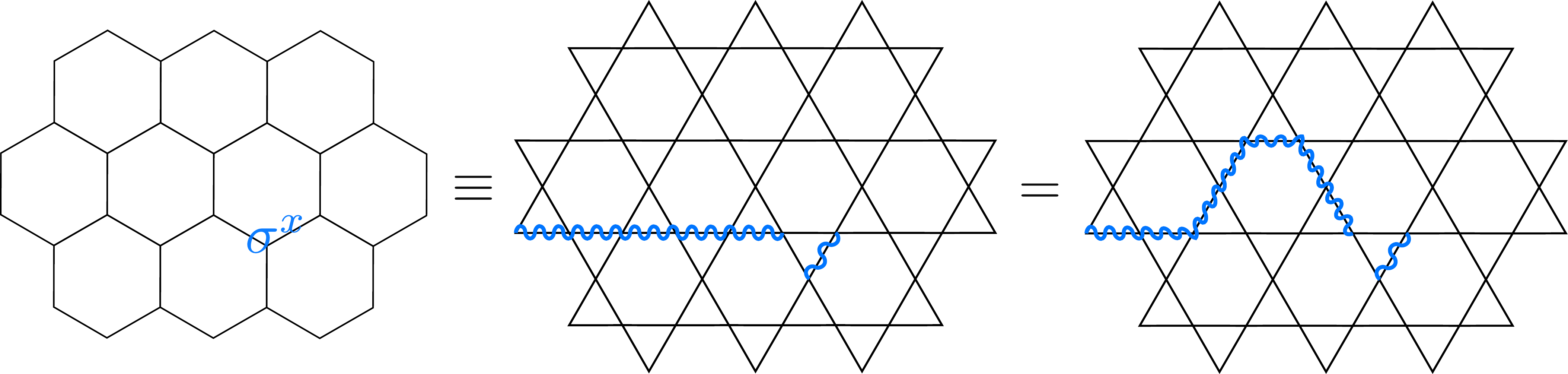}}
\end{align}
These string operators (made up out of $\mathcal Z^\alpha$ and $\mathcal X^\alpha$) run off to infinity---ignoring boundary effects. Crucially, the precise location of the string is unphysical: using the local conditions $W_p=1$ and $\tilde G_p=1$, the strings can be moved arbitrarily, as illustrated above. Only the location of the endpoint is physical. It is straightforward to observe that these operators indeed satisfy the Pauli algebra. 

For these nonlocal variables, one can derive the following identity for nearest-neighbors $\langle i,j \rangle$ on a bond of type $\alpha \in \{x,y,z\}$ on the honeycomb lattice:
\begin{equation}
\sigma^x_i \sigma^x_j = \mathcal X^\alpha_i \mathcal X^\alpha_j, \qquad 
\sigma^y_i \sigma^y_j = -\mathcal Y^\alpha_i \mathcal Y^\alpha_j, \qquad 
\sigma^z_i \sigma^z_j = \mathcal Z^\alpha_i \mathcal Z^\alpha_j.
\end{equation}
Here, $\mathcal Y^\alpha = - \mathcal Z^\alpha X^\alpha$, as described in more detail in Appendix~\ref{app:spinthreehalf}. After a unitary sublattice transformation, this gives the dictionary claimed in Eq.~\eqref{eq:Z2Z2dictionary} of the main text.

One can interpret the above mapping as gauging two non-anomalous 1-form symmetries, to obtain an effective qubit model with two global symmetries, $\prod \sigma^z_i$ and $\prod \sigma^x_i$. In reverse, the above gives a prescription for how to gauge a global $\mathbb Z_2 \times \mathbb Z_2$ symmetry for qubits on the honeycomb lattice, leading to an effective 4-level qudit Hamiltonian on the honeycomb lattice.

\section{Spin-1 quadrupole operators \label{app:quadruopole}}

Let us define:
\begin{align}
\mathcal S^x &= \cos\left(\frac{\theta}{3} \right) S^x + \sin\left(\frac{\theta}{3}\right) \{ S^y,S^z \}, \label{eq:deformedx} \\
\mathcal S^y &= \cos\left(\frac{\theta}{3} \right) S^y + \sin\left(\frac{\theta}{3}\right) \{ S^z,S^x \}, \label{eq:deformedy} \\
\mathcal S^z &= \cos\left(\frac{\theta}{3} \right) S^z + \sin\left(\frac{\theta}{3}\right) \{ S^x,S^y \}. \label{eq:deformedz}
\end{align}

These satisfy the $q$-deformed commutator relation:
\begin{equation}
[\mathcal S^x,\mathcal S^y]_q = i \mathcal S^z \quad \textrm{where } [A,B]_q = qAB-q^{-1} BA \quad \textrm{with } q=e^{i\theta}
\end{equation}
(and similarly for cyclic permutations). For $q=1$, we have that $[A,B]_1 = [A,B]$, in which case the above is the usual (spin-1) relation $[\mathcal S^x,\mathcal S^y]=i \mathcal S^z$, whereas for $q=i$ we have $[A,B]_{i} = i\{A,B\}$, such that for $\theta=\pi/2$ we have that $\{\mathcal S^x, \mathcal S^y\} = \mathcal S^z$. For $\theta=\pi$ we have $q=-1$ such that $[A,B]_{-1} = -[A,B]$, and indeed, one can show that in this case $\mathcal S^\alpha$ is unitarily equivalent to $-S^\alpha$. Hence, the operators for $\theta=2\pi$ are unitarily equivalent to those at $\theta=0$. 
In fact, using the unitary operator
\begin{equation}
U = \exp\left(- i \frac{\pi}{4} S^z \right) \exp\left(i \frac{\theta}{3} \{S^x,S^y\} \right) \exp\left(i \frac{\pi}{4} S^z \right) = \exp\left(i \frac{\theta}{3} \left[ \left( S^y\right)^2 - \left( S^x\right)^2\right] \right) =  \exp\left(-i \frac{\theta}{3} Q^{x^2-y^2} \right),
\end{equation}
the above interpolation is unitary equivalent to:
\begin{equation}
\mathcal S^x = S^x, \quad \mathcal S^y = S^y, \quad \textrm{and} \quad \mathcal S^z = \cos(\theta) S^z + \sin(\theta) \{S^x,S^y\} = \cos(\theta) S^z + \sin(\theta) Q^{xy}. \label{eq:deformedv2}
\end{equation}

This shows that the above interpolation is really only a function of $\theta \mod 2\pi$ (up to unitary transformations). For instance, $(S^x,S^y,Q^{xy})$ (Eq.~\eqref{eq:deformedv2} with $\theta=\pi/2$) is unitarily equivalent to
\begin{equation}
\frac{1}{4}\left( \sqrt{3} S^x + Q^{yz} , \sqrt{3} S^y + Q^{xz} , \sqrt{3} S^z + Q^{xy} \right),
\end{equation}
which is Eqs.~\eqref{eq:deformedx}--\eqref{eq:deformedz} with $\theta=\pi/2$, which in turn is unitarily equivalent to the case with $\theta=\pi/2-2\pi = -3\pi/2$, i.e., $-(Q^{yz},Q^{xz},Q^{xy})$.

Along the whole interpolation, we note that $e^{i\pi \mathcal S^\alpha} = e^{i\pi S^\alpha}$, i.e., the $\pi$-rotations are independent of $\theta$.

Finally, let us remark that the above interpolation (parametrized by $\theta$) relates to the one in Sec.~\ref{subsec:kitaevpath} (parametrized by $\varphi$) by $\varphi = 2 \theta$. Hence, the operators in Sec.~\ref{subsec:kitaevpath} only depend on $\varphi \mod 4\pi$; in fact, since the Hamiltonian in Eq.~\eqref{eq:spin1interpolation} only involves products of two operators, it is insensitive to signs, such that it only depends on $\varphi \mod 2\pi$.

\end{document}